\documentclass[12pt]{article}
\usepackage{amsmath,amsfonts,color,amsthm,amssymb,euscript,epsf,epsfig,color,graphicx,bm}
\usepackage{array}
\usepackage{fancybox}
\usepackage{physics}

\usepackage{graphicx}
\usepackage{etoolbox}
\graphicspath{{figures/}}
\usepackage{diagbox}

\usepackage{amsmath,amsfonts,amssymb,amsthm,color,epsfig,graphicx,hyperref,url}

\usepackage{booktabs}
\usepackage{longtable}
\usepackage{array}
\usepackage{multirow}
\usepackage{wrapfig}
\usepackage{float}
\usepackage{caption, subcaption}
\usepackage{colortbl}
\usepackage{tabu}
\usepackage{threeparttable}
\usepackage{threeparttablex}
\usepackage[normalem]{ulem}
\usepackage{makecell}
\usepackage{mathtools}
\usepackage{afterpage}
\usepackage{adjustbox}
\usepackage{longtable}
\usepackage[T1]{fontenc}

\usepackage{comment} 
\usepackage[utf8]{inputenc}
\usepackage[english]{babel}
\usepackage{hyperref}
\usepackage{url} 

\usepackage[svgnames]{xcolor}
\usepackage{tikz}
\usetikzlibrary{calc,decorations.markings, shapes.misc, decorations.pathmorphing}
\tikzset{cross/.style={cross out, draw, 
		minimum size=2*(#1-\pgflinewidth), 
		inner sep=0pt, outer sep=0pt}} 

\definecolor{lstbgcolor}{rgb}{0.9,0.9,0.9} 
\usepackage{fancyvrb}
\definecolor{GREEN}{rgb}{0.0,0.70,0.24}
\definecolor{BLUE}{rgb}{0.0,0.24,0.70}

\newcommand{\graphscale}{1.4}
\newcommand{\graphscaleb}{1.4}

\newcommand{\be}{\begin{equation}}
\newcommand{\ee}{\end{equation}}
\newcommand{\bea}{\begin{eqnarray}\displaystyle}
\newcommand{\eea}{\end{eqnarray}}

\makeatletter
\@addtoreset{equation}{section}
\makeatother
\renewcommand{\theequation}{\thesection.\arabic{equation}}
\def\one{{\hbox{ 1\kern-.8mm l}}}
\def\zero{{\hbox{ 0\kern-1.5mm 0}}}

  \def\cO{{\cal O}}
  
\def\cS{{\cal S}}  
  
 \def\cZ{{\cal Z}}

\begin{document}

\begin{flushright}
	QMUL-PH-23-08
\end{flushright}

\makeatletter
\@addtoreset{equation}{section}
\makeatother
\renewcommand{\theequation}{\thesection.\arabic{equation}}


\makeatletter
\@addtoreset{equation}{section}
\makeatother
\renewcommand{\theequation}{\thesection.\arabic{equation}}
\def\ls[#1]{ {}_{#1}}

\def\wmu{ \widetilde{\mu} } 
\def\wmuo{ \widetilde{\mu}_{ \ls[1] } }
\def\wmut{  \widetilde{\mu}_{ \ls[2] } }
\def\cS{ \mathcal{S} }

{\LARGE{ 
\centerline{\bf  Permutation invariant Gaussian matrix models  }
\centerline{ \bf for financial correlation matrices   } 
}}  

\vskip.5cm 

\thispagestyle{empty} \centerline{
    {\large \bf George Barnes ${}^{a,}$\footnote{ {\tt g.barnes@qmul.ac.uk}}, }
   {\large \bf  Sanjaye Ramgoolam${}^{a,b,}$\footnote{ {\tt s.ramgoolam@qmul.ac.uk}},    }
               {\large \bf  Michael Stephanou${}^{c,  }$\footnote{ {\tt michael.stephanou@gmail.com }}    }
                                                       }

\vspace{.4cm}
\centerline{{\it ${}^a$ Centre for Theoretical Physics, School of Physical and Chemical Sciences, }}
\centerline{{\it Queen Mary University of London, Mile End Road, London E1 4NS, UK  }}

    \vspace{.2cm}
\centerline{{\it ${}^b$ National Institute for Theoretical Physics, }}
\centerline{{\it School of Physics and Centre for Theoretical Physics, } }
\centerline{{\it University of the Witwatersrand, Wits, 2050, South Africa } }

\vspace{.2cm}
\centerline{{\it ${}^c$ Rand Merchant Bank, }}
\centerline{{\it 1 Merchant Place, Fredman Drive, Johannesburg, 2196, South Africa }}

\vspace{.5truecm}

\thispagestyle{empty}

\centerline{\bf ABSTRACT}

\vskip.2cm 
\noindent
We construct an ensemble of correlation matrices from high-frequency foreign exchange market data, with one matrix for every day for 446 days.  The matrices are symmetric and have vanishing diagonal elements after subtracting the identity matrix. 
For this case, we construct the general permutation invariant Gaussian matrix model, which has 4 parameters characterised using the  representation theory of symmetric groups. 
The permutation invariant polynomial functions of the symmetric, diagonally vanishing matrices have a basis labelled by undirected loop-less graphs. Using the expectation values of the general linear and quadratic permutation invariant functions of the matrices in the dataset, the 4 parameters of the matrix  model are determined. The model then predicts the expectation values of the cubic and quartic polynomials. These predictions are compared to the data to give strong evidence for a good overall fit of the  permutation invariant Gaussian matrix model. The linear, quadratic, cubic and quartic polynomial functions are then used to define low-dimensional feature vectors for the days associated to the matrices. These vectors, with choices informed by the refined structure of small non-Gaussianities, are found to be effective as a tool for anomaly detection in market states:  statistically significant correlations are established between atypical days as defined using these feature vectors, and days with significant economic events as recognized in standard foreign exchange economic calendars. They are also shown to be useful as a tool for ranking pairs of days in terms of their similarity, yielding a strongly statistically significant correlation with a ranking based on a higher dimensional proxy for visual similarity.

\setcounter{page}{0}
\setcounter{tocdepth}{2}

\newpage

\tableofcontents

\section{ Introduction  } 

Permutation invariant Gaussian matrix models have recently been introduced \cite{ramgoolam2019permutation}, with motivations coming from the study of the statistics of ensembles of matrices arising in computational linguistics \cite{Kartsaklis:2017lfq}.  They have been applied to show the existence of approximate Gaussianity in these ensembles \cite{Huber:2022ohf,Ramgoolam:2019ldg}. Permutation invariant polynomials of matrix variables, which are the key observables in the study of Gaussianity, have also been used as tools for lexical semantic tasks in computational linguistics \cite{Ramgoolam:2019ldg}. 

In this paper, we construct and analyze financial correlation matrices for a sequence of days obtained by calculating correlations between price movements in high frequency foreign exchange market data. This is an interesting study in itself and adds to a rather sparse literature on high frequency forex correlation matrices (\cite{gkebarowski2019detecting} is one other recent example). The correlation matrices are symmetric and have vanishing matrix elements along the diagonal (by subtracting the identity matrix). We also briefly note the positive-semidefinite property of these correlation matrices which we do not explicitly model. The permutation invariant matrix model constructed in \cite{ramgoolam2019permutation} uses an integration over general matrices and has $13$ parameters, 2 are coefficients  in the action of linear invariants while $11$  are coefficients of quadratic invariants. For the case of the restricted matrices here, there is a reduction of the $13$-parameter model to a $4$-parameter model.  In this paper we explicitly construct the $4$-parameter model and use it to demonstrate approximate permutation invariant Gaussianity in the ensemble of forex correlation matrices we construct.

The permutation invariant polynomial functions of generic matrices of size $D$ and degree $k$ have a basis labelled by directed graphs with $k$ edges and any number of nodes, when $ D \ge 2k $. This condition is satisfied for the cases of interest in this paper since we consider $k$ up to $8$ and $D =19$. The nodes of the graph correspond to  indices being summed and the edges of the graph correspond to the matrix variables. In the case of permutation invariant polynomial functions of symmetric, diagonally vanishing matrices of degree $k$, there exists a basis labelled by undirected graphs containing no loops (i.e. no edges starting and ending at the same node). See section 3.4 for an explanation of this connection between graphs and polynomial  invariant functions, as well as examples of the  loopless graphs and the corresponding polynomials.

In part the motivation from physics for applying the permutation invariant Gaussian matrix (PIGM) models is to establish universality properties of these permutation invariant Gaussianities in diverse matrix ensembles in data sciences. In the case of traditional random matrix theory (RMT) \cite{Wigner,Dyson}, which is also Gaussian but has a continuous, as opposed to a discrete symmetry, the areas of application extend across many-body quantum physics \cite{Guhr}  and various areas of data science \cite{EdelmanWang}. This diverse range of applications defines a universality class of random matrix statistics. There are already hints of such a universality in existing applications of PIGM models, where different matrix ensembles in computational linguistics constructed from different kinds of  algorithms show approximate Gaussianity. The Gaussianity analysed in \cite{Ramgoolam:2019ldg}  was based on the construction of matrices by linear regression in \cite{Kartsaklis:2017lfq} while \cite{Huber:2022ohf} extended the analysis of \cite{Ramgoolam:2019ldg} and also analysed the matrices constructed by neural network methods in \cite{wijnholds-etal-2020-representation}. 
 
Further motivation for investigating the broad theme of Gaussianity within a financial setting comes from the known statistical properties of correlation matrices. Existing results establish the asymptotic $D^2$-variate Gaussianity of the sampling distribution of $D \times D$ dimensional correlation matrices under fairly general conditions for example (related to finite fourth moments of the underlying observations from which the correlation matrices are constructed, see \cite{anderson2003introduction}, Theorem 3.4.4 and subsequent comments). This differs from our non-asymptotic i.e. finite sample setting, but does imply that the distribution of ensembles of correlation matrices approaches a multivariate Gaussian in the large observation sample limit. 

Other known statistical results include the fact that the non-asymptotic sampling distribution of correlation matrices is the Wishart distribution, under certain specific conditions \cite{potters2020first}. In particular, the Wishart distribution only arises in the case where the observations from which the correlation matrix estimates are constructed are themselves multivariate Gaussian. We do not make this assumption. Thus our setting is one of finite observation samples which may not be Gaussian distributed. It has also been noted that the correlation information encoded in correlation matrices is not sensitive to the ordering of basis vectors (in \cite{marti2020corrgan} for financial correlation matrices for example). This motivates a permutation invariant model for correlation matrices. The Wishart distribution is not permutation invariant in general, in contrast to the PIGM model. To summarise then, in our setting we apply the PIGM model as a phenomenological model for ensembles of correlation matrices. The PIGM model is permutation invariant and is able to capture leading Gaussian structure which forms the basis for quantifying small non-Gaussian corrections.

We note a rich history of applying random matrix theory (RMT) to the study of financial correlation matrices. In particular, the eigenvalue distributions of these matrices have been studied, demonstrating close agreement between the majority of eigenvalues and the eigenvalue distribution as given by the so-called Marchenko-Pastur (M-P) law \cite{marvcenko1967distribution} applied to random correlation matrices \cite{bouchaud1999, bouchaud2000,plerou2002random}. Evidence has also been presented that the largest eigenvalues - those that deviate most strongly from the M-P distribution - are associated with non-random overall market, sector and stock correlation structure (see \cite{plerou2002random} for example). Practical applications of these findings have been developed such as cleaning/de-noising correlation matrices, amongst others \cite{bouchaud2009,bouchaud2017,de2020machine,potters2005financial}.

The PIGM model provides a new approach to studying and describing financial correlation matrices that is distinct from existing approaches based on RMT. It focuses on low degree permutation invariant polynomial functions of matrices (which we refer to as observables)  instead of eigenvalue distributions, which are the focus of traditional RMT. This perspective is based on the postulate that  near-Gaussian permutation invariant sectors of real world matrix data contain useful information. The PIGM model furnishes a parsimonious specification of the probability density function of these matrices using only 4 free parameters for the symmetric, vanishing diagonal model. This is  close to the one or two parameters of the simplest RMT and far smaller than a multi-variate Gaussian distribution for a $D \times D$ matrix, which has order $D^2$ parameters. It provides an analytical solution to the expectation values of permutation invariant products of matrix elements. The empirical higher order observables (cubic, quartic etc.) that agree closely with the model - which is only fit to linear and quadratic observables - reveal consistency with random matrices implied by the fitted model. The empirical higher order observables that depart from theoretical expectations indicate informative structure beyond that encoded in the random model. A vector of observables therefore provides a signature for a particular correlation matrix, which may provide a useful, lower dimensional, permutation invariant representation of correlation matrices. The effectiveness of this representation in anomaly detection and measurement of  similarity are explored as  initial examples.  

It should be noted that the PIGM model can be applied beyond symmetric correlation matrices to general cross-correlation matrices with only permutation invariance (e.g. constructed from returns at different times) through the full 13-parameter PIGM model developed in \cite{ramgoolam2019permutation}. In this article we lay the foundations for these future studies. 

In section \ref{sec: technical summary} we summarise the theoretical results of the paper on permutation invariant Gaussian models. We define general permutation invariant Gaussian matrix models and consider the restriction of these models appropriate to the financial data described in section \ref{sec: data description}, namely that the matrices must be symmetric and have vanishing diagonal elements. We also define the permutation invariant observables of the model and explain a  useful bijection between these observables and loopless graphs, examples of which are given.

Section \ref{sec:PIGMMSDV} contains the bulk of the theory, the primary goal is to solve the most general PIGM model of symmetric matrices with vanishing diagonal. This is achieved with the help of representation theory of the symmetric group and builds on the results of \cite{ramgoolam2019permutation} and \cite{PIG2MM}. We find that these models are characterised by 1 linear and 3 quadratic couplings. Linear and quadratic expectation values of observables can be expressed simply in terms of these coupling parameters, \eqref{eq: one point function} and \eqref{eq: two point function} respectively. Higher order expectation values are simply constructed from these with the application of Wick's theorem.

Section \ref{sec: data description} gives details of the high-frequency forex data used to construct the matrix ensemble studied in the remainder of the paper, as well as the method by which the members of this ensemble are constructed from the underlying data. 

Section \ref{sec: Numerical results} contains a description of the empirical statistical properties of the observables. This includes practical measures of their Gaussianity and comparison of their properties with those predicted by the model presented in section  \ref{sec:PIGMMSDV}. 

In section \ref{sec: applications} we construct  vectors of observables for each correlation matrix. These observable vectors are low dimensional representations of the correlation matrices. There are 31 cubic and quartic observables for general matrix size $D$ (as long as $D \geq 8$, a condition which is generally satisfied in large $D$ applications such as the one here). 

In general, we find that the observable vectors provide a good  representation of the original correlation matrices, performing well in anomaly detection and similarity ranking applications.  The best performances are achieved by selecting  subsets of the cubic and quartic observables, based on the ranking of their small non-Gaussianities, and on the postulate that the more non-Gaussian observables are most informative of economic factors driving atypicality of the days. The performance of observable vectors in these applications compares favourably with standard dimensionality reduction techniques, namely, Principal Component Analysis (PCA). We conclude in section \ref{sec: conclusion} and suggest interesting future research directions arising from this work. 

\section{  Summary of results on the $4$-parameter Gaussian matrix model } \label{sec: technical summary}

Here we summarise the main results and outline the key ideas behind  the construction of the general PIGM model for an ensemble of symmetric matrices which have vanishing matrix elements along the diagonal. This section is intended to provide, for a reader with background in mathematical finance or statistics, an understanding of the key theoretical points of the paper, without getting into the details of the construction of section \ref{sec:PIGMMSDV} which will be more easily accessible to a reader with  knowledge of group representation theory at a level covered in mathematical texts such as in chapters 3, 5 and 7 of \cite{hamermesh}.  We will review the description of probability distributions using a Euclidean action which is Gaussian or near-Gaussian  using the simple case of a one-variable statistics and motivate the measure of non-Gaussianity we use later in the case of permutation invariant matrix distributions. We explain the structure of the 4-parameter  permutation invariant Gaussian matrix model and  the connection between the permutation invariant polynomial functions of the matrices with loopless graphs. 

It is useful to recall that a one-variable Gaussian distribution, with mean $ \mu $ and
standard deviation $\sigma $, for a random variable $x$, 
  is described by a probability density function 
 \bea 
 f(x ) = { 1 \over \sigma \sqrt { 2 \pi }} e^{ - { ( x - \mu )^2 \over 2 \sigma^2 } } .
 \eea
 The moments of the distribution are expectation values $\langle x^k \rangle $, defined as 
 \bea
 \langle x^k \rangle = \int_{ - \infty }^{ \infty }  dx  ~~ f(x) ~~  x^k .  
 \eea
 It is also useful to define, by analogy with statistical physics and quantum mechanical path integrals, the action 
 $S = { ( x - \mu )^2 \over 2 \sigma^2 }  $. The partition function is 
 \bea 
 Z = \int dx ~~ e^{ - S } ,
 \eea
 while the moments are 
 \bea \label{eq: moments def}
 \langle x^k \rangle =  { 1 \over Z }  \int dx  ~~  e^{ - S } ~~ x^k   .
 \eea
The action $S$ is a quadratic function of $x$. 
 
It is often the case that the action of a theory is approximately Gaussian - deviating from Gaussianity by some small higher order terms. The full action of the system $S'$ is then written as a Gaussian piece $S$, plus an additional non-Gaussian piece $\delta S$
\bea
S' = S + \lambda \delta S.
\eea
The smallness of the higher order terms is governed by the interaction strength $\lambda$, whose smallness is required to ensure
\bea
\langle \cO_{\alpha} \rangle_{S} - \langle  \cO_{\alpha} \rangle_{S'} < \sigma_{\langle \cO_{\alpha} \rangle_{S'}},
\eea
i.e. the expectation value of some observable $\cO_{\alpha}$, is largely insensitive to the non-Gaussian contribution to the true action governing the theory, $\delta S$. 

More concretely, take $\mu = 0$ in the simple pure Gaussian, one parameter toy model defined above. The partition function is
\begin{align}
Z = \int \dd x ~~  e^{-S} = \int \dd x ~~  e^{-\frac{x^2}{2 \sigma^2}}.
\end{align}
View this as an approximation to some true physical partition function which includes some small, non-Gaussian perturbation. We explain by way of a simple example which captures the mechanism, that the absolute  differences between expectation values of low order polynomials in the random variables (observables) in the Gaussian model and the perturbed model are small compared to the standard deviation of the observable. The simple example consists of a Gaussian action perturbed by a small quartic correction, so that the perturbed partition function $Z'$ is 
\begin{align}
Z' = \int \dd x e^{-S'}  = \int \dd x e^{-\frac{x^2}{2 \sigma^2} -\frac{\lambda}{4!}x^4} = Z \Big( 1 - \frac{\lambda}{4!} \langle x^4 \rangle +  \dots \Big).
\end{align}

Using \eqref{eq: moments def} we calculate the absolute difference between the fourth moment of each of the theories. In the purely Gaussian case we have
\begin{align}
\langle x^4 \rangle = 3 \sigma^4 \,,
\end{align}
and in the perturbed theory 
\begin{align} \nonumber
\langle x^4 \rangle_{S'} = \frac{Z}{Z'} \Big( \langle x^4 \rangle - \frac{\lambda}{4!} \langle x^8 \rangle + \dots \Big) &\approx \langle x^4 \rangle + \frac{\lambda}{4!} \big( \langle x^4 \rangle^2 - \langle x^8 \rangle \big), \\ \nonumber
&= 3 \sigma^4 + \frac{\lambda}{4!} \big( 9 \sigma^8 - 105 \sigma^8 \big) \\
&= 3 \sigma^4 - \frac{96}{4!} \lambda \sigma^8 \,.
\end{align}
Taking the difference of these values
\begin{align}
\abs{\langle x^4 \rangle - \langle x^4 \rangle_{S'}} \approx \frac{96}{4!} \lambda \sigma^8 \,.
\end{align}
The standard deviation of the fourth moment in the perturbed theory requires 
\begin{align} \nonumber
\langle x^8 \rangle_{S'} = \frac{Z}{Z'} \Bigg( \langle x^8 \rangle - \frac{\lambda}{4!} \langle x^{12} \rangle + \dots \Bigg) &\approx \langle x^8 \rangle + \frac{\lambda}{4!} \big( \langle x^8 \rangle \langle x^4 \rangle - \langle x^{12} \rangle \big) \\ \nonumber
&= 105 \sigma^8 + \frac{\lambda}{4!} \big( 315 \sigma^{12} - 11!! \sigma^{12} \big) \\
&= 105 \sigma^8 - \frac{(11!! - 315)}{4!} \lambda \sigma^{12} \, . 
\end{align}
Which gives
\begin{align} \nonumber
\sigma_{\langle x^4 \rangle_{S'}} &= \sqrt{ \langle x^8 \rangle_{S'} - \langle x^4 \rangle^2_{S'}} + \mathcal{O}(\sqrt{\lambda}) \\ \nonumber
&= \sqrt{105 \sigma^8 - \Big( 3 \sigma^4 \Big)^2} \\
&\approx \sqrt{96} \sigma^4 \, . 
\end{align}
Finally, we see that the absolute difference between the fourth moments normalised by the standard deviation is
\begin{align}
\frac{\abs{\langle x^4 \rangle - \langle x^4 \rangle_{S'}}}{\sigma_{\langle x^4 \rangle_{S'}}} \sim \lambda \sigma^4.
\end{align}
Therefore,  as long as the physical theory is approximately Gaussian, i.e. $\lambda$ is small, its normalised fourth moment is well approximated by that of the purely Gaussian theory.

We postulate that real market effects governing the interactions between currency rates included in this study are modelled analogously by a Gaussian action plus some small non-Gaussian perturbation. The smallness of the non-Gaussian terms allows us to approximate expectation values using a purely Gaussian theory. Evidence for this near-Gaussianity is provided primarily by the smallness of the measured observable deviations from those predicted by a purely Gaussian theory. These are listed in table \ref{tab: abs err} of section \ref{sec: Numerical results}.

Permutation invariant Gaussian matrix models for generic real matrices  of size $D$ are described by a partition function 
\bea 
\cZ = \int \prod_{ i , j =1}^D dM_{ ij}  ~~ e^{ - \mathcal{S}^{\text{PIGMM}} }  \, .
\eea
The integration measure is the standard Euclidean measure on $\mathbb{R}^{ D^2}$. The general Gaussian action is defined by a general quadratic permutation invariant function of the matrix elements $M_{ij}$, depending on $13$ parameters, which is compatible with convergence \cite{ramgoolam2019permutation}. The set of all permutations of $D$ objects forms the symmetric group $S_D$.  The permutations $ \sigma \in S_D $ act as
\bea 
\sigma : M_{ ij} \rightarrow M_{ \sigma (i) \sigma (j) }  \, .
\eea
A precise description of the general parameter space for such a Gaussian action is found using the representation theory of the symmetric group.  A key outcome of the representation theoretic treatment is that there exists a convenient set of variables $S^{ V_A ; \tau_A }_a$, which are linear combinations of the $D^2$ variables $M_{ij}$:  
\bea
S^{ V_A ; \tau_A }_{ a } = \sum_{ i, j = 1 }^D  C^{V_{A} , \tau_A }_{ a ~ ; ~ ij  } M_{ ij}  \, .
\eea
The variables $V_A $ are vector spaces $\{ V_{ [D ]} = V_0  , \, V_{ [ D-1 , 1 ]} = V_H  , \,V_{ [ D-2 , 2 ] } = V_2 , \, V_{ [D-2 ,1,1]} = V_3  \}  $. The meaning of this list will be explained further in section \ref{sec:PIGMMSDV}. For now, in this general introduction aiming to be accessible to a reader with knowledge of linear algebra but no detailed knowledge of representation theory, the only important point is that this is a list of 4 elements. The index $a$ ranges over a set of basis elements for each vector space $V_A$, numbering respectively $\{ 1, \, (D-1)  , \, D ( D-3 )/2 , \, ( D- 1) (D-2)/2 \}$.  The index $\tau_A$ runs over $\{ 2, 3, 1, 1\}$ values respectively. The key result is that the action for the general permutation invariant Gaussian model takes the form 
\bea \label{eq: PIGMM action}
\cS^{\text{PIGMM}} =  - \mu_1 S^{ V_0 ; 1 }  - \mu_2 S^{ V_0 ; 2 }  + \sum_{ A   }  \sum_{ a = 1 }^{ {\rm Dim}  V_A } g_{\tau_A , \tau_A'   }^{(A)}  \sum_{ \tau , \tau' } S^{ V_A ; \tau_A }_{ a } S^{ V_A ; \tau_A' }_{ a } \, , 
\eea
where $\mu_1$ and $\mu_2$ are linear coupling parameters. The parameters $g_{\tau_A , \tau'_A }^{(A)}$ are symmetric matrix parameters of matrix size $2,3,1,1$. Thus they define parameter spaces of dimension $\{ 3,6,1,1\}$. Convergence is guaranteed by the condition that these matrices have positive eigenvalues. Thus the general Gaussian model has 2 parameters for the linear invariants and 11 parameters for the quadratic invariants.

We show in section \ref{sec:PIGMMSDV} that the general permutation invariant Gaussian  matrix model for symmetric matrices is a 9-parameter model, and for symmetric matrices with diagonally vanishing matrix elements the permutation invariant Gaussian model is a 4-parameter model. The 9-dimensional parameter space for the symmetric matrices and the 4-dimensional parameter space for the symmetric diagonally vanishing matrices are subspaces of the 13-dimensional parameter space for generic matrices. The embedding of the 4-dimensional parameter space in the 13-parameter space is described in section \ref{sec:PIGMMSDV}. 

An important ingredient in understanding permutation invariant random matrix models is the structure of the permutation invariant polynomial functions of matrices, which are closely related to graphs. The key point behind this relation is that permutation invariant polynomial functions can be constructed by summing over the matrix indices. For generic matrices,  the two linear and eleven quadratic functions are
\begin{equation} \label{Eqn: Two matrix multi-graph diagrams}
\begin{array}{cccc}
\begin{tikzpicture}[baseline, scale = \graphscale]
\begin{scope}[decoration={markings, mark=at position 0.45 with \arrow{latex}}]
\draw[draw, postaction={decorate}] (0,0)node[circle, fill=black, inner sep=1pt, draw=black] {} to[in=180, out=180] (0,1) to[out=0, in=0] (0,0);
\end{scope}
\end{tikzpicture} \, , \hspace{1em} &
\begin{tikzpicture}[scale = \graphscale]
\begin{scope}[decoration={markings, mark=at position 0.45 with \arrow{latex}}]
\draw[draw, postaction={decorate}] (0,0)node[circle, fill=black, inner sep=1pt, draw=black] {} to[bend left] (1,0)node[circle, fill=black, inner sep=1pt, draw=black] {};
\end{scope}
\end{tikzpicture} \, , \hspace{1em} &
\begin{tikzpicture}[scale = \graphscale]
\begin{scope}[decoration={markings, mark=at position 0.45 with \arrow{latex}}]
\draw[draw, postaction={decorate}] (0,0)node[circle, fill=black, inner sep=1pt, draw=black] {} to[in=180, out=180] (0,1) to[out=0, in=0] (0,0);
\draw[draw, postaction={decorate}] (0,0)node[circle, fill=black, inner sep=1pt, draw=black] {} to[in=180, out=180] (0,.8) to[out=0, in=0] (0,0);
\end{scope}
\end{tikzpicture} \, , \hspace{1em} &
\begin{tikzpicture}[scale = \graphscale]
\begin{scope}[decoration={markings, mark=at position 0.45 with \arrow{latex}}]
\draw[draw, postaction={decorate}] (0,0)node[circle, fill=black, inner sep=1pt, draw=black] {} to[bend left] (1,0);
\draw[draw, postaction={decorate}] (0,0) to[bend right] (1,0)node[circle, fill=black, inner sep=1pt, draw=black] {};
\end{scope}
\end{tikzpicture} \, , \hspace{1em}  \\ [0.1em]
\sum_{i = 1}^D M_{ii} \hspace{1em} &
\sum_{i, j = 1}^D M_{ij} \hspace{1em} &
\sum_{i = 1}^D M_{ii} M_{ii} \hspace{1em} &
\sum_{i, j = 1}^D M_{ij} M_{ij} \hspace{1em} \\
\begin{tikzpicture}[scale = \graphscale]
\begin{scope}[decoration={markings, mark=at position 0.45 with \arrow{latex}}]
\draw[draw, postaction={decorate}] (1,0)node[circle, fill=black, inner sep=1pt, draw=black] {} to[bend right] (0,0);
\draw[draw, postaction={decorate}] (0,0)node[circle, fill=black, inner sep=1pt, draw=black] {} to[bend right] (1,0);
\end{scope}
\end{tikzpicture} \, , \hspace{1em} & 
\begin{tikzpicture}[scale = \graphscale]
\begin{scope}[decoration={markings, mark=at position 0.45 with \arrow{latex}}]
\draw[draw, postaction={decorate}] (0,0)node[circle, fill=black, inner sep=1pt, draw=black] {} to[in=180, out=180] (0,1) to[out=0, in=0] (0,0);
\draw[draw, postaction={decorate}] (.75,0)node[circle, fill=black, inner sep=1pt, draw=black] {} to[in=180, out=180] (.75,1) to[out=0, in=0] (.75,0);
\end{scope}
\end{tikzpicture} \, , \hspace{1em} &
\begin{tikzpicture}[scale = \graphscale]
\begin{scope}[decoration={markings, mark=at position 0.45 with \arrow{latex}}]
\draw[draw, postaction={decorate}] (0,0)node[circle, fill=black, inner sep=1pt, draw=black] {} to[bend left] (1,0)node[circle, fill=black, inner sep=1pt, draw=black] {};
\draw[draw, postaction={decorate}] (0,0) to[in=180, out=180] (0,1) to[out=0, in=0] (0,0);
\end{scope}
\end{tikzpicture} \, , \hspace{1em}  &
\begin{tikzpicture}[scale = \graphscale]
\begin{scope}[decoration={markings, mark=at position 0.45 with \arrow{latex}}]
\draw[draw, postaction={decorate}] (1,0)node[circle, fill=black, inner sep=1pt, draw=black] {} to[bend right] (0,0)node[circle, fill=black, inner sep=1pt, draw=black] {};
\draw[draw, postaction={decorate}] (0,0) to[in=180, out=180] (0,1) to[out=0, in=0] (0,0);
\end{scope}
\end{tikzpicture} \, , \\[0.1em]
\sum_{i, j = 1}^D M_{ji} M_{ij} \hspace{1em} &
\sum_{i, j = 1}^D M_{ii} M_{jj} \hspace{1em} &
\sum_{i, j = 1}^D  M_{ij} M_{ii} \hspace{1em}  &
\sum_{i, j = 1}^D  M_{ii}M_{ji} \hspace{1em} \\
\begin{tikzpicture}[scale = \graphscale]
\begin{scope}[decoration={markings, mark=at position 0.45 with \arrow{latex}}]
\draw[draw, postaction={decorate}] (0,0)node[circle, fill=black, inner sep=1pt, draw=black] {} to[bend left] (1,0)node[circle, fill=black, inner sep=1pt, draw=black] {};
\draw[draw, postaction={decorate}] (1,0) to[bend left] (2,0)node[circle, fill=black, inner sep=1pt, draw=black] {};
\end{scope}
\end{tikzpicture} \, , \hspace{1em}  &
\begin{tikzpicture}[scale = \graphscale]
\begin{scope}[decoration={markings, mark=at position 0.45 with \arrow{latex}}]
\draw[draw, postaction={decorate}] (0,0)node[circle, fill=black, inner sep=1pt, draw=black] {} to[bend left] (1,0)node[circle, fill=black, inner sep=1pt, draw=black] {};
\draw[draw, postaction={decorate}] (0,0) to[bend left] (2,0)node[circle, fill=black, inner sep=1pt, draw=black] {};
\end{scope}
\end{tikzpicture}  \, , \hspace{1em}  &
\begin{tikzpicture}[scale = \graphscale]
\begin{scope}[decoration={markings, mark=at position 0.45 with \arrow{latex}}]
\draw[draw, postaction={decorate}] (0,0)node[circle, fill=black, inner sep=1pt, draw=black] {} to[bend left] (1,0)node[circle, fill=black, inner sep=1pt, draw=black] {};
\draw[draw, postaction={decorate}] (2,0)node[circle, fill=black, inner sep=1pt, draw=black] {} to[bend right] (1,0)node[circle, fill=black, inner sep=1pt, draw=black] {};
\end{scope}
\end{tikzpicture} \, , \hspace{1em} &
\begin{tikzpicture}[scale = \graphscale]
\begin{scope}[decoration={markings, mark=at position 0.45 with \arrow{latex}}]
\draw[draw, postaction={decorate}] (0,0)node[circle, fill=black, inner sep=1pt, draw=black] {} to[bend left] (1,0)node[circle, fill=black, inner sep=1pt, draw=black] {};
\draw[draw, postaction={decorate}] (1.5,0)node[circle, fill=black, inner sep=1pt, draw=black] {} to[in=180, out=180] (1.5,1) to[out=0, in=0] (1.5,0);
\end{scope}
\end{tikzpicture} \, , \\[0.1em]
\sum_{i, j = 1}^D  M_{jk} M_{ij} \hspace{1em} &
\sum_{i, j, k = 1}^D M_{ij} M_{ik} \hspace{1em} &
\sum_{i, j, k = 1}^D M_{kj} M_{ij} \hspace{1em} &
\sum_{i, j, k = 1}^D M_{kk} M_{ij} \hspace{1em} \\
\begin{tikzpicture}[scale = \graphscale]
\begin{scope}[decoration={markings, mark=at position 0.45 with \arrow{latex}}]
\draw[draw, postaction={decorate}] (0,0)node[circle, fill=black, inner sep=1pt, draw=black] {} to[bend left] (1,0)node[circle, fill=black, inner sep=1pt, draw=black] {};
\draw[draw, postaction={decorate}] (1.5,0)node[circle, fill=black, inner sep=1pt, draw=black] {} to[bend left] (2.5,0)node[circle, fill=black, inner sep=1pt, draw=black] {};
\end{scope}
\end{tikzpicture}  \, . \\ [0.1em]
\sum_{i, j, k, l = 1}^D M_{ij} M_{kl}
\end{array}
\end{equation}
The graphs are associated with these polynomials as follows: each vertex corresponds to a summed index and each directed edge to a matrix $M_{ij}$ from the vertex corresponding to index  $i$ to the vertex corresponding to index  $j$. This intuitive connection was used in \cite{Kartsaklis:2017lfq} and it was recognised that it is valid for $D \ge 2k$, with $D$ the size of the matrix and $k$ the degree of the polynomial. A mathematical proof the connection was given in section 2 of \cite{PIG2MM}.  There is also a formulation of the counting in terms of partition algebras as developed in section 3.1 of \cite{barnes2022hidden} and there are related results in  earlier papers \cite{Gabriel:2015mha}\cite{ Gabriel:2015nka}.

For symmetric matrices with vanishing diagonal elements there is 1 linear invariant function and 3 quadratic invariant functions, these are
\begin{equation} \label{Eqn: multi-graph diagrams phys}
\begin{array}{cccc}
\begin{tikzpicture}[baseline, scale = \graphscaleb]
\begin{scope}
\draw[draw, postaction={decorate}] (0,0)node[circle, fill=black, inner sep=1pt, draw=black] {} to[bend left] (1,0)node[circle, fill=black, inner sep=1pt, draw=black] {};
\end{scope}
\end{tikzpicture} \, , \hspace{2cm} &
\begin{tikzpicture}[scale = \graphscaleb]
\begin{scope}[]
\draw[draw=black] (0,0)node[circle, fill=black, inner sep=1pt, draw=black] {} to[bend left] (1,0);
\draw[draw=black] (0,0) to[bend right] (1,0)node[circle, fill=black, inner sep=1pt, draw=black] {};
\end{scope}
\end{tikzpicture} \, ,  \hspace{2cm} &
\begin{tikzpicture}[scale = \graphscaleb]
\begin{scope}[]
\draw[draw=black] (0,0)node[circle, fill=black, inner sep=1pt, draw=black] {} to[bend left] (1,0)node[circle, fill=black, inner sep=1pt, draw=black] {};
\draw[draw=black] (1,0) to[bend left] (2,0)node[circle, fill=black, inner sep=1pt, draw=black] {};
\end{scope}
\end{tikzpicture} \, , \hspace{2cm} &
\begin{tikzpicture}[scale = \graphscaleb]	
\begin{scope}[]
\draw[draw=black] (0,0)node[circle, fill=black, inner sep=1pt, draw=black] {} to[bend left] (1,0)node[circle, fill=black, inner sep=1pt, draw=black] {};
\draw[draw=black] (1.5,0)node[circle, fill=black, inner sep=1pt, draw=black] {} to[bend left] (2.5,0)node[circle, fill=black, inner sep=1pt, draw=black] {};
\end{scope}
\end{tikzpicture} \, . \\
\sum_{i,j}^D M_{ij}   &
\sum_{i,j}^D M_{ij}^2   &
\sum_{i,j,k}^D M_{ij} M_{jk}  &
\sum_{i,j,k,l}^D M_{ij} M_{kl} 
\end{array}
\end{equation}
Graphs are similarly associated with these polynomials, however, the edges are no longer directed - a consequence of the symmetry of the matrices - and they no longer contain loops - a consequence of their vanishing diagonal elements.

For a fixed degree, the permutation invariant polynomial functions form a vector space. As long as the matrix dimension is larger than twice the degree of the polynomial the graphs are in one-to-one correspondence with basis elements of this vector space. In our present financial application this degree condition is always satisfied. A detailed discussion of this condition and the effects of going beyond it are given in \cite{PIG2MM}.

The action of the reduced PIGM model is given by the most general combination of permutation invariant linear and quadratic terms
\begin{align} \label{eq: M action}
\mathcal{S} = \mu \sum_{i, j = 1}^D M_{ij} + \tau_1 \sum_{i, j = 1}^D M_{ij}M_{ij} + \tau_2 \sum_{i, j, k = 1}^D M_{ij}M_{jk} + \tau_3 \sum_{i, j, k, l = 1}^D M_{ij}M_{kl} \, , 
\end{align}
where $\mu$ is the linear coupling strength and $\tau_1, \tau_2$ and $\tau_3$ are the quadratic couplings. Label observables of this theory $\mathcal{O}_{\alpha}(M_{ij})$, where $\alpha$ indexes the particular observable, they are permutation invariant polynomials of the random matrix variables of symmetric matrices with vanishing diagonal. Expectation values of these variables are defined as
\begin{align} \label{eq: observables definition}
\langle \mathcal{O}_{\alpha}(M_{ij}) \rangle = \frac{1}{\mathcal{Z}} \int \dd M \mathcal{O}_{\alpha} e^{- \mathcal{S}}.
\end{align}
In order to solve \eqref{eq: observables definition} for any choice of $\mathcal{O}_{\alpha}$ we must find a change of basis that factorises the RHS. This is possible with the application of appropriate projectors. Given these the action for the 4-parameter model can be written
\begin{align} \label{eq: physical action} \nonumber
\mathcal{S} = &\sum_{i,j,k,l = 1}^D \frac{1}{2} \Big( \tau_{V_0} M_{ij} (P^{\text{phys}; V_0})_{ij, kl} M_{kl} + \tau_{V_H} M_{ij} (P^{\text{phys}; V_H})_{ij, kl} M_{kl} + \tau_{V_2} M_{ij} (P^{\text{phys}; V_2})_{ij, kl} M_{kl} \Big) \\ 
&- \sum_{i, j = 1}^D \mu_{V_0} C^{\text{phys}; V_0}_{ij} M_{ij}.
\end{align}
Where $\mu_{V_0}, \tau_{V_0}, \tau_{V_H}$ and $\tau_{V_2}$ are the couplings in the transformed basis. Performing the projections in \eqref{eq: physical action} allows for the solution of the linear and quadratic expectation values via standard techniques of Gaussian integration. Cubic, quartic and higher expectation values can be calculated with the application of Wick's theorem, which allows them to be expressed as sums of products of linear and quadratic expectation values. Details of Wick's theorem as applied to cubic and quartic expectation values are given in section \ref{sec: Wick's theorem}.

The equations for the 13-parameters appearing in the action \eqref{eq: PIGMM action} in terms of the 4-parameters of the physical action \eqref{eq: physical action} are given in \eqref{eq: linear coupling relations} and \eqref{eq: coupling relations}. As expected both models give consistent results for the expectation values of physical observables.

To determine how well this PIGM model predicts the statistics of the forex correlation data described in section \ref{sec: data description} we first use the correlation matrices to define the Gaussian model i.e. to fix the linear and quadratic couplings of the model. We then calculate the theoretical expectation values $\langle \mathcal{O}_{\alpha}(M) \rangle_{\text{T}}$ defined by \eqref{eq: observables definition} using this action with the application of Wick's theorem. These can then be compared to the experimental expectation values $\langle \mathcal{O}_{\alpha}(M) \rangle_{\text{E}}$ calculated from the financial correlation matrices themselves using the following similarity measure
\begin{align}
\Delta_{\alpha} = \frac{ \big| \langle \mathcal{O}_{\alpha}(M) \rangle_{\text{T}} - \langle \mathcal{O}_{\alpha}(M) \rangle_{\text{E}} \big|}{\sigma_{E, \alpha}(M) },\label{eqn: similarity}
\end{align}
where $\sigma_{E, \alpha}$ is the standard deviation of the expectation value with respect to the ensemble of correlation matrices. This measure of similarity is used to identify the observables which deviate most significantly from Gaussianity. In section \ref{sec: anomaly detection} and \ref{sec: similarity} these least Gaussian observables are shown to be the optimal candidates for data reduction in a variety of anomaly detection and similarity tests.

\section{$4$-parameter Gaussian model: detailed construction  }\label{sec:PIGMMSDV} 

In this section we give a detailed account of the construction  of the 4-parameter matrix model which was  outlined in section \ref{sec: technical summary}. The aim is to model the statistics of the particular ensemble of correlation matrices introduced in section \ref{sec: data description} as well as any matrix ensemble in the same universality class. This class is composed of symmetric square matrices with zeros along the diagonal, for which physical quantities are invariant under simultaneous permutations of the rows and columns. Given the universality of the model we label our matrices $M$ and index them with lowercase indices $1 \leq i, j \leq D$, to distinguish them from the financial correlation matrices specifically, which we label $\hat{\rho}$. We will refer to matrices with zeros on the diagonal as "diagonally vanishing" throughout.

We begin by defining the action of permutations on the matrix variables and establishing their irreducible decomposition under this group action. We then define the most general action of a permutation invariant Gaussian matrix model, containing symmetric matrices with vanishing diagonal. It is parameterised by one linear and three quadratic couplings. Finding projectors that project from the original matrix basis $M_{ij}$ to a basis transforming according to this irreducible decomposition allows us to rewrite this action in a diagonalised form. In turn, this diagonalisation permits the application of standard multi-dimensional Gaussian integration techniques which, along with the application of Wick's theorem, produce analytic formulae for the expectation values of observables as a function of $D$ - the dimension of the matrices.

\subsection{Symmetric group representation theory and matrix variables}
The diagonal action of $S_D$ simultaneously permutes the rows and columns of a matrix
\begin{align} \label{eq: sigma on M}
M_{ij} \rightarrow M_{\sigma(i) \sigma(j)}, \qquad \forall \, \sigma \in S_D.
\end{align}
We refer to the space of symmetric matrices with vanishing diagonal as the physical subspace $V^{\text{phys}}$ of general $D \times D$ matrices and label matrices in the space with a superscript i.e. $M^{\text{phys}} \in V^{\text{phys}}$. They obey the conditions
\begin{align} \label{eq: symm diagonal-less conditions}
M_{ij} = M_{ji}, \qquad M_{ii} = 0, \qquad \, 1 \leq i,j \leq D.
\end{align} 
Since these conditions are $S_D$ equivariant with respect to the action defined in \eqref{eq: sigma on M} the physical subspace is invariant under $S_D$
\begin{align}
M_{ij} \in V^{\text{phys}}  \Rightarrow M_{\sigma(i) \sigma(j)} \in V^{\text{phys}}, \qquad \forall \, \sigma \in S_D.
\end{align}
By physical we mean only to restrict to the non-trivial data of interest. In the correlation matrices described in section \ref{sec: data description} all the data is contained within symmetric matrices with vanishing diagonal.  

The natural representation of the symmetric group $V_D$, is a $D$-dimensional representation with basis $\{ e_1, e_2, \dots, e_D \}$. For each permutation $\sigma \in S_D$ the linear operator acting on this space is defined by
\begin{align}
\rho_{V_D}(\sigma) e_i = e_{\sigma^{-1}(i)}.
\end{align}
The natural representation is a reducible representation and decomposes as
\begin{align} \label{eq: V_D decomp}
V_D \cong V_0 \oplus V_H,
\end{align}
where $V_0$ is the trivial representation and $V_H$ is the Hook representation, both of which are irreducible. This decomposition is given explicitly by forming the following linear combinations of natural basis elements
\bea 
E_0 & = &  { 1 \over \sqrt { D}  }  ( e_1 + e_2 + \cdots + e_D ) \,, \cr 
E_{ 1} & = &  { 1 \over  \sqrt { 2 }  } ( e_1 - e_2 ) \,, \cr 
E_{ 2} & = & { 1 \over  \sqrt{ 6 } }    ( e_1 + e_2 - 2 e_3  ) \,, \cr 
& \vdots &  \cr 
E_{ a } & = & { 1 \over \sqrt{ a ( a+1)   } } ( e_1 + e_2 + \cdots + e_a - a e_{a+1}  ) \, ,   \cr 
& \vdots & \cr 
E_{D-1}  & = & { 1 \over \sqrt{ D ( D + 1) } }  ( e_1 + e_2 + \cdots + e_{D-1}  - ( D-1)  e_D ) \,.   \, 
\eea
$E_0$ is the trivial representation, and the remaining $D-1$ vectors $\{ E_1, E_2. \dots, E_{D-1} \}$ form an irreducible representation of $S_D$ which is $V_H$.

A general matrix $M_{ij}$ transforming under \eqref{eq: sigma on M} forms the representation $V_D \otimes V_D$, the tensor product of two copies of the natural representation of $S_D$. A natural basis for this tensor product is given by
\begin{align} \label{eq: M natural basis}
e_i \otimes e_j, \qquad 1 \leq i, j \leq D,
\end{align}
on which $S_D$ acts as
\begin{align}
\rho_{V_D^{\otimes 2}}(\sigma) e_i \otimes e_j = e_{\sigma^{-1}(i)} \otimes e_{\sigma^{-1}(j)}.
\end{align}
$\{ e_1, e_2, \dots, e_D \}$ is an orthonormal basis of $V_D$ under the inner product
\begin{align} \label{eq: inner prod}
\big( e_i , e_j \big) = \delta_{ij} \, .
\end{align}
This inner product is extended to $V_D \otimes V_D$ as
\begin{align} \label{eq: vd vd inner prod}
\big( e_i \otimes e_j , e_k \otimes e_l \big) = \delta_{ik} \delta_{jl}.
\end{align}
Using \eqref{eq: inner prod} we define the change of basis coefficients
\begin{align} \label{eq: C0}
C_{ 0 ,  i } & \equiv  ( E_0 , e_i ) = \frac{1}{\sqrt{D}},  \\ \label{eq: Ca}
C_{ a  , i }  & \equiv  ( E_a , e_i ). 
\end{align}
$V_D \otimes V_D$ is a reducible representation of $S_D$ which has the following irreducible decomposition
\begin{align} \label{eq: V_D V_D decomp}
V_D \otimes V_D \cong 2 V_0 \oplus 3 V_H \oplus V_2 \oplus V_3.
\end{align}
This decomposition can be deduced using \eqref{eq: V_D decomp} along with the tensor product rule described in section 7.13 of \cite{hamermesh}. Other useful references for symmetric group representation theory include \cite{FultonHarris, Sagan2013}. 
In terms of Young diagrams, described by listing the number of columns in each row in descending order, the irreducible representations appearing on the right hand side of this decomposition are
\begin{align} \nonumber
V_0 &= [ D ], \\ \nonumber 
V_{ H } &= [ D-1 , 1 ], \\ \nonumber
V_2 &= [ D-2, 2 ], \\ 
V_3 &= [ D -2 , 1, 1 ] .
\end{align}

The space of symmetric matrices with vanishing diagonal form a subspace of the representations in \eqref{eq: V_D V_D decomp}. Firstly, symmetric matrices transform as $\text{Sym}^2(V_D)$. This is a reducible representation with the following decomposition
\begin{align} \label{eq: sym V_D}
\text{Sym}^2(V_D) \cong 2 V_0 \oplus 2 V_H \oplus V_2.
\end{align}
The matrix elements along the diagonal $\{ M_{ii} | 1 \leq i \leq D \}$ transform like the natural representation $V_D$. Removing a copy of $V_D \cong V_0 \oplus V_H$ from the symmetric product of $V_D$ in \eqref{eq: sym V_D} gives the following decomposition of the physical subspace
\begin{align} \label{eq: mat decomp}
V^{\text{phys}} \cong \text{Sym}^2(V_D)/V_D \cong  V_0 \oplus V_H \oplus V_2.
\end{align}
The decomposition \eqref{eq: mat decomp} tells us that the enforcement of permutation invariance on the action of a Gaussian theory containing symmetric $D$ dimensional matrices without diagonal permits a single independent linear term (i.e. the number of trivial representations appearing on the RHS). 
Quadratic products of physical matrices transform as
\begin{align} \label{eq: physical quadratic}
V^{\text{phys}} \otimes V^{\text{phys}} \cong \big( V_0 \oplus V_H \oplus V_2 \big) \otimes \big( V_0 \oplus V_H \oplus V_2 \big).
\end{align}
Using the orthogonality property of characters as well as the reality property of irreducible representations of the symmetric group, it can be shown that the tensor product of two irreducible representations contains the trivial representation if and only if the irreducible representations are identical - and in this case the decomposition contains exactly one copy of the trivial representation. This enables us to count the number of independent quadratic terms. We find three independent quadratic contributions to the action corresponding to the following three terms in \eqref{eq: physical quadratic}
\begin{align}
V_0 \otimes V_0 &\cong V_0 + \dots \, , \\
V_H \otimes V_H &\cong V_0 + \dots \, ,  \\
V_2 \otimes V_2 &\cong V_0 + \dots \, .
\end{align}

Much of our task in solving the physical model for symmetric diagonally vanishing matrices amounts to finding a change of basis for $V_D \otimes V_D$ from the original $e_i \otimes e_j$  to one which transforms in the same manner as the irreducible decomposition of $V^{\text{phys}}$, i.e. from the LHS of \eqref{eq: mat decomp} to the RHS. Once found it diagonalises the physical action and consequently permits the calculation of expectation values of observables. The coefficients that define this change of basis are called Clebsch-Gordon coefficients. For each irreducible representation on the RHS of \eqref{eq: mat decomp} define $C^{\text{phys}; V_0}_{ij}, C^{\text{phys}; V_H}_{ij, \, a}, C^{\text{phys}; V_2}_{ij, \, a}$ respectively, where $a$ is a state index running over the dimension of the irreducible representation.

Note that if we only imposed the condition of symmetry $M_{ij} = M_{ ji}$ on the matrices, we would have two linear couplings corresponding to the two copies of $V_0$ in \eqref{eq: sym V_D}. We would have 3 parameters of a $ 2 \times 2 $ symmetric matrix of couplings for quadratic terms arising from the two copies of $V_0$ in \eqref{eq: sym V_D}, 3 parameters of a $2 \times 2 $ symmetric matrix of couplings for quadratic terms arising from the two copies of $V_H$ in \eqref{eq: sym V_D}, and finally one parameter for $V_2$. For symmetric  matrices, therefore, there is a   $7$ parameter permutation invariant matrix model. We will focus, in the following, on the $4$-parameter model which incorporates the symmetry condition $M_{ij} = M_{ ji}$ as well as the condition of vanishing diagonal.

\subsection{Projector for $V^{\text{phys}}$}
The projectors to the trivial representations appearing in the quadratic products of $M_{ij}$ are given by squaring the relevant Clebsch coefficients and summing over intermediate states
\begin{align} \label{eq: Phys V_0 projector}
(P^{\text{phys}; V_0})_{ij, kl} &= C^{\text{phys}; V_0}_{ij} C^{\text{phys}; V_0}_{kl} \, , \\ \label{eq: Phys V_H projector}
(P^{\text{phys}; V_H})_{ij, kl} &= \sum_{a=1}^{D-1} C^{\text{phys}; V_H}_{ij, \, a} C^{\text{phys}; V_H}_{kl, \, a} \, , \\ \label{eq: Phys V_2 projector}
(P^{\text{phys}; V_2})_{ij, kl} &= \sum_{a=1}^{\text{dim}V_2} C^{\text{phys}; V_2}_{ij, \, a} C^{\text{phys}; V_2}_{kl, \, a} \, .
\end{align}
In this section we find explicit formulae for these projectors. The $V_0$ and $V_H$ projectors are constructed by finding the Clebschs on the RHS of \eqref{eq: Phys V_0 projector} and \eqref{eq: Phys V_H projector} explicitly. In the case of the $V_2$ projector things are not so simple as the Clebsch is not so easily calculable, none-the-less we are able to construct the projector using general properties of Clebsch coefficients and other known projectors, without precise knowledge of the $V_2$ Clebsch itself.  

To find $C^{\text{phys}; V_0}_{ij}$ and $C^{\text{phys}; V_H}_{ij, \, a}$ we first write down a representation theory basis for $V_D \otimes V_D$ in terms of the change of basis coefficients given in \eqref{eq: C0} and \eqref{eq: Ca} as was done in \cite{ramgoolam2019permutation} i.e. a basis that transforms like the RHS of \eqref{eq: V_D V_D decomp},
\begin{align} \label{eq: S V0 1}
S^{V_{0}; 1} &\equiv \sum_{i,j = 1}^D C_{0, i} C_{0, j} M_{ij} = \frac{1}{D} \sum_{i,j = 1}^D e_i \otimes e_j \, , \\ 
S^{V_{0}; 2} &\equiv  \frac{1}{\sqrt{D-1}} \sum_{a=1}^{D-1} \sum_{i, j = 1}^D C_{a, i} C_{a, j} M_{ij} =  \frac{1}{\sqrt{D-1}} \sum_{a=1}^{D-1} E_a \otimes E_a \, , \\
S^{V_{H}; 1}_a &\equiv \sum_{i, j =1}^{D}  C_{0, i} C_{a, j} M_{ i j } = \frac{1}{\sqrt{D}} \sum_{i=1}^D e_i \otimes E_a \, , \\
S^{V_{H}; 2}_a &\equiv \sum_{i, j =1}^{D}  C_{a, i} C_{0, j} M_{ i j } =  \frac{1}{\sqrt{D}} \sum_{i=1}^D E_a \otimes e_i \, , \\ \nonumber
S^{V_{H}; 3}_a &\equiv \sqrt{\frac{D}{D-2}} \sum_{b,c =1}^{D-1} \sum_{i,j,k=1}^D C_{a, k} C_{b, k} C_{c, k} C_{b, i} C_{c, j} M_{ij} \\ \label{eq: S VH 3}
&\qquad = \sqrt{\frac{D}{D-2}} \sum_{b,c =1}^{D-1} \sum_{i=1}^D C_{a,i} C_{b,i} C_{c,i} E_b \otimes E_c \, .
\end{align}
We also note the orthogonal decomposition
\begin{align}
V_D \otimes V_D \cong \text{Sym}^2(V_D) \oplus \Lambda^2(V_D) \cong V^{\text{phys}} \oplus V^{\text{diag}} \oplus \Lambda^2(V_D), 
\end{align}
in which $V^{\text{diag}}$ is the subspace of diagonal matrix elements and $\Lambda^2(V_D)$ is the antisymmetric subspace of $V_D \otimes V_D$. 
 
Define further representation variables $S^{\text{diag}; V_0}$ and $S^{\text{diag}; V_H}_a$ composed of the diagonal elements of $M_{ij}$, that transform according to the first and second terms on the RHS of the $V^{\text{diag}}$ decomposition respectively
\begin{align}\label{DiagVars} 
S^{\text{diag}; V_0} &\equiv \frac{1}{\sqrt{D}} \sum_{i=1}^D e_i \otimes e_i, \\
S^{\text{diag}; V_H}_a &\equiv E_a \otimes E_a.
\end{align}
Using the inner product on $V_D \otimes V_D$ in \eqref{eq: vd vd inner prod} We can express these in terms of the representation variables  
\begin{align} \nonumber
S^{\text{diag}; V_0} &= (S^{\text{diag}; V_0}, S^{V_{0}; 1}) S^{V_{0}; 1} + (S^{\text{diag}; V_0}, S^{V_{0}; 2}) S^{V_{0}; 2} \\
&= \frac{1}{\sqrt{D}} S^{V_{0}; 1} + \sqrt{\frac{D-1}{D}} S^{V_{0}; 2} \, ,
\end{align} 
and
\begin{align} \nonumber
S^{\text{diag}; V_H}_a &= \frac{1}{2} \sum_{b = 1}^{D-1} \big( (S^{\text{diag}; V_H}_a, S^{V_{H}; 1}_b) S^{V_{H}; 1}_b + (S^{\text{diag}; V_H}_a, S^{V_{H}; 2}_b) S^{V_{H}; 2}_b \big) + \sum_{b = 1}^{D-1} (S^{\text{diag}; V_H}_a, S^{V_{H}; 3}_b) S^{V_{H}; 3}_b \\ \nonumber
&= \frac{1}{2} \sum_{b = 1}^{D-1} \Big( \sqrt{\frac{2}{D}} \delta_{ab} S^{V_{H}; 1}_b + \sqrt{\frac{2}{D}} \delta_{ab} S^{V_{H}; 2}_b \Big) + \sum_{b = 1}^{D-1} \sqrt{\frac{D-2}{D}} \delta_{ab} S^{V_{H}; 3}_b \\ 
&= \frac{1}{\sqrt{2D}} \big( S^{V_{H}; 1}_a + S^{V_{H}; 2}_a \big) +  \sqrt{\frac{D-2}{D}} S^{V_{H}; 3}_a \, .
\end{align}

The physical variables $S^{\text{phys}; V_0}$ and $S^{\text{phys}; V_H}_a$ span the orthogonal complement of the diagonal variables in the $V_0$ and $V_H$ subspaces (as given in \eqref{DiagVars})  of $\text{Sym}^2(V_D)$
\begin{align} \label{eq: physical var in terms of M 1} \nonumber
S^{\text{phys}; V_0} &=  \sqrt{\frac{D-1}{D}} S^{V_{0}; 1} - \frac{1}{\sqrt{D}} S^{V_{0}; 2} \\ 
&=\sqrt{\frac{D-1}{D}} \sum_{i,j = 1}^D C_{0,i} C_{0,j} M_{ij} - \frac{1}{\sqrt{D(D-1)}} \sum_{a=1}^{D-1} \sum_{i,j =1}^D C_{a,i} C_{a,j} M_{ij} \, , \\ \label{eq: physical var in terms of M 2} \nonumber
S^{\text{phys}; V_H}_a &= \sqrt{\frac{D-2}{2D}} \big( S^{V_{H}; 1}_{a} + S^{V_{H}; 2}_{a} \big) - \sqrt{\frac{2}{D}}  S^{V_{H}; 3}_{a} \\
&= \sqrt{\frac{D-2}{2D^2}} \sum_{i,j=1}^D \big( C_{a,i} + C_{a,j} \big) M_{ij} - \sqrt{\frac{2}{D}} \sum_{i,j=1}^D \sum_{b,c=1}^{D-1} C_{b,i} C_{c,j} C^{HH \rightarrow H}_{b,c \hspace{10pt} a} M_{ij} \, .
\end{align}
From \eqref{eq: physical var in terms of M 1} and \eqref{eq: physical var in terms of M 2} we can read off the Clebsch coefficients needed in the construction of the projectors \eqref{eq: Phys V_0 projector} and \eqref{eq: Phys V_H projector}:
\begin{align} \label{eq: v0 clebsch}
S^{\text{phys}; V_0} = \sum_{i,j} C^{\text{phys}; V_0}_{ij} M_{ij} \, &\Rightarrow \, C^{\text{phys}; V_0}_{ij} = \sqrt{\frac{D-1}{D}} C_{0,i} C_{0,j} - \frac{1}{\sqrt{D(D-1)}} \sum_{a=1}^{D-1} C_{a,i} C_{a,j} \, , \\ \nonumber
S^{\text{phys}; V_H}_a = \sum_{i,j} C^{\text{phys}; V_H}_{ij, \, a} M_{ij} \, &\Rightarrow \, C^{\text{phys}; V_H}_{ij, \, a} =  \sqrt{\frac{(D-2)}{2D^2}} \big( C_{a,i} + C_{a,j} \big) \\ \label{eq: vh clebsch}
& \hspace{3.5cm} - \sqrt{\frac{2}{(D-2)}} \sum_{b,c=1}^{D-1} \sum_{k=1}^D C_{b,i} C_{c,j} C_{a,k} C_{b,k} C_{c,k} \, .
\end{align}

Although we do not know the $V_2$ Clebsch coefficients $C^{HH \rightarrow V_2}_{b,c \hspace{10pt} a}$ appearing in the remaining physical variables
\begin{align} \label{eq: physical var in terms of M 3}
S^{\text{phys}; V_2}_a &= S^{V_2}_a = \sum_{i,j = 1}^D \sum_{b,c =1}^{D-1} C_{b,i} C_{c,j} C^{HH \rightarrow V_2}_{b,c \hspace{10pt} a} M_{ij},
\end{align}
we do know that they possess the usual Clebsch orthogonality property 
\begin{align}
\sum_{b, c = 1}^{D-1} C^{HH \rightarrow V_2}_{b \, c, \hspace{10pt} a_1} C^{HH \rightarrow V_2}_{b \, c, \hspace{10pt} a_2} = \delta_{a_1 a_2} \, .
\end{align}
Also, we are able to write an expression for the $V_2$ projector using the decomposition
\begin{align}
\text{Sym}^2(V_H) = V_0 \oplus V_H \oplus V_2.
\end{align}
This allows us to express the $V_2$ projector in terms of the projectors from $V_H \otimes V_H$ to $\text{Sym}^2(V_H)$, $V_0$ and $V_H$
\begin{align} \label{eq: phys v2 proj}
P^{\text{phys}; V_2} = P^{V_H, V_H \rightarrow V_2} = \big( 1 - P^{V_H, V_H \rightarrow V_0} - P^{V_H, V_H \rightarrow V_H}  \big) P^{V_H, V_H \rightarrow \text{Sym}^2(V_H)}.
\end{align}
The projectors on the RHS of this expression are given by
\begin{align}
\big( P^{V_H, V_H \rightarrow \text{Sym}^2(V_H)} \big)_{ab, cd} &= \frac{1}{2} \big( \delta_{ac} \delta_{bd} - \delta_{ad} \delta_{bc} \big) \, , \\
\big( P^{V_H, V_H \rightarrow V_0} \big)_{ab, cd} &= \frac{1}{D-1} \delta_{a b} \delta_{c d} \, , \\
\big( P^{V_H, V_H \rightarrow V_H} \big)_{ab, cd} &= \frac{D}{D-2} \sum_{e = 1}^{D-1} \sum_{i, j = 1}^D C_{a,i} C_{b,i} C_{e,i} C_{c,j} C_{d,j} C_{e,j} \, .
\end{align}
The derivations of these expressions are given in section 2.2 of \cite{ramgoolam2019permutation}

The action of the physical $V_0, V_H$ projectors, given by the square of the Clebsch coefficients in \eqref{eq: v0 clebsch} and \eqref{eq: vh clebsch}, along with the $V_2$ projector given in \eqref{eq: phys v2 proj} can be found by acting on a generic state $e_i \otimes e_j$ in $V_D \otimes V_D$. Doing so leaves us with the following delta expressions
\begin{align} \label{eq: Phys projectors final v0}
(P^{\text{phys}; V_0})_{ij, kl} &= \frac{1}{D(D-1)} \big( \delta_{ij} - 1 \big) \big( \delta_{kl} - 1 \big), \\ \label{eq: Phys projectors final vh}
(P^{\text{phys}; V_H})_{ij, kl} &= \frac{1}{2(D-1)} \big( 1 - \delta_{ij} \big) \big( 1- \delta_{kl} \big) \big( \delta_{ik} + \delta_{il} + \delta_{jk} + \delta_{jl} - \frac{4}{D} \big), \\ \nonumber
(P^{\text{phys}; V_2})_{ij, kl} &= \frac{1}{D-2} \Big( -D \delta_{ij} \delta_{jk} \delta_{kl} + \delta_{ij} \delta_{ik} + \delta_{ij} \delta_{il} + \delta_{ik} \delta_{il} + \delta_{jk} \delta_{jl} \\ \label{eq: Phys projectors final v2}
&+ \frac{1}{D-1} \big( \delta_{ij} - 1 \big) \big( \delta_{kl} - 1 \big) - \frac{1}{2} \big( \delta_{ik} + \delta_{il} + \delta_{jk} + \delta_{jl} \big) \Big).
\end{align}

The projector from $V_D \otimes V_D$ to the entire physical subspace, projects general $D \times D$ matrices onto the space of symmetric matrices with vanishing diagonal. The action of this projector on a general state $e_k \otimes e_l$ can be written using the inner product \eqref{eq: vd vd inner prod}, 
\begin{align} \nonumber
P^{\text{phys}} e_k \otimes e_l &= \frac{1}{2} \sum_{i < j}^D e_i \otimes e_j (e_i \otimes e_j, e_k \otimes e_l) \\
&= \frac{1}{2} (e_k \otimes e_l + e_l \otimes e_k ) - \delta_{kl} (e_k \otimes e_l) \, .
\end{align}
Such that it can be written as a delta expression that acts by unrestricted summation as
\begin{align}
(P^{\text{phys}})_{ij, kl} = \frac{1}{2} \big( \delta_{ik} \delta_{jl} + \delta_{il} \delta_{jk} \big) - \delta_{ij} \delta_{jk} \delta_{kl}.
\end{align}
As expected, given the orthogonality of the physical projectors \eqref{eq: Phys projectors final v0} - \eqref{eq: Phys projectors final v2} 
\begin{align}
P^{\text{phys}} = P^{\text{phys}; V_0} + P^{\text{phys}; V_H} + P^{\text{phys}; V_2}.
\end{align}

\subsection{The action and physical projectors}
The action of the most general permutation invariant Gaussian matrix model of symmetric matrices with vanishing diagonal is given by
\begin{align} \label{eq: M action}
\mathcal{S} = \tau_1 \sum_{i, j = 1}^D M^{\text{phys}}_{ij}M^{\text{phys}}_{ij} + \tau_2 \sum_{i, j, k = 1}^D M^{\text{phys}}_{ij}M^{\text{phys}}_{jk} + \tau_3 \sum_{i, j, k, l = 1}^D M^{\text{phys}}_{ij} M^{\text{phys}}_{kl} - \mu \sum_{i, j = 1}^D M^{\text{phys}}_{ij},
\end{align}
where $\mu$ is the linear coupling strength and $\tau_1, \tau_2, \tau_3$ are quadratic couplings. Define the partition function
\begin{align}
\mathcal{Z} \equiv \int \dd M e^{- \mathcal{S}},
\end{align}
where the measure is defined as   
\begin{align}
\dd M \equiv \prod_{i < j} \dd M_{ij}.
\end{align}
We are interested in calculating permutation invariant expectation values of operators composed of the $M_{ij}$ variables. These are defined as
\begin{align} \label{eq: expectation values def}
\langle \mathcal{O}_{\alpha}(M) \rangle \equiv \frac{1}{\mathcal{Z}} \int \dd M \mathcal{O}_{\alpha}(M) e^{- \mathcal{S}}.
\end{align}

In order to evaluate \eqref{eq: expectation values def} we must factorise the integrals that appear. The action as written in \eqref{eq: M action} contains $\frac{D(D-1)}{2}$ matrix variables mixed in a non-trivial way. The solution to this problem  exploits the decomposition \eqref{eq: mat decomp}. 
The presence of one copy of $V_0$ means that there is one linear term with coefficient $\mu_{V_0}$ employing the Clebsch-Gordan coefficient for $V_0$ in $ V^{ \text{phys}} $. There are three quadratic terms in the action  which employ the projectors for each of the three irreducible representations 
\begin{align} \label{eq: action with physical projectors} \nonumber
\mathcal{S} = &\sum_{i,j,k,l} \frac{1}{2} \Big( \tau_{V_0} M_{ij} (P^{\text{phys}; V_0})_{ij, kl} M_{kl} + \tau_{V_H} M_{ij} (P^{\text{phys}; V_H})_{ij, kl} M_{kl} + \tau_{V_2} M_{ij} (P^{\text{phys}; V_2})_{ij, kl} M_{kl} \Big) \\ 
&- \sum_{i,j} \mu_{V_0} C^{\text{phys}; V_0}_{ij} M_{ij}.
\end{align}
Applying these projectors amounts to rewriting the action in terms of the representation theory variables $ S^{\text{phys} ; V_0}, S^{\text{phys} ; V_H}_a ,  S^{\text{phys} ; V_2}_a $ as given in equations 
\eqref{eq: v0 clebsch}, \eqref{eq: vh clebsch} and   \eqref{eq: physical var in terms of M 3}  : 
\begin{align}
\mathcal{S} =&- \mu_{V_0} S^{\text{phys} ; V_0} + \frac{\tau_{V_0}}{2} S^{\text{phys} ; V_0} S^{\text{phys} ; V_0} +  \frac{\tau_{V_H}}{2} \sum_{a=1}^{D-1} S^{\text{phys} ; V_H}_a S^{\text{phys} ; V_H}_a + \frac{\tau_{V_2}}{2} \sum_{a=1}^{\text{dim}V_2} S^{\text{phys} ; V_2}_a S^{\text{phys} ; V_2}_a.
\end{align}
With this change of variables the partition function reads
\begin{align}
\mathcal{Z} \equiv \int \dd S e^{- \mathcal{S}},
\end{align}
with
\begin{align}
\dd S = \dd S^{\text{phys} ; V_0} \prod_{a_1=1}^{D-1} \dd S^{\text{phys} ; V_H}_{a_1} \prod_{a_2=1}^{\text{dim} V_2} \dd S^{\text{phys} ; V_2}_{a_2}. 
\end{align}
The factorised expression permits the application of standard techniques of Gaussian integration. Substituting the expressions for the projectors into \eqref{eq: action with physical projectors} and performing the summations we can also write the action in terms of the full $V_D \otimes V_D$ representation variables
\begin{align} \label{eq: physical action wrt S vars} \nonumber
\mathcal{S} = &- \mu_{V_0} \Big( \sqrt{\frac{D-1}{D}} S^{V_0 ; 1} - \frac{1}{\sqrt{D}} S^{V_0 ; 2} \Big) + \frac{\tau_{V_0}}{2} \Big( \sqrt{\frac{D-1}{D}} S^{V_0 ; 1} - \frac{1}{\sqrt{D}} S^{V_0 ; 2}  \Big)^2  \\ 
&+  \frac{\tau_{V_H}}{2} \sum_{a=1}^{D-1} \Big( \sqrt{\frac{D-2}{2D}} (S^{V_H ; 1}_a + S^{V_H ; 2}_a) - \sqrt{\frac{2}{D}} S^{V_H ; 3}_a \Big)^2 + \frac{\tau_{V_2}}{2} \sum_{a=1}^{\text{dim}V_2} S^{V_2}_a S^{V_2}_a.
\end{align}

\subsection{Observables and correlators}
The observables of our theory $\mathcal{O}_{\alpha}$ are the permutation invariant functions of the matrix variables $M_{ij}$, they obey
\begin{align}
	\mathcal{O}_{\alpha}(M_{ij}) = \mathcal{O}_{\alpha}(M_{\sigma(i) \sigma(j)}), \qquad \forall \, \sigma \in S_D.
\end{align}
The physical permutation invariant observables of order $k$ are in one-to-one correspondence with undirected, loopless multigraphs with $k$ edges.  Each matrix describes an edge connecting vertices labelled by the row and column indices of the matrix. Requiring the matrices to be symmetric is equivalent to considering undirected edges. The further requirement that the matrices have vanishing diagonal entries is equivalent to restricting to loopless multigraphs.

Below we list the complete set of quadratic, cubic and quartic graphs of general matrices that survive the projection to the physical subspace. 
\begin{align}
M^{\text{phys}}_{ij} = ( P^{\text{phys}} )_{ij, kl} M_{kl}.
\end{align}
The counting of these graphs organised by number of edges is given by the OEIS sequence A050535. The three quadratic observables are
\begin{equation} \label{Eqn: quadratic matrix multi-graph diagrams} 
\begin{array}{ccc}
\begin{tikzpicture} [baseline]
\begin{scope}[]
\draw[draw=black] (0,0)node[circle, fill=black, inner sep=1pt, draw=black] {} to[bend left] (1,0);
\draw[draw=black] (0,0) to[bend right] (1,0)node[circle, fill=black, inner sep=1pt, draw=black] {};
\end{scope}
\end{tikzpicture} \hspace{1cm} &
\begin{tikzpicture}	
\begin{scope}[]
\draw[draw=black] (0,0)node[circle, fill=black, inner sep=1pt, draw=black] {} to[bend left] (1,0)node[circle, fill=black, inner sep=1pt, draw=black] {};
\draw[draw=black] (1,0) to[bend left] (2,0)node[circle, fill=black, inner sep=1pt, draw=black] {};
\end{scope}
\end{tikzpicture}  &
\begin{tikzpicture}	
\begin{scope}[]
\draw[draw=black] (0,0)node[circle, fill=black, inner sep=1pt, draw=black] {} to[bend left] (1,0)node[circle, fill=black, inner sep=1pt, draw=black] {};
\draw[draw=black] (1.5,0)node[circle, fill=black, inner sep=1pt, draw=black] {} to[bend left] (2.5,0)node[circle, fill=black, inner sep=1pt, draw=black] {};
\end{scope}
\end{tikzpicture} \\
\sum_{i,j} M_{ij}^2 \, ,  &
\sum_{i,j,k} M_{ij} M_{jk} \, ,  &
\sum_{i,j,k,l} M_{ij} M_{kl} \, .  
\end{array}
\end{equation}
The eight cubic observables are
\begin{equation} \nonumber
\begin{array}{ccc}
\begin{tikzpicture} [baseline]
\begin{scope}[]
\draw[draw=black] (0,0)node[circle, fill=black, inner sep=1pt, draw=black] {} to[bend left] (1,0);
\draw[draw=black] (0,0) to[bend right] (1,0)node[circle, fill=black, inner sep=1pt, draw=black] {};
\draw[draw=black] (0,0) to (1,0);
\end{scope}
\end{tikzpicture} &
\begin{tikzpicture}	
\begin{scope}[]
\draw[draw=black] (0,0)node[circle, fill=black, inner sep=1pt, draw=black] {} to[bend left] (1,0)node[circle, fill=black, inner sep=1pt, draw=black] {};
\draw[draw=black] (1,0) to[bend left] (2,0)node[circle, fill=black, inner sep=1pt, draw=black] {};
\draw[draw=black] (0,0) to [bend right] (1,0);
\end{scope}
\end{tikzpicture}  &
\begin{tikzpicture}	
\begin{scope}[]
\draw[draw=black] (0,0)node[circle, fill=black, inner sep=1pt, draw=black] {} to[bend left] (1,0)node[circle, fill=black, inner sep=1pt, draw=black] {};
\draw[draw=black] (1,0) to[bend left] (2,0)node[circle, fill=black, inner sep=1pt, draw=black] {};
\draw[draw=black] (0,0) to [bend left] (2,0);
\end{scope}
\end{tikzpicture}  \\
\mathcal{O}_1 = \sum_{i,j} M_{ij}^3 \, , &
\mathcal{O}_2 = \sum_{i,j,k} M_{ij}^2 M_{jk} \, , &
\mathcal{O}_3 = \sum_{i,j,k} M_{ij} M_{ik} M_{jk} \, , 
\end{array}
\end{equation}

\begin{equation} \nonumber 
\begin{array}{ccc}
\begin{tikzpicture} [baseline]
\begin{scope}[]
\draw[draw=black] (0,0)node[circle, fill=black, inner sep=1pt, draw=black] {} to[bend left] (1,0)node[circle, fill=black, inner sep=1pt, draw=black] {};
\draw[draw=black] (1.5,0)node[circle, fill=black, inner sep=1pt, draw=black] {} to[bend left] (2.5,0)node[circle, fill=black, inner sep=1pt, draw=black] {};
\draw[draw=black] (0,0) to [bend right] (1,0);
\end{scope}
\end{tikzpicture}  &
\begin{tikzpicture}	
\begin{scope}[]
\draw[draw=black] (0,0)node[circle, fill=black, inner sep=1pt, draw=black] {} to[bend left] (1,0)node[circle, fill=black, inner sep=1pt, draw=black] {};
\draw[draw=black] (1,0) to[bend left] (2,0)node[circle, fill=black, inner sep=1pt, draw=black] {};
\draw[draw=black] (2,0) to[bend left] (3,0)node[circle, fill=black, inner sep=1pt, draw=black] {};
\end{scope}
\end{tikzpicture}  &
\begin{tikzpicture}	
\begin{scope}[]
\draw[draw=black] (0,0)node[circle, fill=black, inner sep=1pt, draw=black] {} to[bend left] (1,0)node[circle, fill=black, inner sep=1pt, draw=black] {};
\draw[draw=black] (0,0) to[bend left] (2,0)node[circle, fill=black, inner sep=1pt, draw=black] {};
\draw[draw=black] (0,0) to[bend left] (3,0)node[circle, fill=black, inner sep=1pt, draw=black] {};
\end{scope}
\end{tikzpicture} \\
\mathcal{O}_4 = \sum_{i,j,k,l} M_{ij}^2 M_{kl} \, , &
\mathcal{O}_5 = \sum_{i,j,k,l} M_{ij} M_{jk} M_{kl} \, , &
\mathcal{O}_6 = \sum_{i,j,k,l} M_{ij} M_{ik} M_{il} \, , 
\end{array}
\end{equation}

\begin{equation} \label{Eqn: cubic matrix multi-graph diagrams} 
\begin{array}{cc}
\begin{tikzpicture} [baseline]
\begin{scope}[]
\draw[draw=black] (0,0)node[circle, fill=black, inner sep=1pt, draw=black] {} to[bend left] (1,0)node[circle, fill=black, inner sep=1pt, draw=black] {};
\draw[draw=black] (1,0) to[bend left] (2,0)node[circle, fill=black, inner sep=1pt, draw=black] {};
\draw[draw=black] (2.5,0)node[circle, fill=black, inner sep=1pt, draw=black] {} to[bend left] (3.5,0)node[circle, fill=black, inner sep=1pt, draw=black] {};
\end{scope}
\end{tikzpicture} &
\begin{tikzpicture}	
\begin{scope}[]
\draw[draw=black] (0,0)node[circle, fill=black, inner sep=1pt, draw=black] {} to[bend left] (1,0)node[circle, fill=black, inner sep=1pt, draw=black] {};
\draw[draw=black] (1.5,0)node[circle, fill=black, inner sep=1pt, draw=black] {} to[bend left] (2.5,0)node[circle, fill=black, inner sep=1pt, draw=black] {};
\draw[draw=black] (3,0)node[circle, fill=black, inner sep=1pt, draw=black] {} to[bend left] (4,0)node[circle, fill=black, inner sep=1pt, draw=black] {};
\end{scope}
\end{tikzpicture} \\
\mathcal{O}_7 = \sum_{i,j,k,l,m} M_{ij} M_{jk} M_{lm} \, , &
\mathcal{O}_8 = \sum_{\substack{i,j,k,l, \\ m,n}} M_{ij} M_{kl} M_{mn} \, .
\end{array}
\end{equation}
The 23 quartic observables are
\begin{equation} \label{Eqn: quartic matrix multi-graph diagrams 1} \nonumber
\begin{array}{cccc}
\begin{tikzpicture} [baseline]
\begin{scope}[]
\draw[draw=black] (0,0)node[circle, fill=black, inner sep=1pt, draw=black] {} to [bend left = 75] (1,0);
\draw[draw=black] (0,0) to[bend right] (1,0)node[circle, fill=black, inner sep=1pt, draw=black] {};
\draw[draw=black] (0,0) to [bend left] (1,0);
\draw[draw=black] (0,0) to [bend right = 75] (1,0);
\end{scope}
\end{tikzpicture} &
\begin{tikzpicture}	
\begin{scope}[]
\draw[draw=black] (0,0)node[circle, fill=black, inner sep=1pt, draw=black] {} to[bend left] (1,0)node[circle, fill=black, inner sep=1pt, draw=black] {};
\draw[draw=black] (1,0) to[bend left] (2,0)node[circle, fill=black, inner sep=1pt, draw=black] {};
\draw[draw=black] (0,0) to [bend right] (1,0);
\draw[draw=black] (1,0) to [bend right] (2,0);
\end{scope}
\end{tikzpicture}  &
\begin{tikzpicture}	
\begin{scope}[]
\draw[draw=black] (0,0)node[circle, fill=black, inner sep=1pt, draw=black] {} to[bend left] (1,0)node[circle, fill=black, inner sep=1pt, draw=black] {};
\draw[draw=black] (1,0) to[bend left] (2,0)node[circle, fill=black, inner sep=1pt, draw=black] {};
\draw[draw=black] (1,0) to (2,0);
\draw[draw=black] (1,0) to [bend right] (2,0);
\end{scope}
\end{tikzpicture} &
\begin{tikzpicture}	
\begin{scope}[]
\draw[draw=black] (0,0)node[circle, fill=black, inner sep=1pt, draw=black] {} to[bend left] (1,0)node[circle, fill=black, inner sep=1pt, draw=black] {};
\draw[draw=black] (1,0) to[bend left] (2,0)node[circle, fill=black, inner sep=1pt, draw=black] {};
\draw[draw=black] (0,0) to [bend left] (2,0);
\draw[draw=black] (1,0) to [bend right] (2,0);
\end{scope}
\end{tikzpicture}  \\
\mathcal{O}_9 = \sum_{i,j} M_{ij}^4 \, , &
\mathcal{O}_{10} = \sum_{i,j,k} M_{ij}^2 M_{jk}^2 \, , &
\mathcal{O}_{11} = \sum_{i,j,k,l} M_{ij} M_{jk}^3 \, , &
\mathcal{O}_{12} = \sum_{i,j,k,l,m} M_{ij} M_{ik} M_{jk}^2 \, , 
\end{array}
\end{equation}

\begin{equation} \label{Eqn: quartic matrix multi-graph diagrams 2} \nonumber
\begin{array}{ccc}
\begin{tikzpicture} [baseline]
\begin{scope}[]
\draw[draw=black] (0,0)node[circle, fill=black, inner sep=1pt, draw=black] {} to[bend left] (1,0)node[circle, fill=black, inner sep=1pt, draw=black] {};
\draw[draw=black] (1,0)node[circle, fill=black, inner sep=1pt, draw=black] {} to[bend right] (3,0)node[circle, fill=black, inner sep=1pt, draw=black] {};
\draw[draw=black] (1,0) to [bend left] (3,0);
\draw[draw=black] (1,0) to [bend left] (2,0) node[circle, fill=black, inner sep=1pt, draw=black] {};
\end{scope}
\end{tikzpicture} &
\begin{tikzpicture}	
\begin{scope}[]
\draw[draw=black] (0,0)node[circle, fill=black, inner sep=1pt, draw=black] {} to[bend left] (1,0)node[circle, fill=black, inner sep=1pt, draw=black] {};
\draw[draw=black] (2,0)node[circle, fill=black, inner sep=1pt, draw=black] {} to[bend left] (3,0)node[circle, fill=black, inner sep=1pt, draw=black] {};
\draw[draw=black] (2,0) to [bend right] (3,0);
\draw[draw=black] (2,0) to (3,0);
\end{scope}
\end{tikzpicture}  &
\begin{tikzpicture}	
\begin{scope}[]
\draw[draw=black] (0,0)node[circle, fill=black, inner sep=1pt, draw=black] {} to[bend left] (1,0)node[circle, fill=black, inner sep=1pt, draw=black] {};
\draw[draw=black] (2,0)node[circle, fill=black, inner sep=1pt, draw=black] {} to[bend left] (3,0)node[circle, fill=black, inner sep=1pt, draw=black] {};
\draw[draw=black] (1,0) to [bend right] (2,0);
\draw[draw=black] (1,0) to [bend left] (2,0);
\end{scope}
\end{tikzpicture} \\
\mathcal{O}_{14} = \sum_{i,j,k,l} M_{ij} M_{kl}^3 \, , &
\mathcal{O}_{15} = \sum_{i,j,k,l} M_{ij} M_{jk}^2 M_{kl} \, , &
\mathcal{O}_{16} = \sum_{i,j,k,l} M_{ij} M_{jk} M_{kl}^2 \, , 
\end{array}
\end{equation}

\begin{equation} \label{Eqn: quartic matrix multi-graph diagrams 3} \nonumber
\begin{array}{ccc}
\begin{tikzpicture} [baseline]	
\begin{scope}[]
\draw[draw=black] (0,0)node[circle, fill=black, inner sep=1pt, draw=black] {} to[bend left] (1,0)node[circle, fill=black, inner sep=1pt, draw=black] {};
\draw[draw=black] (2,0)node[circle, fill=black, inner sep=1pt, draw=black] {} to[bend left] (3,0)node[circle, fill=black, inner sep=1pt, draw=black] {};
\draw[draw=black] (1,0) to [bend left] (2,0);
\draw[draw=black] (2,0) to [bend right] (3,0);
\end{scope}
\end{tikzpicture}  &
\begin{tikzpicture} 
\begin{scope}[]
\draw[draw=black] (0,0)node[circle, fill=black, inner sep=1pt, draw=black] {} to[bend left] (1,0)node[circle, fill=black, inner sep=1pt, draw=black] {};
\draw[draw=black] (2,0)node[circle, fill=black, inner sep=1pt, draw=black] {} to[bend left] (3,0)node[circle, fill=black, inner sep=1pt, draw=black] {};
\draw[draw=black] (0,0) to [bend left] (2,0);
\draw[draw=black] (1,0) to [bend left] (2,0);
\end{scope}
\end{tikzpicture} &
\begin{tikzpicture}	
\begin{scope}[]
\draw[draw=black] (0,0)node[circle, fill=black, inner sep=1pt, draw=black] {} to[bend left] (1,0)node[circle, fill=black, inner sep=1pt, draw=black] {};
\draw[draw=black] (2,0)node[circle, fill=black, inner sep=1pt, draw=black] {} to[bend left] (3,0)node[circle, fill=black, inner sep=1pt, draw=black] {};
\draw[draw=black] (0,0) to [bend right] (1,0);
\draw[draw=black] (2,0) to [bend right] (3,0);
\end{scope}
\end{tikzpicture}  \\
\mathcal{O}_{13} = \sum_{i,j,k,l} M_{ij} M_{kj} M_{lj}^2 \, , &
\mathcal{O}_{17} = \sum_{i,j,k,l} M_{ij} M_{jk} M_{ik} M_{kl} \, , &
\mathcal{O}_{18} = \sum_{i,j,k,l} M_{ij}^2 M_{kl}^2 \, , 
\end{array}
\end{equation}

\begin{equation} \label{Eqn: quartic matrix multi-graph diagrams 4} \nonumber
\begin{array}{ccc}
\begin{tikzpicture} [baseline]
\begin{scope}[]
\draw[draw=black] (0,0)node[circle, fill=black, inner sep=1pt, draw=black] {} to[bend left] (1,0)node[circle, fill=black, inner sep=1pt, draw=black] {};
\draw[draw=black] (2,0)node[circle, fill=black, inner sep=1pt, draw=black] {} to[bend left] (3,0)node[circle, fill=black, inner sep=1pt, draw=black] {};
\draw[draw=black] (1,0) to [bend left] (2,0);
\draw[draw=black] (0,0) to [bend left] (3,0);
\end{scope}
\end{tikzpicture} &
\begin{tikzpicture}	
\begin{scope}[]
\draw[draw=black] (0,0)node[circle, fill=black, inner sep=1pt, draw=black] {} to[bend left] (2,0)node[circle, fill=black, inner sep=1pt, draw=black] {};
\draw[draw=black] (1,0)node[circle, fill=black, inner sep=1pt, draw=black] {} to[bend left] (2,0);
\draw[draw=black] (2,0) to[bend left] (3,0)node[circle, fill=black, inner sep=1pt, draw=black] {};
\draw[draw=black] (2,0) to[bend left] (4,0)node[circle, fill=black, inner sep=1pt, draw=black] {};
\end{scope}
\end{tikzpicture}  &
\begin{tikzpicture} 
\begin{scope}[]
\draw[draw=black] (0,0)node[circle, fill=black, inner sep=1pt, draw=black] {} to[bend left] (3,0)node[circle, fill=black, inner sep=1pt, draw=black] {};
\draw[draw=black] (1,0)node[circle, fill=black, inner sep=1pt, draw=black] {} to[bend left] (2,0)node[circle, fill=black, inner sep=1pt, draw=black] {};
\draw[draw=black] (2,0) to[bend left] (3,0);
\draw[draw=black] (2,0) to[bend right] (4,0)node[circle, fill=black, inner sep=1pt, draw=black] {};
\end{scope}
\end{tikzpicture} \\
\mathcal{O}_{19} = \sum_{i,j,k,l} M_{ij} M_{jk} M_{kl} M_{li} \, , &
\mathcal{O}_{20} = \sum_{i,j,k,l,m} M_{ik} M_{jk} M_{lk} M_{mk} \, , &
\mathcal{O}_{21} = \sum_{i,j,k,l,m} M_{il} M_{jk} M_{lk} M_{mk}  \, , 
\end{array}
\end{equation}

\begin{equation} \label{Eqn: quartic matrix multi-graph diagrams 5} \nonumber
\begin{array}{ccc}
\begin{tikzpicture} [baseline]
\begin{scope}[]
\draw[draw=black] (0,0)node[circle, fill=black, inner sep=1pt, draw=black] {} to[bend left] (1,0)node[circle, fill=black, inner sep=1pt, draw=black] {};
\draw[draw=black] (2,0)node[circle, fill=black, inner sep=1pt, draw=black] {} to[bend left] (3,0)node[circle, fill=black, inner sep=1pt, draw=black] {};
\draw[draw=black] (3,0) to[bend left] (4,0)node[circle, fill=black, inner sep=1pt, draw=black] {};
\draw[draw=black] (2,0) to[bend left] (4,0);
\end{scope}
\end{tikzpicture}  &
\begin{tikzpicture}	
\begin{scope}[]
\draw[draw=black] (0,0)node[circle, fill=black, inner sep=1pt, draw=black] {} to[bend left] (1,0)node[circle, fill=black, inner sep=1pt, draw=black] {};
\draw[draw=black] (0,0) to[bend right] (1,0);
\draw[draw=black] (2,0)node[circle, fill=black, inner sep=1pt, draw=black] {} to[bend left] (3,0)node[circle, fill=black, inner sep=1pt, draw=black] {};
\draw[draw=black] (3,0) to[bend left] (4,0)node[circle, fill=black, inner sep=1pt, draw=black] {};
\end{scope}
\end{tikzpicture} &
\begin{tikzpicture}
\begin{scope}[]
\draw[draw=black] (0,0)node[circle, fill=black, inner sep=1pt, draw=black] {} to[bend left] (1,0)node[circle, fill=black, inner sep=1pt, draw=black] {};
\draw[draw=black] (2,0)node[circle, fill=black, inner sep=1pt, draw=black] {} to[bend left] (3,0)node[circle, fill=black, inner sep=1pt, draw=black] {};
\draw[draw=black] (3,0) to[bend left] (4,0)node[circle, fill=black, inner sep=1pt, draw=black] {};
\draw[draw=black] (3,0) to[bend right] (4,0);
\end{scope}
\end{tikzpicture} \\
\mathcal{O}_{22} = \sum_{i,j,k,l,m} M_{ij} M_{kl} M_{lm} M_{mk} \, , &
\mathcal{O}_{23} = \sum_{i,j,k,l,m} M_{ij}^2 M_{kl} M_{lm} \, , &
\mathcal{O}_{24} = \sum_{i,j,k,l,m} M_{ij} M_{kl} M_{lm}^2 \, , 
\end{array}
\end{equation}

\begin{equation} \label{Eqn: quartic matrix multi-graph diagrams 6} \nonumber
\begin{array}{cc}
\begin{tikzpicture} [baseline]
\begin{scope}[] 
\draw[draw=black] (0,0)node[circle, fill=black, inner sep=1pt, draw=black] {} to[bend left] (1,0)node[circle, fill=black, inner sep=1pt, draw=black] {};
\draw[draw=black] (1,0) to[bend left] (2,0)node[circle, fill=black, inner sep=1pt, draw=black] {};
\draw[draw=black] (2,0) to[bend left] (3,0)node[circle, fill=black, inner sep=1pt, draw=black] {};
\draw[draw=black] (3,0) to[bend left] (4,0)node[circle, fill=black, inner sep=1pt, draw=black] {};
\end{scope}
\end{tikzpicture}  &
\begin{tikzpicture}	
\begin{scope}[]
\draw[draw=black] (0,0)node[circle, fill=black, inner sep=1pt, draw=black] {} to[bend left] (1,0)node[circle, fill=black, inner sep=1pt, draw=black] {};
\draw[draw=black] (2,0)node[circle, fill=black, inner sep=1pt, draw=black] {} to[bend left] (3,0)node[circle, fill=black, inner sep=1pt, draw=black] {};
\draw[draw=black] (2,0) to[bend left] (4,0)node[circle, fill=black, inner sep=1pt, draw=black] {};
\draw[draw=black] (2,0) to[bend left] (5,0)node[circle, fill=black, inner sep=1pt, draw=black] {};
\end{scope}
\end{tikzpicture} \\
\mathcal{O}_{25} = \sum_{i,j,k,l,m} M_{ij} M_{jk} M_{kl} M_{lm} \, , &
\mathcal{O}_{26} = \sum_{i,j,k,l,m,n} M_{ij} M_{kl} M_{km} M_{kn} \, , 
\end{array}
\end{equation}

\begin{equation} \label{Eqn: quartic matrix multi-graph diagrams 7} \nonumber
\begin{array}{cc}
\begin{tikzpicture} [baseline]
\begin{scope}[]
\draw[draw=black] (0,0)node[circle, fill=black, inner sep=1pt, draw=black] {} to[bend left] (1,0)node[circle, fill=black, inner sep=1pt, draw=black] {};
\draw[draw=black] (1,0) to[bend left] (2,0)node[circle, fill=black, inner sep=1pt, draw=black] {};
\draw[draw=black] (3,0)node[circle, fill=black, inner sep=1pt, draw=black] {} to[bend left] (4,0)node[circle, fill=black, inner sep=1pt, draw=black] {};
\draw[draw=black] (4,0) to[bend left] (5,0)node[circle, fill=black, inner sep=1pt, draw=black] {};
\end{scope}
\end{tikzpicture} &
\begin{tikzpicture}	
\begin{scope}[]
\draw[draw=black] (0,0)node[circle, fill=black, inner sep=1pt, draw=black] {} to[bend left] (1,0)node[circle, fill=black, inner sep=1pt, draw=black] {};
\draw[draw=black] (0,0) to[bend right] (1,0);
\draw[draw=black] (2,0)node[circle, fill=black, inner sep=1pt, draw=black] {} to[bend left] (3,0)node[circle, fill=black, inner sep=1pt, draw=black] {};
\draw[draw=black] (4,0)node[circle, fill=black, inner sep=1pt, draw=black] {} to[bend left] (5,0)node[circle, fill=black, inner sep=1pt, draw=black] {};
\end{scope}
\end{tikzpicture} \\
\mathcal{O}_{27} = \sum_{i,j,k,l,m,n} M_{ij} M_{jk} M_{lm} M_{ln} \, , &
\mathcal{O}_{28} = \sum_{i,j,k,l,m,n} M_{ij}^2 M_{kl} M_{mn} \, , 
\end{array}
\end{equation}

\begin{equation} \label{Eqn: quartic matrix multi-graph diagrams 8} \nonumber 
\begin{array}{ccc}
\begin{tikzpicture} [baseline]	
\begin{scope}[]
\draw[draw=black] (0,0)node[circle, fill=black, inner sep=1pt, draw=black] {} to[bend left] (1,0)node[circle, fill=black, inner sep=1pt, draw=black] {};
\draw[draw=black] (2,0)node[circle, fill=black, inner sep=1pt, draw=black] {} to[bend left] (3,0)node[circle, fill=black, inner sep=1pt, draw=black] {};
\draw[draw=black] (3,0) to[bend left] (4,0)node[circle, fill=black, inner sep=1pt, draw=black] {};
\draw[draw=black] (4,0) to[bend left] (5,0)node[circle, fill=black, inner sep=1pt, draw=black] {};
\end{scope}
\end{tikzpicture} &
\begin{tikzpicture} 
\begin{scope}[]
\draw[draw=black] (0,0)node[circle, fill=black, inner sep=1pt, draw=black] {} to[bend left] (1,0)node[circle, fill=black, inner sep=1pt, draw=black] {};
\draw[draw=black] (1,0) to[bend left] (2,0)node[circle, fill=black, inner sep=1pt, draw=black] {};
\draw[draw=black] (3,0)node[circle, fill=black, inner sep=1pt, draw=black] {} to[bend left] (4,0)node[circle, fill=black, inner sep=1pt, draw=black] {};
\draw[draw=black] (5,0)node[circle, fill=black, inner sep=1pt, draw=black] {} to[bend left] (6,0)node[circle, fill=black, inner sep=1pt, draw=black] {};
\end{scope}
\end{tikzpicture} \\
\mathcal{O}_{29} = \sum_{i,j,k,l,m,n} M_{ij} M_{kl} M_{lm} M_{mn} \, , &
\mathcal{O}_{30} = \sum_{i,j,k,l,m,n,o} M_{ij} M_{jk} M_{lm} M_{no} \, , 
\end{array}
\end{equation}

\begin{equation} \label{Eqn: quartic matrix multi-graph diagrams 9} 
\begin{array}{c}
\begin{tikzpicture} [baseline]	
\begin{scope}[]
\draw[draw=black] (0,0)node[circle, fill=black, inner sep=1pt, draw=black] {} to[bend left] (1,0)node[circle, fill=black, inner sep=1pt, draw=black] {};
\draw[draw=black] (2,0)node[circle, fill=black, inner sep=1pt, draw=black] {} to[bend left] (3,0)node[circle, fill=black, inner sep=1pt, draw=black] {};
\draw[draw=black] (4,0)node[circle, fill=black, inner sep=1pt, draw=black] {} to[bend left] (5,0)node[circle, fill=black, inner sep=1pt, draw=black] {};
\draw[draw=black] (6,0)node[circle, fill=black, inner sep=1pt, draw=black] {} to[bend left] (7,0)node[circle, fill=black, inner sep=1pt, draw=black] {};
\end{scope}
\end{tikzpicture}  \\
\mathcal{O}_{31} = \sum_{i,j,k,l,m,n,o,p} M_{ij} M_{kl} M_{mn} M_{op} \, .
\end{array}
\end{equation}

\subsubsection{Expectation values}
We can write the partition function in the following form
\begin{align} \label{eq: action and its solution} \nonumber
\mathcal{Z} &= \int \dd S \exp( \sum_{\Lambda, a} \mu_{\Lambda, a } S^{\text{phys} ; \Lambda}_a - \frac{1}{2} \sum_{\Lambda, a} S^{\text{phys} ; \Lambda}_a \tau_{\Lambda} S^{\text{phys} ; \Lambda}_a ) \\ 
&= \frac{(2 \pi)^{\frac{D(D-1)}{2}}}{(\text{det} \tau)^{\frac{1}{2}}} \exp( \frac{1}{2} \sum_{\Lambda} \sum_a \mu_{\Lambda, a} \tau_{\Lambda}^{-1} \mu_{\Lambda, a}),
\end{align}
in which we have included linear couplings for all $S^{\text{phys}}$ variables. This is the usual trick employed to generate expectation values from the partition function, included so that we are able to source expectation values of all observables by taking derivatives with respect to the linear couplings. Of course, in order to recover the permutation invariant model all but the $V_0$ linear coupling are set to zero. The integral on the LHS is performed using standard techniques from multidimensional Gaussian integration.

Expectation values of the physical variables $\mathcal{O}(S)$ are defined by
\begin{align}
\langle \mathcal{O}(S) \rangle \equiv \frac{\int \dd S \mathcal{O}(S) e^{- \mathcal{S}} }{\int \dd S e^{- \mathcal{S}}}.
\end{align}
Calculating these by taking derivatives of the RHS of \eqref{eq: action and its solution}, with respect to $\mu_{\Lambda, a}$ we find for example, the linear expectation values are given by
\begin{align} \nonumber
\langle S^{\text{phys} ; \Lambda}_a \rangle &= \frac{1}{\mathcal{Z}} \int \dd S S^{\text{phys} ; \Lambda}_a e^{-\mathcal{S}} = \frac{1}{\mathcal{Z}} \frac{\partial \mathcal{Z}}{\partial \mu_{\Lambda, a}} \Big|_{\mu_{\Lambda, a} \neq 0\, \text{iff} \, \Lambda = V_0} \\
&= \tau_{\Lambda}^{-1} \mu_{\Lambda} \delta(\Lambda, V_0).
\end{align}
That is, the only non-zero linear expectation value is
\begin{align} \label{eq: 1 pt of phys vars}
\langle S^{\text{phys} ; V_0} \rangle = \tau_{V_0}^{-1} \mu_{V_0}.
\end{align}
For later convenience we define
\begin{align}
\widetilde{\mu}_{V_0} \equiv \tau_{V_0}^{-1} \mu_{V_0}.
\end{align}

Similarly, we can calculate the two-point function by taking two derivatives of the RHS of \eqref{eq: action and its solution}
\begin{align} \label{eq: 2 pt of phys vars} \nonumber
\langle S^{\text{phys} ; \Lambda_1}_a S^{\text{phys} ; \Lambda_2}_b \rangle &= \frac{1}{\mathcal{Z}} \int \dd S S^{\text{phys} ; \Lambda_1}_a S^{\text{phys} ; \Lambda_2}_b e^{-\mathcal{S}} = \frac{1}{\mathcal{Z}} \frac{\partial}{\partial \mu_{\Lambda_2, b}} \frac{\partial \mathcal{Z}}{\partial \mu_{\Lambda_1, a}} \Big|_{\mu_{\Lambda, a} \neq 0\, \text{iff} \, \Lambda = V_0} \\ \nonumber
&= \frac{1}{\mathcal{Z}} \frac{\partial}{\partial \mu_{\Lambda_2, b}} (\tau_{\Lambda_1})^{-1}_{c d} \mu_{\Lambda_1, c} \delta_{a c} \mathcal{Z} \Big|_{\mu_{\Lambda, a} \neq 0\, \text{iff} \, \Lambda = V_0}  \\ \nonumber
&= \tau_{\Lambda_1}^{-1} \delta_{ab} \delta(\Lambda_1, \Lambda_2) + \langle S^{\text{phys} ; \Lambda_1}_a \rangle \langle S^{\text{phys} ; \Lambda_2}_b \rangle \\
&= \tau_{\Lambda_1}^{-1} \delta_{ab} \delta(\Lambda_1, \Lambda_2) + \widetilde{\mu}^2_{V_0} \delta(\Lambda_1, V_0) \delta(\Lambda_2, V_0) \, . 
\end{align}

Defining 
\begin{align} \label{eq: F definition}
F(i, j) \equiv \sum_{a=1}^{D-1} C_{a,i } C_{a, j} = \big( \delta_{ij} - \frac{1}{D} \big),
\end{align}
we can find the one-point function of the $M^{\text{phys}}_{ij}$ by writing it in terms of the physical $S$ variables and applying \eqref{eq: 1 pt of phys vars} 
\begin{align} \nonumber
\langle M^{\text{phys}}_{ij}  \rangle &= C^{\text{phys}; V_0}_{ij} \langle S^{\text{phys} ; V_0} \rangle \\ 
&= \Big( \sqrt{\frac{D-1}{D^3}} - \frac{1}{\sqrt{D(D-1)}} F(i, j) \Big) \widetilde{\mu}_{V_0}.
\end{align}
Similarly, we can find the two-point function of the $M^{\text{phys}}_{ij}$ variables by writing each $M^{\text{phys}}$ in terms of the physical $S$ variables and then using \eqref{eq: 1 pt of phys vars} and \eqref{eq: 2 pt of phys vars} to evaluate each of the resulting expectation values
\begin{align} \label{eq: two point function} \nonumber
\langle M^{\text{phys}}_{ij}  &M^{\text{phys}}_{kl}  \rangle = \Big( \sqrt{\frac{D-1}{D^3}} - \frac{1}{\sqrt{D(D-1)}} F(i, j) \Big) \Big( \sqrt{\frac{D-1}{D^3}} - \frac{1}{\sqrt{D(D-1)}} F(k, l) \Big) \widetilde{\mu}^2_{V_0} \\ \nonumber
&+ \frac{1}{D} \Big( \frac{1}{D-1} F(i, j) F(k, l) - \frac{1}{D} \big( F(i, j) + F(k, l) \big) + \frac{D-1}{D^2} \Big) \tau^{-1}_{V_0} \\ \nonumber
&+ \frac{1}{2(D-2)} \big( 1- \delta_{ij} \big) \big( 1- \delta_{kl} \big) \big( F(i, k) + F(j, k) + F(i, l) + F(j,l) \big) \tau^{-1}_{V_H} \\ \nonumber
&+ \Bigg( \frac{1}{2} F ( i , k ) F ( j , l ) + \frac{1}{2} F ( i , l ) F ( j , k ) - \frac{D}{D-2} \sum_{ p , q =1}^D F ( i , p ) F (  j , p ) F ( k , q ) F ( l , q ) F ( p , q ) \\
&- \frac{1}{D-1} F ( i , j ) F ( k , l )  \Bigg) \tau^{-1}_{V_2}.
\end{align}
Evaluating these expressions for the linear and quadratic permutation invariant observables gives
\begin{align} \label{eq: one point function}
\sum_{i,j} \langle M^{\text{phys}}_{ij}  \rangle = \sqrt{D(D-1)} \widetilde{\mu}_{V_0},
\end{align}
and
\begin{align}
&\sum_{i,j} \langle M^{\text{phys}}_{ij} M^{\text{phys}}_{ij} \rangle = \widetilde{\mu}^2_{V_0} + \tau^{-1}_{V_0} + (D-1) \tau^{-1}_{V_H} + \frac{D(D-3)}{2} \tau^{-1}_{V_2}, \\
&\sum_{i,j,k} \langle M^{\text{phys}}_{ij} M^{\text{phys}}_{jk} \rangle_{\text{conn}} = (D-1) \widetilde{\mu}^2_{V_0} + (D-1) \tau^{-1}_{V_0} + \frac{(D-1)(D-2)}{2} \tau^{-1}_{V_H}, \\ 
&\sum_{i,j,k,l} \langle M^{\text{phys}}_{ij} M^{\text{phys}}_{kl} \rangle_{\text{conn}} = D (D-1) \widetilde{\mu}^2_{V_0} + D(D-1) \tau^{-1}_{V_0}.
\end{align}

\subsubsection{Wick's theorem} \label{sec: Wick's theorem}
Since our theory is Gaussian, Wick's theorem allows us to calculate higher point expectation values from the linear and quadratic expectation values. The theorem states that a $k$-point expectation value is equal to the sum of all possible ways of breaking that expectation value into expectation values of order one and two. For instance, defining the connected piece of the quadratic expectation value as
\begin{align}
\langle M^{\text{phys}}_{ij} M^{\text{phys}}_{kl} \rangle_{\text{conn}} &\equiv \langle M^{\text{phys}}_{ij} M^{\text{phys}}_{kl} \rangle - \langle M^{\text{phys}}_{ij} \rangle \langle M^{\text{phys}}_{kl} \rangle,
\end{align}
cubic expectation values are given by
\begin{align} \nonumber
\langle M^{\text{phys}}_{ij} &M^{\text{phys}}_{kl} M^{\text{phys}}_{mn} \rangle = \langle M^{\text{phys}}_{ij} M^{\text{phys}}_{kl} \rangle_{\text{conn}} \langle M^{\text{phys}}_{mn} \rangle +  \langle M^{\text{phys}}_{ij} M^{\text{phys}}_{mn} \rangle_{\text{conn}} \langle M^{\text{phys}}_{kl} \rangle \\ 
&+ \langle M^{\text{phys}}_{kl} M^{\text{phys}}_{mn} \rangle_{\text{conn}} \langle M^{\text{phys}}_{ij} \rangle +  \langle M^{\text{phys}}_{ij} \rangle \langle M^{\text{phys}}_{kl} \rangle \langle M^{\text{phys}}_{mn} \rangle.
\end{align} 


\subsection{Embedding within 13-parameter PIGM model}
Previous work has solved the most general Gaussian matrix models for general $D \times D$ matrices \cite{ramgoolam2019permutation}. These models are defined by 2 linear and 11 quadratic coupling parameters. We expect the $1 + 3$ parameter model considered in this paper to be embedded in the larger $2 + 11$ parameter model.

Indeed, comparing the action of the physical model \eqref{eq: physical action wrt S vars} written in terms of $S$ variables \eqref{eq: S V0 1} - \eqref{eq: S VH 3} (rewritten here for convenience)
\begin{align} \nonumber
\mathcal{S} = &- \mu_{V_0} \Big( \sqrt{\frac{D-1}{D}} S^{V_0 ; 1} - \frac{1}{\sqrt{D}} S^{V_0 ; 2} \Big) + \frac{\tau_{V_0}}{2} \Big( \sqrt{\frac{D-1}{D}} S^{V_0 ; 1} - \frac{1}{\sqrt{D}} S^{V_0 ; 2}  \Big)^2  \\ 
&+  \frac{\tau_{V_H}}{2} \sum_{a=1}^{D-1} \Big( \sqrt{\frac{D-2}{2D}} (S^{V_H ; 1}_a + S^{V_H ; 2}_a) - \sqrt{\frac{2}{D}} S^{V_H ; 3}_a \Big)^2 + \frac{\tau_{V_2}}{2} \sum_{a=1}^{\text{dim}V_2} S^{V_2}_a S^{V_2}_a.
\end{align}
to that of the $2 + 11$ parameter model written in terms of the same variables 
\bea\label{action} 
&& \cS^{\text{PIGMM}}  =  - \sum_{ \alpha =1  }^2  \mu^{ V_0}_{ \alpha } S^{ V_0 ; \alpha } 
         +  { 1 \over  2 }  \sum_{ \alpha , \beta = 1 }^{ 2 } S^{ V_0 ;  \alpha } ( \Lambda_{V_0})_{ \ls[ \alpha \beta] } S^{ V_0 ;  \beta  } + { 1 \over 2 } \sum_{ a = 1 }^{ D-1  } \sum_{  \alpha , \beta = 1 }^{ 3 } S^{ H ;  \alpha }_{ a  } (\Lambda_H)_{ \ls[\alpha \beta]  } S^{ H ;  \beta }_{ a   } \cr 
         && 
          + { 1 \over 2 } \Lambda_{ V_2} \sum_{ a =1 }^{  ( D-1) ( D-2) /2 }  S^{ V_2 }_{ a }  S^{ V_2 }_{ a }             +  { 1 \over 2 }  \Lambda_{ V_3 } \sum_{ a =1 }^{ D (D-3)/2 }   S^{ V_3 }_{ a }  S^{ V_3}_{   a } \, . 
\eea
we see the point at which the $2+11$ model reduces to the physical model is given by 
\begin{align} \label{eq: linear coupling relations}
&\mu^{V_0} = \mu_{V_0} 
\begin{bmatrix}
\sqrt{\frac{D-1}{D}} & - \frac{1}{\sqrt{D}}
\end{bmatrix}, 
\end{align}
and
\begin{align} \label{eq: coupling relations} \nonumber
&\Lambda_{V_0} = \tau_{V_0} 
\begin{bmatrix}
\frac{D-1}{D} & - \frac{\sqrt{D-1}}{D} \\
- \frac{\sqrt{D-1}}{D} & \frac{1}{D}
\end{bmatrix}, \quad
&&\Lambda_{V_H} = \tau_{V_H} 
\begin{bmatrix}
 \frac{D-2}{2D} & \frac{D-2}{2D} & -2 \frac{\sqrt{D-2}}{D} \\
\frac{D-2}{2D} & \frac{D-2}{2D} & -2 \frac{\sqrt{D-2}}{D} \\
-2 \frac{\sqrt{D-2}}{D} & -2 \frac{\sqrt{D-2}}{D} & \frac{2}{D}
\end{bmatrix}, \\
&\Lambda_{V_2} = \tau_{V_2},  \quad
&&\Lambda_{V_3} = 0.
\end{align}

\section{Daily correlation matrices from high-frequency forex data} \label{sec: data description}

The high-frequency forex data that we analyse pertain to 19 of the most liquidly traded currency pairs and cover the date range from 1 April 2020 to 31 January 2022. The data is sourced from TrueFX \cite{truefx} and is comprised of all updates of the best price quotes at which any market participant is willing to buy (top-of-book bid quotes) or sell (top-of-book offer quotes). Market participants providing such price quotes include banks, brokers and asset managers on the Integral OCX platform\footnote{This platform is an ECN i.e. an Electronic Communication Network}. The data precision is in milliseconds for time stamps and fractions of a pip\footnote{1 pip = $10^{-4}$} for prices. We exclude United States currency settlement holidays (days where no settlements of prior transactions are made) due to the central importance of the US Dollar to FX trading. We also exclude the 24th, 25th, 26th, 31st of December and the 1st, 2nd of January due to reduced liquidity.  In total, around one billion pricing updates were analysed. For each currency pair, the mid-price series, $p^{(I)}_{j}$, is calculated from the bid and offer quotes as follows:
\begin{equation}
p^{(I)}_{j} = (b^{(I)}_{j} + a^{(I)}_{j})/2 \quad I \in \{1,\dots,19\}, \, j \in \{1,\dots,n^{I}\},
\end{equation}
where $b^{(I)}$ and $a^{(I)}$ are contemporaneous bid and offer quotes respectively, $I \in \{1,\dots,19\}$ indexes the currency pairs, $j$ indexes the quotes and $n^{I}$ corresponds to the number of quotes for the currency pair $I$ per day. See table \ref{tab:currency_pairs} for a mapping of these indices to actual currency pair names. These mid-prices are then sampled on a regular time grid using the last-tick methodology, where the most recent quotes in each currency pair are used to calculate the mid-price for that time interval. The regularly sampled mid-prices are then:
\begin{equation}
p^{(I)}_{t_{(1)}},\dots, p^{(I)}_{t_{(n)}}, \quad t_{(i+1)}-t_{(i)} = 5\mbox{ minutes}, \quad i \in \{1,\dots,n\},
\end{equation}
where $t_{(i)}, \, i \in \{1,\dots,n\}$ are the time stamps on a regularly sampled grid and $n$ is the number of 5 minute intervals per day. If we denote the time stamp of each quote as $\tau_{j},\, j \in \{1,\dots,n^{I}\}$, then the quote used for each 5 minute interval can be described as, 
\begin{equation} 
	p^{(I)}_{t_{(i)}} = p^{(I)}_{{\max} \{ 1 \leq j \leq n^{I} | \tau_{j} \leq t_{(i)}\}}.
\end{equation}
We note that the choice of 5 minutes as a time interval is common in high frequency financial correlation analyses. We obtain the (log) mid-price returns via:
\begin{equation}
r^{(I)}_{(i)} = \log \frac{p^{(I)}_{t_{(i+1)}}}{p^{(I)}_{t_{(i)}}}, \quad i \in \{1,\dots,n-1\}. \label{logret}
\end{equation}
Note that the first time interval of each day, for all currency pairs, begins at 00:00:00.000 UTC/GMT and ends at 00:04:59:59.999 UTC/GMT. The last time interval begins at 23:55:00.000 UTC/GMT and ends at 23:59:59.999 UTC/GMT. The advantage of determining calendar date based on UTC/GMT is that the major FX trading sessions are all captured on the same calendar date, namely Asia, then Europe, then North America. There are $n=288$ five minute intervals per day. The time intervals are not only regular, but also aligned across all the currency pairs. See table \ref{tab:currency_pair_updates} for the statistics on the number of quotes per time interval for each currency pair, aggregated across all days. See table \ref{tab:currency_pair_rets} for the descriptive statistics of the regularly sampled (log) returns per currency pair, again aggregated across all days in the data set. It is readily apparent from the descriptive statistics in table \ref{tab:currency_pair_rets} that the means of the (log) return distributions are very close to zero and that the standard deviations vary  between currency pairs. In addition, the returns have high kurtosis consistent with the expected behaviour of price movements of financial instruments with a calendar time clock. The only currency pair that has a markedly asymmetric distribution is USD/TRY as evidenced by a large negative skewness (i.e. a left-skewed distribution). The large volatility, kurtosis and negative skewness of the USD/TRY distribution can be related to the sharp depreciation of the Turkish Lira during the Turkish currency and debt crisis which occurred during the period of analysis.
\afterpage{
{\footnotesize
\begin{longtable}[t]{lrrrrrrr}
\toprule
Currency Pair & Mean & Std Dev. & Q1 & Med. & Q3\\
\midrule
AUD/JPY & 419 & 374.5 & 187 & 319 & 531\\
AUD/NZD & 304 & 279.3 & 135 & 231 & 383\\
AUD/USD & 404 & 396.9 & 165 & 296 & 511\\
CAD/JPY & 272 & 257.0 & 113 & 197 & 348\\
CHF/JPY & 287 & 271.7 & 118 & 212 & 368\\
EUR/CHF & 269 & 295.5 & 90 & 173 & 338\\
EUR/GBP & 308 & 316.7 & 103 & 206 & 407\\
EUR/JPY & 506 & 435.8 & 202 & 387 & 685\\
EUR/PLN & 277 & 487.4 & 47 & 125 & 296\\
EUR/USD & 512 & 506.4 & 181 & 376 & 683\\
GBP/JPY & 553 & 471.5 & 240 & 433 & 724\\
GBP/USD & 492 & 472.6 & 174 & 361 & 668\\
NZD/USD & 269 & 269.3 & 114 & 199 & 333\\
USD/CAD & 373 & 371.7 & 152 & 271 & 470\\
USD/CHF & 237 & 255.4 & 82 & 161 & 305\\
USD/JPY & 322 & 319.9 & 135 & 234 & 399\\
USD/MXN & 447 & 434.1 & 143 & 324 & 610\\
USD/TRY & 205 & 580.4 & 9 & 36 & 138\\
USD/ZAR & 445 & 498.5 & 139 & 325 & 588\\
\bottomrule\\
\caption{Descriptive statistics of number of quote updates per 5 minute time interval.}\label{tab:currency_pair_updates}
\end{longtable}
}
}
\afterpage{
{\footnotesize
\begin{longtable}[t]{lrrrrr}
\toprule
Currency Pair & Mean ($\mbox{x } 10^{-4}$) & Std Dev. ($\mbox{x } 10^{-4}$) & Skewness & Kurtosis\\
\midrule
AUD/JPY & 0.0 & 3.8 & 0.0 & 24.4\\
AUD/NZD & 0.0 & 2.1 & 0.2 & 36.6\\
AUD/USD & 0.0 & 3.8 & 0.0 & 15.3\\
CAD/JPY & 0.0 & 3.1 & 0.1 & 16.2\\
CHF/JPY & 0.0 & 2.4 & 0.0 & 12.5\\
EUR/CHF & 0.0 & 1.8 & -0.6 & 80.5\\
EUR/GBP & 0.0 & 2.6 & -0.2 & 37.5\\
EUR/JPY & 0.0 & 2.4 & 0.1 & 16.8\\
EUR/PLN & 0.0 & 2.7 & 0.0 & 31.1\\
EUR/USD & 0.0 & 2.4 & -0.2 & 32.9\\
GBP/JPY & 0.0 & 3.1 & 0.0 & 20.0\\
GBP/USD & 0.0 & 3.1 & 0.0 & 14.9\\
NZD/USD & 0.0 & 3.8 & 0.1 & 19.8\\
USD/CAD & 0.0 & 2.7 & 0.0 & 18.8\\
USD/CHF & 0.0 & 2.5 & -0.2 & 19.8\\
USD/JPY & 0.0 & 2.1 & 0.2 & 17.3\\
USD/MXN & 0.0 & 5.5 & 0.0 & 18.8\\
USD/TRY & 0.1 & 11.7 & -7.3 & 795.7\\
USD/ZAR & 0.0 & 6.4 & -0.1 & 17.2\\
\bottomrule\\
\caption{Summary statistics of regular 5-minute (log) mid-price returns.}\label{tab:currency_pair_rets}
\end{longtable}
}
}

\subsection{Correlation matrix methodology} \label{sec: correlation matrix methodology}

\begin{figure}[t]
\centering
\includegraphics[width=110mm]{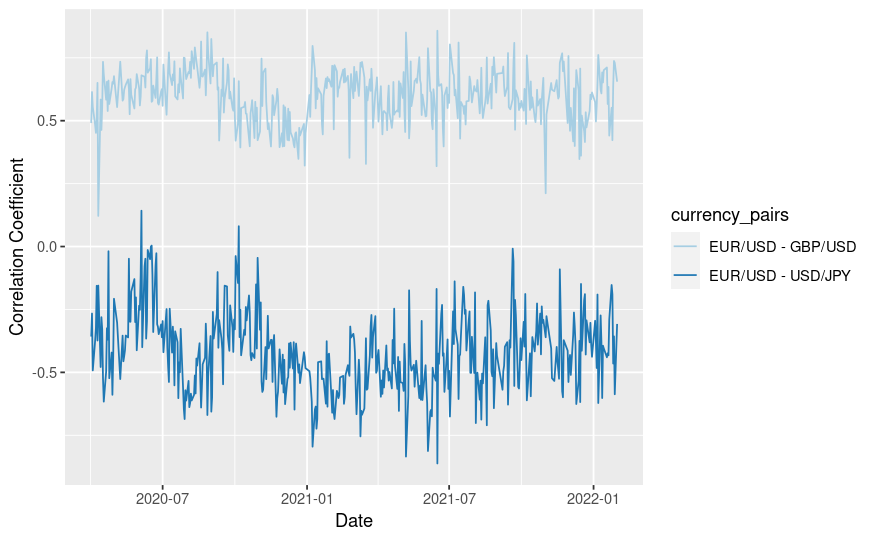}
\caption{\label{fig:corr_curr} Examples of realized daily correlation estimates over time.}
\end{figure}

In statistics, various measures of association between two random variables have been defined. In our context, we apply certain measures of correlation to ascertain the degree to which currency (log) returns are concordant or discordant. Intuitively, this should capture an important aspect of the relationship of one currency pair with another. Calculating correlations on high frequency financial data is complicated by two main effects. The first is the fact that observations occur irregularly in time and moreover, asynchronously across instruments. The second is the presence of microstructure noise due to various factors such as bid-ask bounce (relevant mainly for transaction based data), minimum tick intervals, latency effects etc. See \cite{ait2010high} and the references therein for more detail on these two complicating issues and various approaches to address them. In this article we utilize the correlation estimator \eqref{eq: correlation matrices} on (log) mid-price returns. This estimator is referred to as the realized correlation estimator in the finance literature and is defined as,
\begin{equation} \label{eq: correlation matrices}
\hat{\rho}^{IJ} = \frac{\sum_{i=1}^{n-1} r^{(I)}_{(i)} r^{(J)}_{(i)} }{\sqrt{ (\sum_{i=1}^{n-1} r^{(I)}_{(i)})^2 (\sum_{i=1}^{n-1} r^{(J)}_{(i)})^2 }} , \quad I,J \in \{1,\dots,19\},
\end{equation}
where $I,J$ are currency pair indices. This estimator captures the normalized, aggregated co-movement (i.e. covariance) of two series of returns over a given time period (one day in our case). It is well established that the realized correlation estimator is, in general, sensitive to the issues described above. However, it is widely acknowledged in the literature that the impact of these issues can be mitigated by sampling regularly at a lower frequency i.e. 5 - 15 minute intervals. We utilize 5 minute time intervals in particular, as is common in analysing high frequency financial data. We have also verified empirically that the correlation results are not very sensitive to the choice of the time interval length (beyond a certain length). We do acknowledge however that the procedure we have applied is not likely to be the most efficient and discards some information (see \cite{ait2010high} or \cite{malliavin2002fourier, 10.1214/08-AOS633} for approaches that are likely to be more efficient for example). However, the simplicity of the realised correlation estimator is appealing and it allows us to make contact with asymptotic Gaussian sampling properties as discussed in the introduction. The main focus of the present article is to explore the phenomenological modelling of ensembles of correlation matrices and not particular correlation estimators. The impact of using more sophisticated and potentially more efficient estimators in our context can be explored in future research. 
Note that the resultant $\hat{\rho}^{IJ}$ correlation matrix is a symmetric, real matrix with $19(19-1)/2 = 171$ independent entries. This figure accounts for the fact that the diagonal elements are fixed and equal and do not contribute to the degrees of freedom (we subtract the identity to get a correlation matrix with vanishing diagonal elements). As mentioned previously, we are concerned with the ensemble statistics of such matrices. There are several ways to construct such an ensemble. We choose to calculate the correlation matrix for each trading day (aligned with UTC/GMT boundaries), and study the sampling distribution of the matrices. In particular, we study, ${\hat{\rho}_{A}^{IJ}}$, where $A\in \{1,\dots, N_D \}$ indexes trading dates. In our data, there are 446 unique trading days, i.e. $N_D=446$. We plot examples of the evolution of two elements of these correlation matrices over time in figure \ref{fig:corr_curr}.

\section{ Matrix theory and matrix data: near-Gaussianity  } \label{sec: Numerical results}

In this section we apply the Gaussian model of section \ref{sec:PIGMMSDV} to the ensemble of correlation matrices defined in section \ref{sec: data description}. We show that the vast majority of cubic and quartic observables closely match the predictions of the Gaussian theory. 

We form a compact representation of the original correlation matrix data by defining a vector of observables for each correlation matrix. The observable vectors are shown to perform well in a variety of anomaly detection and similarity ranking tasks in section \ref{sec: applications}. Optimal performance is achieved in these tasks by constructing observable vectors from the least Gaussian observables.

\subsection{Theory/experiment deviations normalized by standard deviations of the observables } \label{subsec: theory exp deviations}

\begin{figure}[t]
\centering
\includegraphics[width=120mm]{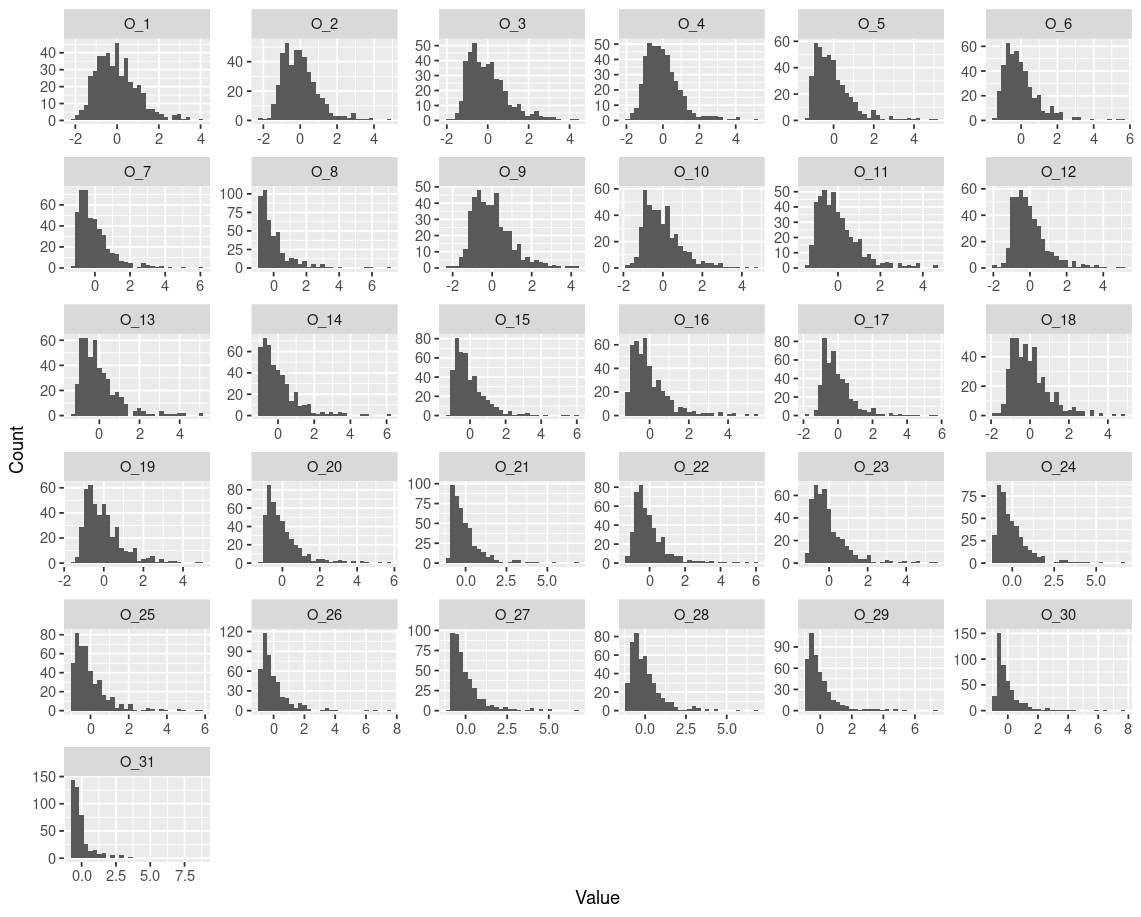}
\caption{\label{fig:hist_observables} Histograms of the standardized values of each of the observables (one value per correlation matrix i.e. per day).}
\end{figure}

\begin{figure}[t]
\includegraphics[width=160mm]{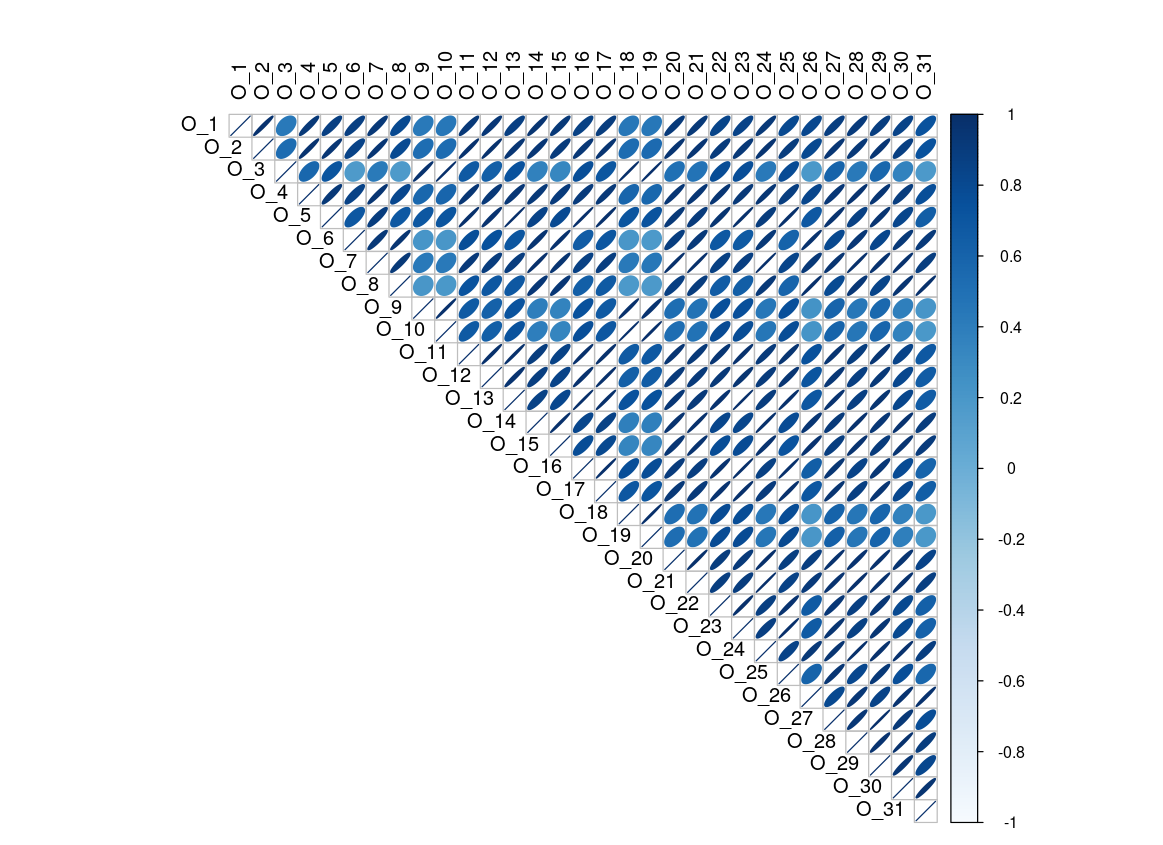}
\caption{\label{fig:corr_observables} Pearson product-moment correlation of the observables.}
\end{figure}

We begin by elucidating some empirical statistical properties of the observables - the permutation invariant polynomials of the correlation matrix elements. These are listed in table \ref{tab: abs err} and the distributions of their standardised values over the 446 days are plotted in the histograms in figure \ref{fig:hist_observables}. Many appear roughly Gaussian, while others exhibit a right/positive skew along with heavier tails than the normal distribution. The estimated Pearson product-moment correlation of the observable elements is plotted in figure \ref{fig:corr_observables}. It is noteworthy that all correlations are positive and that most correlations are very strongly positive. The strength of the correlations is particularly relevant in our choice of statistical distance measure in the anomaly detection analysis presented in section \ref{sec: anomaly detection}. Indeed, this motivated utilising the Mahalanobis distance which typically performs well even in the presence of such correlations.

Equipped with the Gaussian model and its solution, given in section \ref{sec:PIGMMSDV}, and the financial data described in section \ref{sec: data description} we now perform a variety of tests to assess how well this model describes the statistics of the data. Firstly, we calculate the normalised absolute error
\bea \label{eq: obs measure}
\Delta_{\alpha} = \frac{ | \langle  \cO_{\alpha} \rangle_{\text{T}} - \langle \cO_{\alpha} \rangle_{\text{E}} |}{\sigma_{\text{E}, \alpha}},
\eea
between the experimental cubic and quartic observable average values 
\bea 
\langle \cO_{\alpha} \rangle_{\text{E}}  = { 1 \over n_A } \sum_{ A=1 }^{ N_D } 
  \cO_{\alpha} ( \rho^{ij}_A )  \, ,
\eea
and the Gaussian model's prediction of those expectation values $\langle \cO_{\alpha} \rangle_{\text{T}}$. In both equations $\alpha$ labels the observable and in \eqref{eq: obs measure} we have normalised by the standard deviation of the experimental observable values.

 As we argued in section \ref{sec: technical summary} we expect these normalised errors to be small where the underlying data is approximately Gaussian. The normalised absolute error for each observable is listed in the third column of table \ref{tab: abs err}. In general these are in very good agreement: only 4 observables differ from the theoretical prediction by more than 1 standard deviation, and the average normalised absolute error of the cubic and quartic observables is 0.42 standard deviations. This can be regarded as strong evidence for Gaussianity in the permutation invariant sector of the FX-rate correlation matrix data.

The model defined in section \ref{sec:PIGMMSDV} can be used to predict the standard deviation of cubic and quartic observables. We call this the theoretical standard deviation, and define it for each observable $\alpha$ as
\begin{align}
\sigma_{\text{T}, \alpha} \equiv \sqrt{ | \langle (\mathcal{O}_{\alpha})^2 \rangle_{\text{T}} - \langle \mathcal{O}_{\alpha} \rangle^2_{\text{T}} |} \, .
\end{align}
In itself this is an interesting quantity to compare to the experimental observable standard deviations $\sigma_{\text{E}, \alpha}$. The ratio of the two standard deviations is shown for each observable in the fourth column of table \ref{tab: abs err}. The values of $\sigma_{\text{T}, 3}$ and $\sigma_{\text{T}, 22}$ provided by the model are much smaller than the observed values. This is consistent with the finding in column three of table \ref{tab: abs err}, in which we see the expectation values of these observables deviating the most from the model. It is these large deviations from Gaussianity that lend these observables their power in the construction of lower-dimensional representations of the correlation matrices (see section \ref{sec: dim red}).

We briefly note an alternative approach to estimating the theoretical standard deviation, also employed in \cite{Ramgoolam:2019ldg} to give good theoretical predictions of the experimental standard deviations of observables. This estimate is obtained by calculating the absolute difference between $\langle \mathcal{O}_{\alpha} \rangle_{\text{T}}$ and $\langle \mathcal{O}_{\alpha} \rangle_{\text{T'}}$, where $\langle \mathcal{O}_{\alpha} \rangle_{\text{T'}}$ is the expectation value evaluated with the quadratic couplings that parametrise the model shifted by one standard deviation. Taking the average of this difference over all 8 possible permutations of sign for the shifts of the 3 parameters gives us our estimate of the standard deviation. This method was used to estimate the standard deviations of $\mathcal{O}_{12}$ and $\mathcal{O}_{19}$ due to the prohibitive computational demands of calculating the octic expectation values $\langle \mathcal{O}_{12}^2 \rangle$ and $\langle \mathcal{O}_{19}^2 \rangle$.

\subsection{ Day capture and balanced accuracy of theoretical typicality prediction for days}

The normalised errors presented in the previous section are encouragingly small, but rather abstract. In order to get a more intuitive sense of the agreement between the data and the Gaussian model, and with an eye toward developing useful applications, we consider a more practical measure. This second measure we call the day capture and define it as the proportion of days for which the value of an observable lies within 2 standard deviations of the mean value. If the observables were exactly Gaussian distributed we would expect the proportion of days captured to be close to 95.4\%, in line with the expectation of a one-variable Gaussian. This practical measure seems a sensible one given the approximately Gaussian distributions in figure \ref{fig:hist_observables}. First, we list the empirical day capture of each of the observables. These values are presented in the second column of table \ref{tab: balanced accuracy day capture}, all of which are very close to the expected $95.4\%$.

\afterpage{
{\footnotesize
\begin{longtable}[hbt!]{llcc}
\toprule
\textbf{Label} & \textbf{Observable} & $\bm{\Delta_{\alpha}}$ & $\bm{\sigma}_{\textbf{E}}/\bm{\sigma}_{\textbf{T}}$ \\
\midrule
$\mathcal{O}_{1} $ & $\sum_{i,j} \hat{\rho}_{ij}^3                                                              $ & 0.02 & 0.90  \\
$\mathcal{O}_{2} $ & $\sum_{i,j,k} \hat{\rho}_{ij}^2 \hat{\rho}_{jk}                                            $ & 0.33 & 1.49  \\
$\mathcal{O}_{3} $ & $\sum_{i,j,k} \hat{\rho}_{ij} \hat{\rho}_{jk} \hat{\rho}_{ki}                              $ & 2.04 & 9.39  \\
$\mathcal{O}_{4} $ & $\sum_{i,j,k,l} \hat{\rho}_{ij}^2 \hat{\rho}_{kl}                                          $ & 0.01 & 1.06  \\
$\mathcal{O}_{5} $ & $\sum_{i,j,k,l} \hat{\rho}_{ij} \hat{\rho}_{jk} \hat{\rho}_{kl}                            $ & 0.97 & 3.36  \\
$\mathcal{O}_{6} $ & $\sum_{i,j,k,l} \hat{\rho}_{ij} \hat{\rho}_{ik} \hat{\rho}_{il}                            $ & 0.33 & 0.73  \\
$\mathcal{O}_{7} $ & $\sum_{i,j,k,l,m} \hat{\rho}_{ij} \hat{\rho}_{jk} \hat{\rho}_{lm}                          $ & 0.12 & 1.05  \\
$\mathcal{O}_{8} $ & $\sum_{i,j,k,l,m,n} \hat{\rho}_{ij} \hat{\rho}_{kl} \hat{\rho}_{mn}                        $ & 0.01 & 0.65  \\
\midrule
$\mathcal{O}_{9} $ & $\sum_{i,j} \hat{\rho}_{ij}^4                                                              $ & 0.54 & 1.10  \\
$\mathcal{O}_{10}$ & $\sum_{i,j,k} \hat{\rho}_{ij}^2 \hat{\rho}_{jk}^2                                          $ & 0.42 & 2.42  \\
$\mathcal{O}_{11}$ & $\sum_{i,j,k} \hat{\rho}_{ij} \hat{\rho}_{jk}^3                                            $ & 0.05 & 1.30  \\
$\mathcal{O}_{12}$ & $\sum_{i,j,k} \hat{\rho}_{ij} \hat{\rho}_{ik} \hat{\rho}_{jk}^2                            $ & 0.88 & 3.27* \\
$\mathcal{O}_{13}$ & $\sum_{i,j,k,l} \hat{\rho}_{ij} \hat{\rho}_{kj} \hat{\rho}_{lj}^2                          $ & 0.19 & 1.67  \\
$\mathcal{O}_{14}$ & $\sum_{i,j,k,l} \hat{\rho}_{ij} \hat{\rho}_{kl}^3                                          $ & 0.04 & 0.64  \\
$\mathcal{O}_{15}$ & $\sum_{i,j,k,l} \hat{\rho}_{ij} \hat{\rho}_{jk}^2 \hat{\rho}_{kl}                          $ & 0.21 & 1.07  \\
$\mathcal{O}_{16}$ & $\sum_{i,j,k,l} \hat{\rho}_{ij} \hat{\rho}_{jk} \hat{\rho}_{kl}^2                          $ & 0.49 & 2.45  \\
$\mathcal{O}_{17}$ & $\sum_{i,j,k,l} \hat{\rho}_{ij} \hat{\rho}_{jk} \hat{\rho}_{ik} \hat{\rho}_{kl}            $ & 1.00 & 7.44  \\
$\mathcal{O}_{18}$ & $\sum_{i,j,k,l} \hat{\rho}_{ij}^2 \hat{\rho}_{kl}^2                                        $ & 0.07 & 2.06  \\
$\mathcal{O}_{19}$ & $\sum_{i,j,k,l} \hat{\rho}_{ij} \hat{\rho}_{jk} \hat{\rho}_{kl} \hat{\rho}_{li}            $ & 1.24 & 8.57* \\
$\mathcal{O}_{20}$ & $\sum_{i,j,k,l, m} \hat{\rho}_{ik} \hat{\rho}_{jk} \hat{\rho}_{lk} \hat{\rho}_{mk}         $ & 0.36 & 0.79  \\
$\mathcal{O}_{21}$ & $\sum_{i,j,k,l, m} \hat{\rho}_{il} \hat{\rho}_{jk} \hat{\rho}_{lk} \hat{\rho}_{mk}         $ & 0.39 & 1.67  \\
$\mathcal{O}_{22}$ & $\sum_{i,j,k,l, m} \hat{\rho}_{ij} \hat{\rho}_{kl} \hat{\rho}_{lm} \hat{\rho}_{mk}         $ & 1.31 & 10.1  \\
$\mathcal{O}_{23}$ & $\sum_{i,j,k,l, m} \hat{\rho}_{ij}^2 \hat{\rho}_{kl} \hat{\rho}_{lm}                       $ & 0.05 & 1.63  \\
$\mathcal{O}_{24}$ & $\sum_{i,j,k,l, m} \hat{\rho}_{ij} \hat{\rho}_{kl} \hat{\rho}_{lm}^2                       $ & 0.24 & 0.81  \\
$\mathcal{O}_{25}$ & $\sum_{i,j,k,l,m} \hat{\rho}_{ij} \hat{\rho}_{jk} \hat{\rho}_{kl} \hat{\rho}_{lm}          $ & 0.79 & 5.98  \\
$\mathcal{O}_{26}$ & $\sum_{i,j,k,l,m,n} \hat{\rho}_{ij} \hat{\rho}_{kl} \hat{\rho}_{km} \hat{\rho}_{kn}        $ & 0.12 & 0.54  \\
$\mathcal{O}_{27}$ & $\sum_{i,j,k,l,m,n} \hat{\rho}_{ij} \hat{\rho}_{jk} \hat{\rho}_{lm} \hat{\rho}_{ln}        $ & 0.10 & 1.47  \\
$\mathcal{O}_{28}$ & $\sum_{i,j,k,l,m,n} \hat{\rho}_{ij}^2 \hat{\rho}_{kl} \hat{\rho}_{mn}                      $ & 0.02 & 0.58  \\
$\mathcal{O}_{29}$ & $\sum_{i,j,k,l,m,n} \hat{\rho}_{ij} \hat{\rho}_{kl} \hat{\rho}_{lm} \hat{\rho}_{mn}        $ & 0.65 & 2.10  \\
$\mathcal{O}_{30}$ & $\sum_{i,j,k,l,m,n,o} \hat{\rho}_{ij} \hat{\rho}_{jk} \hat{\rho}_{lm} \hat{\rho}_{no}      $ & 0.13 & 0.67  \\
$\mathcal{O}_{31}$ & $\sum_{i,j,k,l,m,n,o,p} \hat{\rho}_{ij} \hat{\rho}_{kl} \hat{\rho}_{mn} \hat{\rho}_{op}    $ & 0.03 & 0.38  \\
\midrule
& Average & 0.42 & 2.50 \\
\bottomrule\\
\caption{For each observable in the first two columns the third column lists the absolute difference between the experimental value and theoretical prediction normalised by the experimental standard deviation. The fourth column lists the ratio of the experimental and theoretical standard deviations. The * values were obtained using an estimate of $\sigma_{\text{T}}$ described at the end of section \ref{subsec: theory exp deviations}.}\label{tab: abs err}
\end{longtable}
}
}

Secondly, as a test of the Gaussian model we define the theoretical day capture: this is the proportion of days falling within 2 theoretical standard deviations of the theoretical expectation value of an observable. To test the model's predicted day capture rates we calculate the balanced accuracy of the theoretical day capture for each observable. The process for which we now describe.

Consider the following binary classification problem: classify a day as typical (positive) if it falls within 2 standard deviations of the observable mean as calculated from the data, and a day as atypical (negative) if it falls outside this range. Then define the True Positive Rate (TPR) and True Negative Rate (TNR) as follows with reference to the quantities defined in table \ref{tab: condition definitions}

\begin{align}
\text{TPR} =  \frac{\text{TP}}{\text{TP} + \text{FN}} \, , \qquad \text{TNR} =  \frac{\text{TN}}{\text{TN} + \text{FP}} \, .
\end{align}
Translating back into the terminology of day capture we have
\begin{align} \nonumber
\text{TPR} &= \frac{\text{Correctly predicted typical days}}{\text{Total number of typical days}} \, ,  \\
\text{TNR} &= \frac{\text{Correctly predicted atypical days}}{\text{Total number of atypical days}} \, . 
\end{align}
From these quantities we define the Balanced Accuracy, which averages the positive and negative performance of the model
\begin{align}
\text{Balanced Accuracy} = \frac{\text{TPR} + \text{TNR}}{2}.
\end{align}

\begin{table}
\begin{tabular}{|c|l||*{2}{c|}}\hline
\backslashbox{Actual condition}{Predicted condition}
&\makebox[3.5cm]{Positive (typical)}&\makebox[3.5cm]{Negative (atypical)}\\\hline\hline
Positive (typical) & True Positive (TP) & False Negative (FN) \\\hline
Negative (atypical) & False Positive (FP) & True Negative (TN) \\\hline
\end{tabular}
\caption{\label{tab: condition definitions} Definitions of TP, TN, FP and FN in generic binary classification. In our case positive is a day with an observable value falling within two standard deviations of the mean and negative is a day with an observable value falling outside this range. }
\end{table} 

The Balanced Accuracy of the day capture of each observable is listed in the third column of table \ref{tab: balanced accuracy day capture}. Many observables have a Balanced Accuracy of, or very close to 1. The average Balanced Accuracy of the theoretical day capture over all observables is 0.8, which is generally considered to be a good score in data sciences. 

\afterpage{
{\footnotesize
\begin{longtable}[hbt!]{lllc}
\toprule 
\textbf{Label} & $\bm{\mu}_{\textbf{E}} \bm{\pm} \bm{2} \bm{\sigma}_{\textbf{E}}$ & $\bm{\mu}_{\textbf{T}} \bm{\pm} \bm{2} \bm{\sigma}_{\textbf{T}}$ & \textbf{Balanced Accuracy}  \\
\midrule
$\mathcal{O}_{1}$&   95.74 & 97.31 & 0.82 \\
$\mathcal{O}_{2}$&   95.29 & 85.87 & 0.95 \\
$\mathcal{O}_{3}$&   95.07 & 0.22 & 0.50 \\
$\mathcal{O}_{4}$&   95.74 & 95.07 & 1.00 \\
$\mathcal{O}_{5}$&   95.52 & 41.70 & 0.72 \\
$\mathcal{O}_{6}$&   95.74 & 98.43 & 0.68 \\
$\mathcal{O}_{7}$&   95.96 & 94.62 & 0.99 \\
$\mathcal{O}_{8}$&   95.74 & 98.43 & 0.68 \\
\midrule     
$\mathcal{O}_{9}$&   94.84 & 92.83 & 0.81 \\
$\mathcal{O}_{10}$ & 94.39 & 71.08 & 0.88 \\
$\mathcal{O}_{11}$ & 95.29 & 93.50 & 0.99 \\
$\mathcal{O}_{12}$ & 95.07 & 45.74 & 0.74*\\
$\mathcal{O}_{13}$ & 95.29 & 88.57 & 0.96 \\
$\mathcal{O}_{14}$ & 95.96 & 97.98 & 0.75 \\
$\mathcal{O}_{15}$ & 95.74 & 94.39 & 0.99 \\
$\mathcal{O}_{16}$ & 95.07 & 73.32 & 0.89 \\
$\mathcal{O}_{17}$ & 95.29 & 20.85 & 0.61 \\
$\mathcal{O}_{18}$ & 94.84 & 76.68 & 0.90 \\
$\mathcal{O}_{19}$ & 95.07 & 8.07 & 0.54*\\
$\mathcal{O}_{20}$ & 94.84 & 97.31 & 0.76 \\
$\mathcal{O}_{21}$ & 96.19 & 86.10 & 0.95 \\
$\mathcal{O}_{22}$ & 95.96 & 1.79 & 0.51 \\
$\mathcal{O}_{23}$ & 95.52 & 91.03 & 0.98 \\
$\mathcal{O}_{24}$ & 96.41 & 96.41 & 1.00 \\
$\mathcal{O}_{25}$ & 95.52 & 38.57 & 0.70 \\
$\mathcal{O}_{26}$ & 95.74 & 99.10 & 0.61 \\
$\mathcal{O}_{27}$ & 95.52 & 92.38 & 0.98 \\
$\mathcal{O}_{28}$ & 95.96 & 98.65 & 0.67 \\
$\mathcal{O}_{29}$ & 95.96 & 76.01 & 0.90 \\
$\mathcal{O}_{30}$ & 95.07 & 97.53 & 0.75 \\
$\mathcal{O}_{31}$ & 95.74 & 99.33 & 0.58 \\
\midrule
Average Balanced Accuracy & & & 0.80 \\
\bottomrule

\caption{For each cubic and quartic observable the second and third columns list the experimental day capture the theoretical day capture respectively. The fourth column gives the Balanced Accuracy of the theoretical model's day capture at $\pm 2 \sigma$. The * values were calculated using estimates of the theoretical standard deviation described at the end of section \ref{subsec: theory exp deviations}.}\label{tab: balanced accuracy day capture}
\end{longtable}
}
}

\begin{figure}[!htb] 
\centering
\includegraphics[width=160mm]{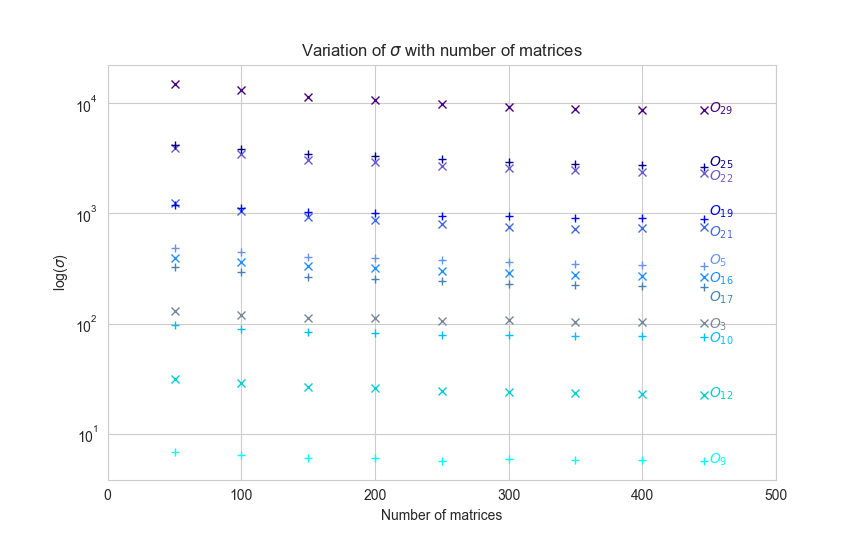}
\caption{Standard deviation of a collection of observables for the first x days in the sample.}
\label{fig: std dev vs days}
\end{figure}

\begin{figure}[!htb] 
\centering
\includegraphics[width=160mm]{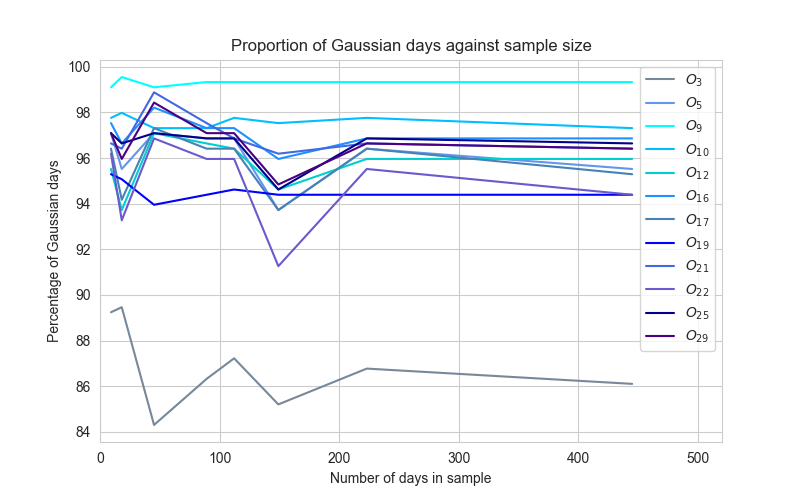}
\caption{Day capture of a selection of observables plotted for a range of sample sizes.}
\label{fig: gaussian days robustness}
\end{figure}

This agreement between theory and experiment for day capture is robust to changes in the size of the sample used for the analysis. To begin with figure \ref{fig: std dev vs days} shows the relationship between sample size and standard deviation. The standard deviation remains constant no matter how many days we include in the sample.\footnote{This is excepting a small spike at the start of the sample period, caused by a genuine increase in market volatility. If the sample size is varied but each sample constructed from a random, i.e. non-sequential, collection of days this spike disappears.} Secondly, in figure \ref{fig: gaussian days robustness} we plot the proportion of days in the full sample that fall within 3 standard deviations of the mean for a selection of observables, in which the mean and standard deviation have been calculated using a subset of the data with sample sizes ranging from 9 to the full 446 days. It is clear from this plot that the day capture for each observable is independent of the sample size. 

\subsection{ Absolute errors relative to standard deviations and standard errors of observables } 

Thus far, we have analysed the differences between theoretical observable mean values and experimental observable mean values, normalised by the experimental standard deviation of the observable values. We have found that 27 out of 31 observables deviate less than 1 experimental standard deviation\footnote{The four observables that differ by more than one standard deviation are $\mathcal{O}_3 ,\mathcal{O}_{17} , \mathcal{O}_{19}$ and $\mathcal{O}_{22}$. }. These normalised differences have a physical interpretation. Small normalised differences in this case are suggestive of small coupling constants for higher order corrections in the action (see section \ref{sec: technical summary}) which is evidence for near-Gaussianity of the experimental data generating process. Furthermore, we have found that "physical" tests of the theoretical model versus experiment such as calculating the proportion of days captured and balanced accuracy of a typicality classifier provide additional evidence for the pure Gaussian model being a good approximation. 

Another possible choice for normalising the differences between the theoretical observable mean values and the experimental observable mean values is the experimental standard error. The standard error in this case is the standard deviation of the experimental observable mean values. We denote the standard error, $\sigma_{\bar{x}}$. Given the definition of the mean estimator, i.e. $\bar{x} = 1/n\sum_{i=1}^{n} x_{i}$, the standard error is equal to the standard deviation of the permutation invariant polynomial values for each matrix, $\sigma_{\text{E}}$, divided by $\sqrt{n}$ i.e. $\sigma_{\bar{x}} = \sigma_{\text{E}}/\sqrt{n}$. The standard error is useful in determining whether the differences between the theoretical observable mean values and experimental observable mean values are plausibly due to sampling variation. The larger the sample size, the smaller the departures between theory and experiment that can be distinguished from sampling variation i.e. "experimental error". In particular, genuine departures correspond to large standard errors e.g. bigger than 3 standard errors. Such departures can be interpreted as highly statistically significant differences.

In our data set, which has a fairly large sample size of 446, we have observed that 13 out of 31 observables have a difference between the theoretical observable mean and experimental observable mean value of fewer than 3 standard errors (i.e. 18 out of 31 observables exhibit a departure of more than 3 standard errors). To further explore the statistical significance of differences in the theoretical versus experimental observable mean values we have also calculated the percentile bootstrap confidence intervals of the experimental observable mean values. The bootstrap procedure involved re-sampling from the original set of observable values, uniformly with replacement, to construct 1000 bootstrap samples of the same size as the original sample (i.e. 446). The mean of each such bootstrap sample was then calculated. Given the asymptotic normality of the mean estimator, it is expected that 99.7th percentile bootstrap confidence intervals will approximately correspond to 3 standard errors on either side of the original experimental mean estimate. This is reflected in our results, which reveal 12 of 31 observables with theoretical observable mean values within this confidence interval and 19 of 31 observables with theoretical mean values outside the interval. This is in close agreement with the aforementioned basic standard error results where we had 13 of 31 observables with theoretical observable mean values lying within 3 standard errors of the experimental observable mean values.

It is  also worth noting  that statistically significant differences may nevertheless be small in terms of relative error, which we recall is defined as 
\bea 
\frac{ | \langle \cO_{\alpha} \rangle_{\text{T}} -  \langle  \cO_{\alpha} \rangle_{\text{E}} |}{   \langle \cO_{\alpha} \rangle_{\text{E}}  } \, . 
\eea
 Indeed we have observed that 20 out of 31 observables have a relative error of less than $30\%$ when comparing theoretical versus experimental mean observable values. 

The key point from this section is that when we consider the absolute error of observables in comparison to the standard deviation, a measure motivated by consideration of perturbative corrections to the toy  Gaussian model, we have 27 of 31 observables which are within 1 standard deviation (all are within 3 standard deviations). On the other hand 6 of 31 are within 1 standard error (13 are within 3 standard errors). This suggests that developing computations of expectation values in theoretical models which contain small cubic and quartic terms, as guided by the data, is likely to give statistically significant improvement, given our current sample sizes,  of the agreement between theoretical and experimental expectation values of observables. This is technically more intricate than computing in the Gaussian model and is left for future investigation. 


\section{Applications of matrix theory to matrix data: anomaly detection and similarity measures }\label{sec: applications}

In establishing approximate Gaussianity in the matrix data of correlation matrices observable vectors, formed using lists of permutation invariant polynomial functions labelled by graphs, provide the key bridge between permutation invariant Gaussian matrix theory and the matrix data. The observable vectors associated with the correlation matrices can be regarded as particular, lower-dimensional representations of the correlation matrices.  The observable vectors themselves are random vectors, for which the statistics entailed by the PIGM model are a good approximation in general. A natural question to ask is whether the observable vectors provide a more compact representation of correlation matrices which accentuate statistical "signal" in the data as opposed to noise. Such a representation would be closely linked to an accurate characterization of the market state and applications could include classification/regression models, clustering analysis, anomaly/outlier detection etc.  In this section, we will consider two tasks:   anomaly detection and similarity measurement, which we describe in more detail in subsections \ref{sec: anomaly detection} and \ref{sec: similarity} respectively. 

\subsection{Anomaly detection and economically significant dates } \label{sec: anomaly detection}

 In this section we will demonstrate that the observable vectors do indeed constitute a promising representation for anomaly/outlier detection. The task of anomaly/outlier detection pertains to identifying observations that differ significantly from the majority of the data set. In our context, we seek to identify unusual and noteworthy observable vectors, each of which is associated with the correlation matrix of a particular date. To verify that a meaningful result has been obtained, we need a notion of unusual and noteworthy dates in the forex market as a reference. The natural approach we take is to consider special dates in the forex trading calendar corresponding to the highest impact economic news announcements. These announcements often lead to a flurry of trading activity along with associated price movements, market volatility and changes in the relationships between currency pairs.

\subsubsection{Anomaly detection algorithm}

A common approach to detecting anomalous/outlier observations, is to utilize a statistical distance measure to determine the distance of each random vector from the mean vector (a natural multivariate measure of centrality), see \cite{ghorbani2019mahalanobis} for example. One can equivalently think of this as determining the length of the random vectors as measured from the origin for centered data. We will utilize the Mahalanobis distance measure to assess these distances. The Mahalanobis distance is similar to the Euclidean distance, when the Euclidean distance is applied to vectors where each element has been scaled by the respective standard deviation, but it better handles the fact that different elements of the vector are correlated in general. Geometrically, surfaces of constant Euclidean distance are spheres, whereas surfaces of constant Mahalanobis distance are ellipsoids in general. The Mahalanobis distance better treats the case of highly correlated elements in the random vectors, as is the case for the observable vectors (as noted in section \ref{subsec: theory exp deviations}, see figure \ref{fig:corr_observables} in particular). Concretely, given a multivariate probability distribution $F$ on $\mathbb{R}^{N}$ (i.e. generating random vectors $\vec{y} \in \mathbb{R}^{N}$), with mean vector $\vec{\mu}$ and covariance matrix $\Sigma$, the Mahalanobis distance of a point $\vec{x} \in \mathbb{R}^{N}$ from the mean $\vec{\mu}$ is defined as,
\begin{equation}
	d(\vec{x},\vec{\mu})  = \sqrt{(\vec{x}-\vec{\mu})\Sigma^{-1} (\vec{x}-\vec{\mu})}, 
\end{equation} 
where $\mu_{i} = E(y_{i})$ and $\Sigma_{ij} = E\left[(y_{i} - \mu_{i})(y_{j} - \mu_{j})\right]$.
The utility of the Mahalanobis distance in anomaly detection is thus in identifying points that are far from the mean, accounting for the covariance structure implied by the distribution $F$. If the distribution $F$ is a multivariate normal distribution for example, there is a particularly direct link between the Mahalanobis distance at a point $\vec{x}$ and the probability density at $\vec{x}$. Distant points have exponentially lower probability density in this case. The Mahalanobis distance can be fruitfully applied even when the distribution $F$ is not known to be a multivariate normal distribution however and we will not need to make this assumption. 

\subsubsection{Economically significant dates}

It is well known amongst forex trading practitioners that there are certain currencies and certain types of economic announcements that typically have the highest impact on the forex markets (see \cite{fxevents} for example). These currencies are associated with the countries or blocs with the largest economies, namely the US Dollar, Chinese Renminbi, Japanese Yen, European Union Euro (Germany is the largest economy in the EU at time of writing) and Great British Pound.  

Four of the most important classes of economic announcements are the following,
\begin{itemize}
	\item Central bank meetings and announcements relating to interest rate decisions etc. These include the FOMC (Federal Open Market Committee), ECB (European Central Bank), BoE (Bank of England), PBoC (People's Bank of China) and BoJ (Bank of Japan) meetings associated with the United States, European Union, Great Britain, China and Japan respectively.
	\item Unemployment data releases. One of the most important examples of this is the US Non-Farm Payrolls release.
	\item Consumer price index releases. The most important release in this category is the US consumer price index release.
	\item Unplanned forex news including special central bank meetings and speeches, political speeches etc.
\end{itemize}
In our subsequent analyses we utilize the economic calendar sourced from \cite{fxstreet} of high impact events and filter for only those events pertaining to the aforementioned currencies (and associated economies) and economic announcements specifically. The exact strings used for filtering the events based on event name are: "ECB Press Conference", "BoE MPC", "FOMC Press Conference", "BoJ Press Conference", "PBoC Interest Rate Decision", "Nonfarm Payrolls", "Consumer Price Index ex Food \& Energy", "European Council Meeting", "EU Leaders Special Summit" and "ECB Special Strategy Meeting". During the period 2020-04-01 to 2022-01-31 one or more of these high impact events occurred on approximately 27$\%$ of business days. 
\begin{figure}[t]
\centering
\includegraphics[width=120mm]{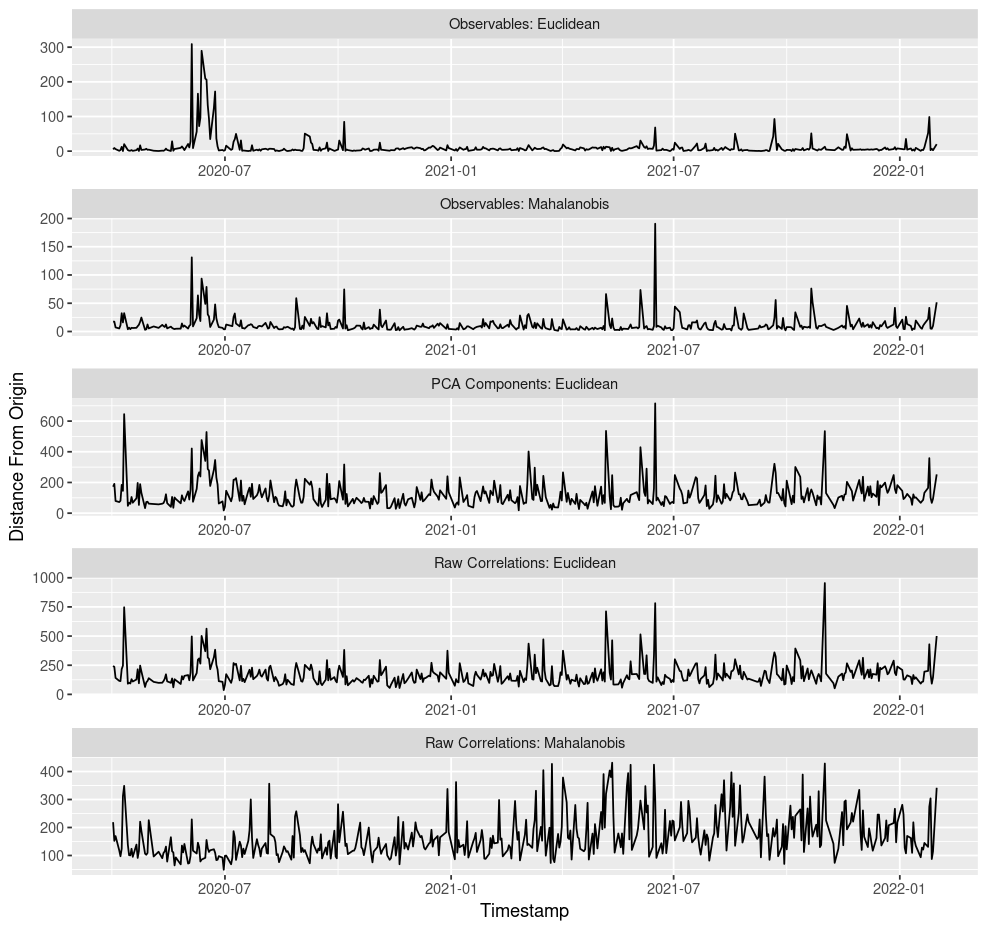}
\caption{\label{fig:distances} Distances of observable vectors and raw correlation vectors from the origin using the standardized Euclidean and Mahalanobis metrics.}
\end{figure}

\subsubsection{Dimensionality reduction} \label{sec: dim red}

The construction of lower-dimensional representations of the correlation matrices - namely the observable vectors - is effectively a dimensionality reduction procedure. As is common with such procedures (e.g. Principal Component Analysis (PCA) ), there is a trade-off between reducing dimensionality and preserving information content. Balancing these trade-offs through a good choice of the number of components often leads to better results in various applications. In PCA, the cumulative variance of the first principal components is typically used as an organizing quantity to select how many such components to include. In our case, we take the normalized magnitude of the differences between the empirical expectation values of the cubic and quartic observables and the theoretical predictions of the PIGM model \eqref{eqn: similarity}, as an organizing quantity for determining which observables to retain. The thesis is that the empirical higher order observables that depart from theoretical expectations indicate additional information beyond the linear and quadratic structure encoded into the PIGM model. We have empirically determined that the 12 "least Gaussian" observables yield optimal anomaly detection results (i.e. statistical significance and odds-ratios). Notably, the results broadly improve as more observables are added starting from a small number of observables, reach a peak and then decline somewhat as more observables are added. These "least Gaussian" observables are listed in table \ref{tab:observable_least_gaussian}.
\afterpage{
{\footnotesize
\begin{longtable}[hbt!]{lcc}
\toprule
\textbf{Obervable Label} & \textbf{Observable Def.} & \textbf{Observable Order}  \\
\midrule
$\mathcal{O}_3$ & $\sum_{i,j,k} \hat{\rho}_{ij} \hat{\rho}_{jk} \hat{\rho}_{ki}                 $ &  Cubic \\
$\mathcal{O}_5$ & $\sum_{i,j,k,l} \hat{\rho}_{ij} \hat{\rho}_{jk} \hat{\rho}_{kl}               $ &  Cubic \\
\midrule
$\mathcal{O}_9$ & $\sum_{i,j} \hat{\rho}_{ij}^4                               $ &   Quartic \\
$\mathcal{O}_{10}$ & $\sum_{i,j,k} \hat{\rho}_{ij}^2 \hat{\rho}_{jk}^2                    $ &  Quartic\\
$\mathcal{O}_{12}$ & $\sum_{i,j,k} \hat{\rho}_{ij} \hat{\rho}_{ik} \hat{\rho}_{jk}^2               $ &  Quartic \\
$\mathcal{O}_{16}$ & $\sum_{i,j,k,l} \hat{\rho}_{ij} \hat{\rho}_{jk} \hat{\rho}_{kl}^2             $ &  Quartic \\
$\mathcal{O}_{17}$ & $\sum_{i,j,k,l} \hat{\rho}_{ij} \hat{\rho}_{jk} \hat{\rho}_{ik} \hat{\rho}_{kl}        $ &  Quartic \\
$\mathcal{O}_{19}$ & $\sum_{i,j,k,l} \hat{\rho}_{ij} \hat{\rho}_{jk} \hat{\rho}_{kl} \hat{\rho}_{li}        $ &  Quartic \\
$\mathcal{O}_{21}$ & $\sum_{i,j,k,l, m} \hat{\rho}_{il} \hat{\rho}_{jk} \hat{\rho}_{lk} \hat{\rho}_{mk}     $ &  Quartic\\
$\mathcal{O}_{22}$ & $\sum_{i,j,k,l, m} \hat{\rho}_{ij} \hat{\rho}_{kl} \hat{\rho}_{lm} \hat{\rho}_{mk}     $ &  Quartic \\
$\mathcal{O}_{25}$ & $\sum_{i,j,k,l,m} \hat{\rho}_{ij} \hat{\rho}_{jk} \hat{\rho}_{kl} \hat{\rho}_{lm}      $ &  Quartic \\
$\mathcal{O}_{29}$ & $\sum_{i,j,k,l,m,n} \hat{\rho}_{ij} \hat{\rho}_{kl} \hat{\rho}_{lm} \hat{\rho}_{mn}    $ &  Quartic \\
\bottomrule\\
\caption{The 12 cubic and quartic observables that have the largest normalized difference from the PIGM predictions.}\label{tab:observable_least_gaussian}
\end{longtable}
}
}
We have also observed that useful information remains in the other observables however. The cubic and quartic observables that are best predicted by the PIGM model still have reasonable effectiveness in anomaly detection for example, as do random subsets of observables and the complete set of observables. This aligns with our overarching findings that the PIGM model is a good fit overall and captures meaningful statistical structure. There does appear to be additional information captured in the least well fit observables however as supported by the results below.

\subsubsection{Longest observable vectors and economically significant dates}

We assess how strongly the lengths of the observable vectors constructed from the observables in table \ref{tab:observable_least_gaussian} are associated with the presence or absence of  economically significant events. To investigate the utility of the observable representation, we also compare to the results obtained for the original correlation matrices, applying both standardized Euclidean and Mahalanobis distance measures. Finally, we compare the observable representation to a representation obtained by applying PCA to the original correlation matrices. In particular, we select the smallest number of principal components that captures at least 70$\%$ of the variance, corresponding to the first 10 principal components in this case (this value also matches the number of components to retain as determined by the elbow method  \cite{jolliffe2002principal}). Only the standardized Euclidean metric is applied to the PCA vector since the principal components are uncorrelated and thus the Mahalanobis distance yields equivalent results. The methodology is as follows.

\begin{enumerate}
	\item Calculate the standardized Euclidean and Mahalanobis vector lengths for the observable vector associated with each date in the dataset (representing distance from the mean observable vector or equivalently the origin in this case). The maximal dimension of the vector space  considered here is $N = 31$ while the optimal number of least Gaussian observables, as stated earlier, is $12$. 
	
	\item Calculate the standardized Euclidean and Mahalanobis vector lengths for the correlation feature vector associated with each date in the dataset. The correlation feature vector for each date is comprised of the 171 pairwise correlations calculated between all 19 currency pairs. We term these features raw correlations.
	\item Calculate the standardized Euclidean vector lengths for the PCA feature vector associated with each date in the dataset.
	\item Rank the dates in the dataset by Euclidean and Mahalanobis vector length, in descending order for the observable vectors, raw correlation and PCA feature vectors.
	\item Assess whether the top 25, 50 and 100 dates have a statistically significantly higher number of economically significant events than the bottom 25, 50 and 100 dates ordered by distance in a descending manner. In addition, we calculate the ratio between the odds of observing a economically significant news event in the top/most anomalous 25, 50, 100 dates and the odds of such an event occurring in the bottom/most typical 25, 50, 100 dates. This odds-ratio (OR) is defined as,
\begin{equation}
	\mbox{OR}=\frac{P_{T}/(1-P_{T})}{P_{B}/(1-P_{B})},\label{odds_ratio}
\end{equation}
where $P_{B}$ corresponds to the proportion of the 25, 50, 100 closest dates to the origin that are associated with economically significant events. Similarly, $P_{T}$ corresponds to the proportion of the 25, 50, 100 furthest dates from the origin that are associated with economically significant events.
\end{enumerate}
The distances for the respective metrics and features are presented in figure \ref{fig:distances}.
Notably, the observable feature vectors appear to have more distinct outlier days and less noise. In addition, we note that the PCA vector lengths yield a fairly similar pattern to the observable vector lengths with the Mahalanobis distance.  The results of comparing the top 25, 50, 100 dates by distance from the origin with the bottom 25, 50, 100 dates respectively are collected in table \ref{tab:hypothesis_tests}. The best contrast of the number of economic events appearing in the furthest days from the origin compared to the closest days from the origin respectively is given by the Mahalanobis distance evaluated on observable vectors. Indeed, for the Mahalanobis distance on observable vectors, there is a higher degree of statistical significance and higher odds-ratios than the other combinations of metric and features in almost all cases. The high odds-ratios imply that the odds of a economically significant event occurring in the anomalous groups (most distant) are much higher than the odds in the typical groups (least distant). This provides evidence that the observable vectors are a good, low-dimensional characterization of the market state and accentuate meaningful financial "signal". 

An additional note relates to the correlation matrices and associated observable vectors for February 2022. During this period, there were several extremely anomalous dates (with extreme vector lengths), coinciding with the beginning of the war in Ukraine. These had the effect of masking the anomalous nature of earlier events and reducing the sensitivity of the detection algorithm. This is a well known consequence of applying the Mahalanobis distance to anomaly detection, known as the masking effect \cite{ghorbani2019mahalanobis}. We therefore excluded February 2022 from all our analyses. The analysis conducted thus far can be regarded as pertaining to in-sample anomaly detection. We also conducted an out-of-sample analysis using the same observables and number of principal components for PCA as the in-sample analysis, now for the date range 2022-03-01 to 2023-03-31. The results are collected in table  \ref{tab:hypothesis_tests_out_sample} and reveal that the in-sample anomaly detection results generalize well, and the Mahalanobis distance calculated on observable vectors continues to out-perform the other alternatives in the majority of cases. Two other robustness checks that were conducted include rerunning the analysis with correlation matrices constructed with 10 and 15 minute sampling intervals for the log returns (as opposed to 5 minutes) as well as testing subsets of the most important economic announcements. The results of these analyses were qualitatively similar to those already presented.       

{\footnotesize
\begin{longtable}[t]{llrrrrr}
\hline
Metric & Features & Subset Size & $P_{T}$ & $P_{B}$ & p-value & Odds-Ratio\\
\hline
Euclidean & Observables & 25 & 0.40 & 0.20 & 10.8$\times10^{-2}$ & 2.67\\
Euclidean & Observables & 50 & 0.38 & 0.24 & 9.7$\times10^{-2}$ & 1.94\\
Euclidean & Observables & 100 & 0.38 & 0.28 & 8.8$\times10^{-2}$ & 1.58\\
\hline
Mahalanobis & Observables & 25 & 0.44 & 0.16 & 3.1$\times10^{-2}$* & 4.12\\
Mahalanobis & Observables & 50 & 0.38 & 0.16 & 1.2$\times10^{-2}$* & 3.22\\
Mahalanobis & Observables & 100 & 0.39 & 0.13 & 0.0$\times10^{-2}$*** & 4.28\\
\hline
Euclidean & PCA Correlations & 25 & 0.60 & 0.32 & 4.4$\times10^{-2}$* & 3.19\\
Euclidean & PCA Correlations & 50 & 0.46 & 0.24 & 1.8$\times10^{-2}$* & 2.70\\
Euclidean & PCA Correlations & 100 & 0.41 & 0.23 & 0.5$\times10^{-2}$** & 2.33\\
\hline
Euclidean & Raw Correlations & 25 & 0.52 & 0.32 & 12.6$\times10^{-2}$ & 2.30\\
Euclidean & Raw Correlations & 50 & 0.44 & 0.28 & 7.2$\times10^{-2}$ & 2.02\\
Euclidean & Raw Correlations & 100 & 0.39 & 0.22 & 0.7$\times10^{-2}$** & 2.27\\
\hline
Mahalanobis & Raw Correlations & 25 & 0.24 & 0.16 & 36.3$\times10^{-2}$ & 1.66\\
Mahalanobis & Raw Correlations & 50 & 0.32 & 0.14 & 2.8$\times10^{-2}$* & 2.89\\
Mahalanobis & Raw Correlations & 100 & 0.28 & 0.20 & 12.3$\times10^{-2}$ & 1.56\\
\hline
\caption{In-sample anomaly detection results. In the table above, the proportions, $P_{B}, P_{T}$ and the odds-ratio, OR, are as defined in equation \eqref{odds_ratio}. The p-value is obtained using Fisher's exact one-sided test. The * symbol following a p-value indicates significance at the 0.05 level, ** indicates significance at the 0.01 level and *** indicates significance at the 0.001 level. }\label{tab:hypothesis_tests}
\end{longtable}
}

{\footnotesize
\begin{longtable}[t]{llrrrrr}
\hline
Metric & Features & Subset Size & $P_{T}$ & $P_{B}$ & p-value & Odds-Ratio\\
\hline
Euclidean & Observables & 25 & 0.44 & 0.20 & 6.4$\times10^{-2}$ & 3.14\\
Euclidean & Observables & 50 & 0.44 & 0.24 & 2.8$\times10^{-2}$* & 2.49\\
Euclidean & Observables & 100 & 0.29 & 0.23 & 21.0$\times10^{-2}$ & 1.37\\
\hline
Mahalanobis & Observables & 25 & 0.56 & 0.16 & 0.4$\times10^{-2}$** & 6.68\\
Mahalanobis & Observables & 50 & 0.50 & 0.10 & 0.0$\times10^{-2}$*** & 9.00\\
Mahalanobis & Observables & 100 & 0.37 & 0.12 & 0.0$\times10^{-2}$*** & 4.31\\
\hline
Euclidean & PCA Correlations & 25 & 0.48 & 0.20 & 3.6$\times10^{-2}$* & 3.69\\
Euclidean & PCA Correlations & 50 & 0.48 & 0.14 & 0.0$\times10^{-2}$*** & 5.67\\
Euclidean & PCA Correlations & 100 & 0.42 & 0.13 & 0.0$\times10^{-2}$*** & 4.85\\
\hline
Euclidean & Raw Correlations & 25 & 0.48 & 0.20 & 3.6$\times10^{-2}$* & 3.69\\
Euclidean & Raw Correlations & 50 & 0.48 & 0.16 & 0.1$\times10^{-2}$*** & 4.85\\
Euclidean & Raw Correlations & 100 & 0.42 & 0.14 & 0.0$\times10^{-2}$*** & 4.45\\
\hline
Mahalanobis & Raw Correlations & 25 & 0.28 & 0.16 & 24.8$\times10^{-2}$ & 2.04\\
Mahalanobis & Raw Correlations & 50 & 0.36 & 0.20 & 5.9$\times10^{-2}$ & 2.25\\
Mahalanobis & Raw Correlations & 100 & 0.31 & 0.25 & 21.6$\times10^{-2}$ & 1.35\\
\hline
\caption{Out-of-sample anomaly detection results. In the table above, the proportions, $P_{B}, P_{T}$ and the odds-ratio, OR, are as defined in equation \eqref{odds_ratio}. The p-value is obtained using Fisher's exact one-sided test. The * symbol following a p-value indicates significance at the 0.05 level, ** indicates significance at the 0.01 level and *** indicates significance at the 0.001 level. }\label{tab:hypothesis_tests_out_sample}
\end{longtable}
}

\subsection{Visual similarity} \label{sec: similarity}

\begin{figure}[ht!]
\centering
\includegraphics[width=110mm]{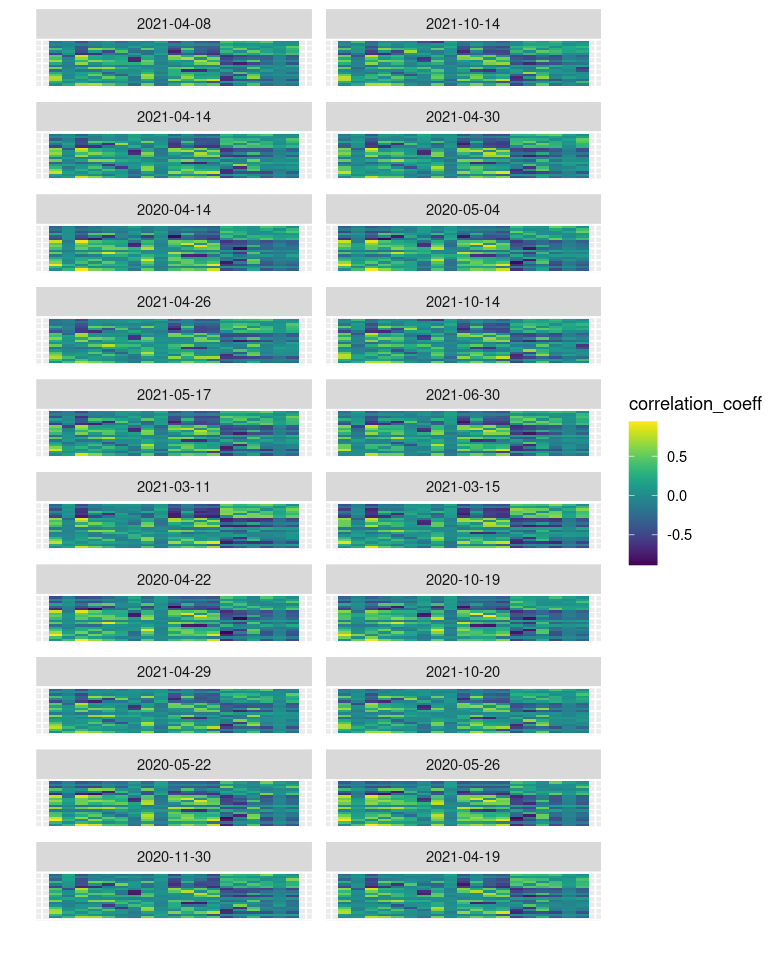}
\caption{\label{fig:closest_dates} Closest 10  pairs of correlation matrices, by Mahalanobis distance on observable vectors, where each row of the figure corresponds to a pair of correlation matrices that were compared.}
\end{figure}

\begin{figure}[ht!]
\centering
\includegraphics[width=110mm]{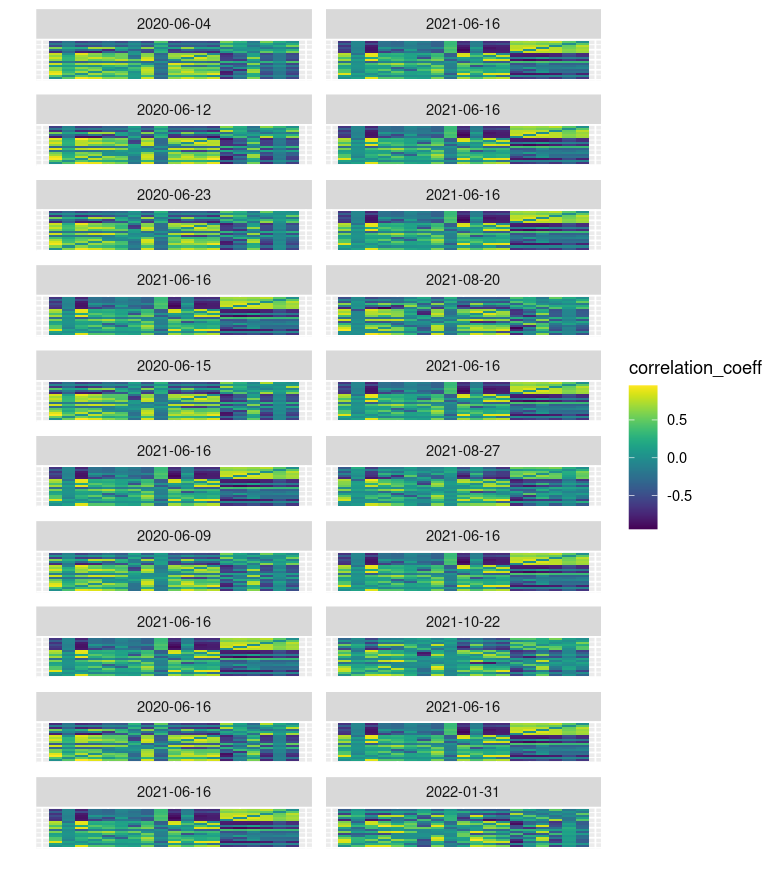}
\caption{\label{fig:furthest_dates} Furthest 10 correlation matrices pairs, by Mahalanobis distance on observable vectors, where each row of the figure corresponds to a pair of correlation matrices that were compared.}
\end{figure}

Equipped with the observable vector representation of correlation matrices and the Mahalanobis distance metric, another interesting investigation is to assess how visually similar the correlation matrices are that correspond to the pairs of dates with the closest observable vectors and contrast this with the furthest pairs. The procedure involves calculating the Mahalanobis distance between each pair of dates, of which there are $446(445)/2 = 99,235$ such pairs, and then sorting by distance in ascending order. The closest 10 pairs are visualized in figure \ref{fig:closest_dates}. The most distant 10 pairs are visualized in figure \ref{fig:furthest_dates}.
It is readily apparent that the proximity of the dates in the space of observables aligns quite closely with the visual similarity or dissimilarity of the correlation matrices.

Finally, we quantitatively assess whether there is a concordance between the closeness of the visual representations of correlation matrices with the closeness of the observable vectors. In particular, we determine the Spearman correlation between the distance ranking of pairs of dates as determined in two ways,
\begin{enumerate}
	\item Using the Euclidean distance to measure distance between pairs of correlation matrices as a proxy for visual similarity (this is essentially the widely used mean squared error metric for comparing image similarity, known to be closely related to perceived visual similarity).
    \item Using the Mahalanobis distance between pairs of observable vectors. 
\end{enumerate}
The result is a Spearman correlation of approximately 0.7 across 99,235 pairs of dates. Spearman correlation coefficients of between 0.7 and 1.0 are regarded as very strongly positive, with 1.0 representing a perfect monotonic relationship. Using the exact Spearman correlation test, the p-value is smaller than $2.2 \times 10^{-16}$ and the hypothesis of the Spearman correlation being zero can be strongly rejected in favor of the alternate hypothesis that the Spearman correlation is positive. This implies that despite the much lower dimensionality of the observable vectors, they do indeed capture the essence of what we would regard as visually similar correlation matrices. 

\section{Discussion and conclusion}
\label{sec: conclusion}

We have developed the most general 4-parameter permutation invariant Gaussian matrix models appropriate for ensembles of matrices which are symmetric and  diagonally vanishing, as appropriate for financial correlation matrices. We have used the models to find near-Gaussianity in ensembles of matrices, one matrix for every day over a period, constructed from high-frequency foreign exchange price quotes. The near-Gaussianity was found to be robust against changes in how the ensemble was constructed : we varied the time intervals between the quote updates used to construct the daily averages as well as the number of days used in our ensemble. The near-Gaussianity is used to motivate a data-reduction technique based on the use of low degree permutation invariant functions of matrices (observables)  as characteristics of the entities represented by the matrices in the ensemble, in  this case the days in the period under consideration. The small non-Gaussianities of each  observable were used to rank the observables in order of decreasing non-Gaussianity and to  find an optimal number of least Gaussian observables for data analysis. The degree of non-Gaussianity is thus being used as an analog of
the magnitude of singular values in principal component analysis (PCA). The sets of observables considered, either the full set of observables up to quartic degree or the subsets with optimal number of least Gaussian observables,  are much smaller than the number of matrix elements in the matrices. We found successful results in anomaly detection based on the observables to find the most atypical and the most typical days in the ensemble. We demonstrated statistically significant matching between these typicality/atypicality results extracted from the data of financial correlation matrices and corresponding results based on human economic judgement of significant events affecting foreign exchange markets. We propose that the success of the use of a set of  least Gaussian observables in anomaly detection should be interpreted as indicating that these ensembles of daily foreign exchange matrices capture an economic reality best described by the Gaussian model perturbed by specific small cubic and quartic couplings in the action. The non-Gaussianities capture system-specific non-universalities while the overall approximate Gaussianity is a universal characteristic which holds across diverse systems, as indeed already evidenced in ensembles of words 
\cite{Ramgoolam:2019ldg, Huber:2022ohf}.  Related discussions of Gaussianity, universality and non-universality appear in random matrix theory literature  \cite{Edelman_2016,Luo2007,tao2011random}. 

Future directions include evaluating  how the parameters of the PIGM model change as the details of the correlation calculation methodology change. This should provide insight into the sensitivity of these parameters to different methodologies. The choice of correlation calculation methodology affects effective sample size along with statistical efficiency and  susceptibility of correlation estimates to market microstructure noise for example. Another interesting avenue of exploration related to statistical methodology is the investigation of the impact of using maximum likelihood estimation in place of the method of moments approach that was utilised in fitting the PIGM model in this paper. While we expect the approaches to yield qualitatively similar results, there could be subtle differences.  

A distinct research direction involves studying the effect of different choices of time-scale for constructing the ensemble of correlation matrices, e.g. one hour, twelve hours, two days etc., on the linear and quadratic parameters. In addition, it should be illuminating to investigate  whether the agreement of empirical observables with PIGM predictions (fit on linear and quadratic observables only) changes with correlation time-scale. Indeed, the evolution of the magnitude of cubic and quartic corrections with time-scale should assist in answering the question of whether the non-Gaussianity increases or decreases with time-scale. This should also help to disentangle the roles of  finite sample effects in the observed non-Gaussinaities. 

There are also several practical directions to pursue with particular relevance to finance. The first involves clustering the correlation matrices in observable vector representation and elucidating how the cluster structure changes with correlation time-scale. This should provide insight into the states of the market at different time-scales and how they relate to each other (along the lines of \cite{marsili2002dissecting} and \cite{hendricks2016detecting} for example). Modelling transitions between market states is a related application. A further area of interest is to study the effectiveness of so-called nowcasting of market state, where the market state is identified as it is formed. These applications could have great utility in risk management for example. Assessing the effectiveness of the PIGM observables as features in machine learning algorithms for predicting quantities such as future price returns would inform viability in trading strategy applications.

Finally, generating realistic, random samples of financial correlation matrices has been recognized as having value in a number of contexts including enhancing trading strategies and stress testing portfolios of assets \cite{marti2020corrgan}. We are presently investigating drawing samples from the PIGM model. In this case we do not expect sampled matrices to be positive-semidefinite in general, where positive-semidefiniteness is a property of all valid correlation matrices. However this is readily addressed using approaches such as those in \cite{higham2002computing}. We are also contrasting to existing approaches \cite{marti2020corrgan}. It will be fascinating to determine whether there is a trade-off between the exceptional parsimony of the PIGM model (i.e. 4 parameters) and the ability to capture the known stylized facts of financial correlation matrices - particularly in contrast to the approach in \cite{marti2020corrgan} which utilizes a generative adversarial neural network with several orders of magnitude more parameters than the PIGM model. 

\section{Statement and acknowledgements}

The views expressed in this article are those of the authors and do not necessarily reflect the views of Rand Merchant Bank. Rand Merchant Bank does not make any representations or give any warranties as to the correctness, accuracy or completeness of the information presented; nor does Rand Merchant Bank assume liability for any losses arising from errors or omissions in the information in this article. We would like to thank Steve Abel, Manuel Accettuli Huber, Stephon Alexander, Adrian Bevan,  Graham Brown, Adrian Padellaro, Mehrnoosh Sadrzadeh, Alex Stapleton and Steve Thomas for useful  discussions on this project. SR is supported by the STFC consolidated grant ST/P000754/1 “String
Theory, Gauge Theory and Duality”. S.R. acknowledges the support of the Institut Henri
Poincar\'e (UAR 839 CNRS-Sorbonne Universit\'e), and LabEx CARMIN (ANR-10-LABX-59-
01). SR also acknowledges the support of the Perimeter Institute for Theoretical Physics
during a visit in the final stages of completion of the paper. Research at Perimeter Institute
is supported by the Government of Canada through Industry Canada and by the Province of
Ontario through the Ministry of Economic Development and Innovation. SR also acknowledges the hospitality of the high energy physics group of Brown University during the completion of this project.

\appendix

\section{Currency Pair Mapping}

{\footnotesize
\begin{longtable}{rl}
\toprule
Index (I) & Currency Pair\\
\midrule
1 & AUD/JPY\\
2 & AUD/NZD\\
3 & AUD/USD\\
4 & CAD/JPY\\
5 & CHF/JPY\\
6 & EUR/CHF\\
7 & EUR/GBP\\
8 & EUR/JPY\\
9 & EUR/PLN\\
10 & EUR/USD\\
11 & GBP/JPY\\
12 & GBP/USD\\
13 & NZD/USD\\
14 & USD/CAD\\
15 & USD/CHF\\
16 & USD/JPY\\
17 & USD/MXN\\
18 & USD/TRY\\
19 & USD/ZAR\\
\bottomrule\\
\caption{Currency pair mapping.}\label{tab:currency_pairs}
\end{longtable}
}

\section{Inner product calculations} \label{app: inner prods}

We present here some inner products which are useful in arriving at equations for the physical variables 
$S^{\text{phys}; V_0}$ and $S^{\text{phys}; V_H}_a$  obtained in \eqref{eq: physical var in terms of M 1}  and \eqref{eq: physical var in terms of M 2} of section \ref{sec:PIGMMSDV}. 

Using the inner product on the orthonormal basis of the natural representation
\begin{align}
(e_i, e_j) = \delta_{ij}.
\end{align}
We calculate the following inner products used to determine the Clebsch coefficients from $V_D \otimes V_D$ to the physical $V_0$ and $V_H$ irreducible representations
\begin{align}
(S^{\text{diag}; V_0}, S^{V_{0}; 1}), \quad (S^{\text{diag}; V_0}, S^{V_{0}; 2}), \quad (S^{\text{diag}; V_H}_{a_1}, S^{V_{H}; 1,2}_{a_2}), \quad (S^{\text{diag}; V_H}_{a_1}, S^{V_{H}; 3}_{a_2}).
\end{align} 
The normalised representation theory states are given by
\begin{align}
S^{V_{0}; 1} &= \frac{1}{D} \sum_{i,j = 1}^D e_i \otimes e_j, \\ 
S^{V_{0}; 2} &= \frac{1}{\sqrt{D-1}} \sum_{a=1}^{D-1} E_a \otimes E_a, \\
S^{\text{diag}; V_0} &= \frac{1}{\sqrt{D}} \sum_{i=1}^D e_i \otimes e_i, \\
S^{V_{H}; 1,2}_a &= \frac{1}{\sqrt{2D}} \sum_{i=1}^D \big( e_i \otimes E_a + E_a \otimes e_i \big), \\
S^{V_{H}; 3}_a &= \frac{1}{2} \sqrt{\frac{D}{D-2}} \sum_{b,c =1}^{D-1} \sum_{i=1}^D C_{a,i} C_{b,i} C_{c,i} \big( E_b \otimes E_c + E_c \otimes E_b \big), \\
S^{\text{diag}; V_H}_a &= E_a \otimes E_a.
\end{align}
The non-zero inner products are given by the following
\begin{align} \nonumber 
(S^{\text{diag}; V_0}, S^{V_{0}; 1}) &= D^{- \frac{3}{2}} \sum_{i,j,k = 1}^D (e_i \otimes e_i ,  e_j \otimes e_k) = D^{- \frac{3}{2}} \sum_{i,j,k = 1}^D \delta_{ij} \delta_{ik} = \sum_{i = 1}^D D^{- \frac{3}{2}} \\
&= \frac{1}{\sqrt{D}},
\end{align}
\begin{align} \nonumber
(S^{\text{diag}; V_0} , S^{V_{0}; 2}) &= \frac{1}{\sqrt{D}\sqrt{D-1}} \sum_{i = 1}^D \sum_{a=1}^{D-1} (e_i \otimes e_i , E_a \otimes E_a) \\ \nonumber
&= \frac{1}{\sqrt{D}\sqrt{D-1}} \sum_{i,j,k = 1}^D \sum_{a=1}^{D-1} C_{a,j} C_{a,k} (e_i \otimes e_i , e_j \otimes e_k) \\ \nonumber
&= \frac{1}{\sqrt{D}\sqrt{D-1}} \sum_{i,j,k = 1}^D \sum_{a=1}^{D-1} C_{a,j} C_{a,k} \delta_{ij} \delta_{ik} \\ \nonumber
&= \frac{1}{\sqrt{D}\sqrt{D-1}} \sum_{j = 1}^D \sum_{a=1}^{D-1} C_{a,j} C_{a,j} \\ \nonumber
&=  \sum_{a=1}^{D-1} \frac{1}{\sqrt{D}\sqrt{D-1}} \\
&= \frac{\sqrt{D-1}}{\sqrt{D}},
\end{align}
\begin{align} \nonumber
(S^{\text{diag}; V_H}_{a_1}, S^{V_{H}; 1,2}_{a_2}) &=  \frac{1}{\sqrt{2D}} \sum_{i,j,k=1}^D C_{a_1, j} C_{a_2, k}  \big( e_j \otimes e_j , ( e_i \otimes e_k + e_k \otimes e_i) \big) \\ \nonumber
&= \frac{1}{\sqrt{2D}} \sum_{i,j,k=1}^D C_{a_1, j} C_{a_2, k} \big( \delta_{ij} \delta_{jk} + \delta_{jk} \delta_{ij} \big) \\ \nonumber
&= \sqrt{\frac{2}{D}} \sum_{i,j=1}^D C_{a_1, j} C_{a_2, j} \delta_{ij} \\ 
&= \sqrt{\frac{2}{D}} \delta_{a_1 a_2},
\end{align}
\begin{align} \nonumber
(S^{\text{diag}; V_H}_{a_1}, S^{V_{H}; 3}_{a_2} ) &=  \frac{1}{2} \sqrt{\frac{D}{D-2}} \sum_{b,c =1}^{D-1} \sum_{i,j,k,l=1}^D C_{a_1,i} C_{b,i} C_{c,i} C_{a_2, j} C_{b, k} C_{c, l} \big( e_j \otimes e_j , ( e_k \otimes e_l + e_l \otimes e_k ) \big) \\ \nonumber
&=  \sqrt{\frac{D}{D-2}} \sum_{b,c =1}^{D-1} \sum_{i,j,k,l=1}^D C_{a_1,i} C_{b,i} C_{c,i} C_{a_2, j} C_{b, k} C_{c, l}  \delta_{jk} \delta_{jl}  \\ \nonumber
&=  \sqrt{\frac{D}{D-2}} \sum_{b,c =1}^{D-1} \sum_{i,j=1}^D C_{a_1,i} C_{b,i} C_{c,i} C_{a_2, j} C_{b, j} C_{c, j}  \\ \nonumber
&=  \sqrt{\frac{D}{D-2}} \sum_{i,j=1}^D C_{a_1,i} C_{a_2, j} \Big(\delta_{ij} - \frac{1}{D} \Big) \Big(\delta_{ij} - \frac{1}{D} \Big) \\ \nonumber
&=  \sqrt{\frac{D}{D-2}} \sum_{i,j=1}^D C_{a_1,i} C_{a_2, j} \Big(\delta_{ij} \Big( 1 - \frac{2}{D} \Big) + \frac{1}{D^2} \Big) \\ \nonumber
&=  \sqrt{\frac{D}{D-2}} \Bigg[ \sum_{i=1}^D C_{a_1,i} C_{a_2, i} \Big( 1 - \frac{2}{D} \Big) + \frac{1}{D^2}  \sum_{i,j=1}^D C_{a_1,i} C_{a_2, j} \Bigg] \\
&=  \sqrt{\frac{D}{D-2}} \Big( 1 - \frac{2}{D} \Big) \delta_{a_1, a_2}.
\end{align}

\section{Expectation value calculations}
The calculation of analytic expressions in $D$ for observable expectation values is a three step process:
\begin{enumerate}
\item{Apply Wick's theorem on the observable expectation value to produce an expression consisting of a product of one and two-point functions.}
\item{Substitute in the expressions \eqref{eq: one point function} and \eqref{eq: two point function} for the one and two-point functions respectively and sum over all unique indices in the observable.}
\item{Evaluate the sums over products of $V_H$ projectors, $F$, defined by \eqref{eq: F definition}.}
\end{enumerate} 
The three subsections below explain the Python code that performs each of the three steps listed above.

\subsection{Apply Wick's theorem}
Each graph basis observable of order $k$ is written as a two-dimensional array of length $k$ and width 2, each sub-array represents a matrix appearing in the observable and contains the indices appearing on that matrix. For example,
\begin{align} \nonumber
\cO_1 &= \sum_{i_1, i_2 = 1}^D M_{i_1 i_2} M_{i_1 i_2} M_{i_1 i_2} \\ \nonumber
\cO_1 &= \sum_{i_1, i_2, i_3 = 1}^D M_{i_1 i_2} M_{i_1 i_2} M_{i_2 i_3} \\
\cO_1 &= \sum_{i_1, i_2 , i_3 = 1}^D M_{i_1 i_2} M_{i_2 i_3} M_{i_3 i_1} 
\end{align}
are represented by the following 2d arrays:
\begin{figure}[H]
\begin{flushleft}
\includegraphics[width=160mm]{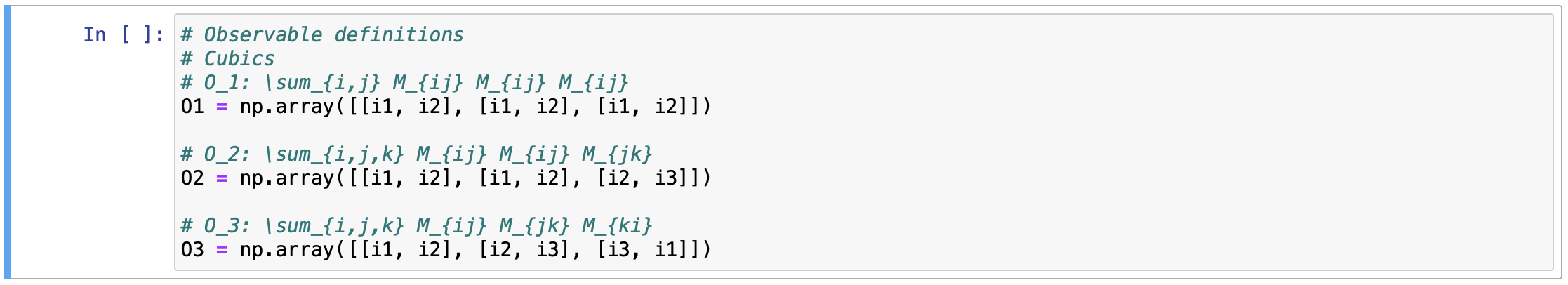}
\end{flushleft}
\end{figure}

In this form the observables are fed into a function $all\_contr$ that produces a list of all Wick contractions of the fields contained within the observable.
\begin{figure}[H]
\begin{flushleft}
\includegraphics[width=160mm]{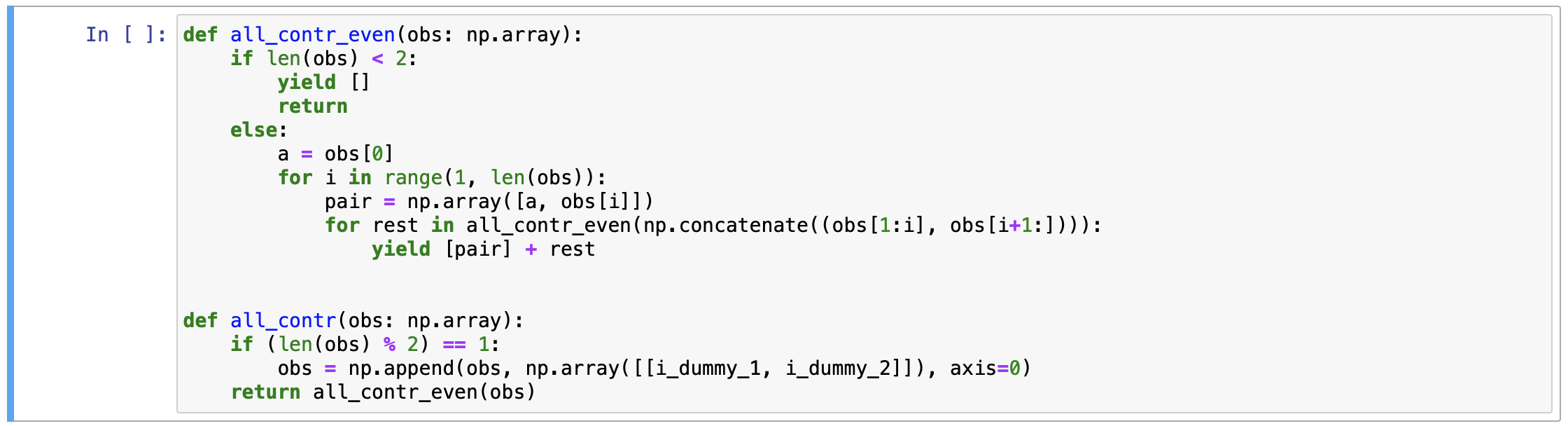}
\end{flushleft}
\end{figure}
\noindent This function operates recursively. It pairs two elements of the original observable before feeding the remaining non-paired elements back to itself, producing a further pair, and so on. Once there are no more elements to pair a different initial pairing is taken and this process is repeated. 

\subsection{Substitute one and two-point expressions}
We define a one-point function given by \eqref{eq: one point function}
\begin{figure}[H]
\begin{flushleft}
\includegraphics[width=160mm]{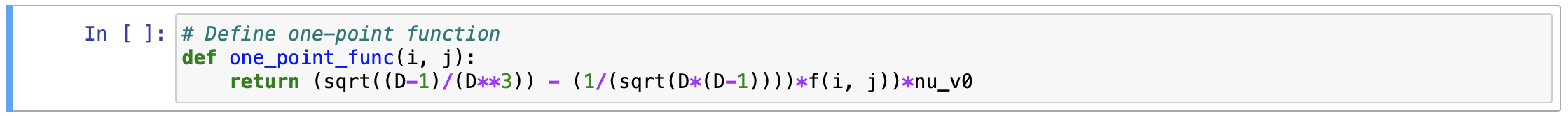}
\end{flushleft}
\end{figure}
\noindent and a two-point function given by \eqref{eq: two point function} 
\begin{figure}[H]
\begin{flushleft}
\includegraphics[width=160mm]{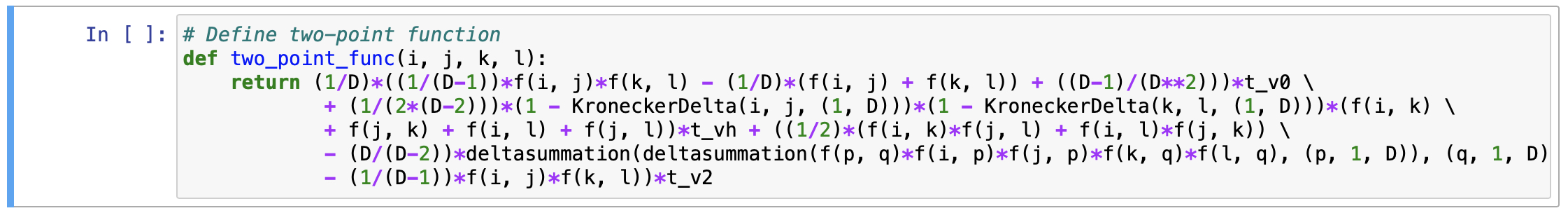}
\end{flushleft}
\end{figure}
\noindent where "$nu\_v0$", "$t\_v0$", "$t\_vh$" and "$t\_v2$" are the inverse couplings $\widetilde{\mu}^2_{V_0}$, $\tau_{V_0}^{-1}$, $\tau_{V_H}^{-1}$ and $\tau_{V_2}^{-1}$ respectively. Each Wick contracted expression produced by $all\_contr$ is evaluated using the above functions.

\subsection{Evaluate sums of $F$ products}
After substituting the one and two-point functions \eqref{eq: one point function} and \eqref{eq: two point function} into the Wick expanded expression of the observable expectation value we are in general left with some complicated sum over many products of $F$'s. These can be evaluated by first substituting the definition
\begin{align}
F(i,j) = \delta_{ij} - \frac{1}{D}
\end{align}
and then using the functions KroneckerDelta and deltasummation within the python library SymPy. These functions allow for the definition of Kronecker deltas with symbolic indicies taking a range of integer values, and the simplification of products of deltas when summed over the relevant range respectively. For example the following code demonstrates some basic properties of the Kronecker
\begin{figure}[H]
\begin{flushleft}
\includegraphics[width=160mm]{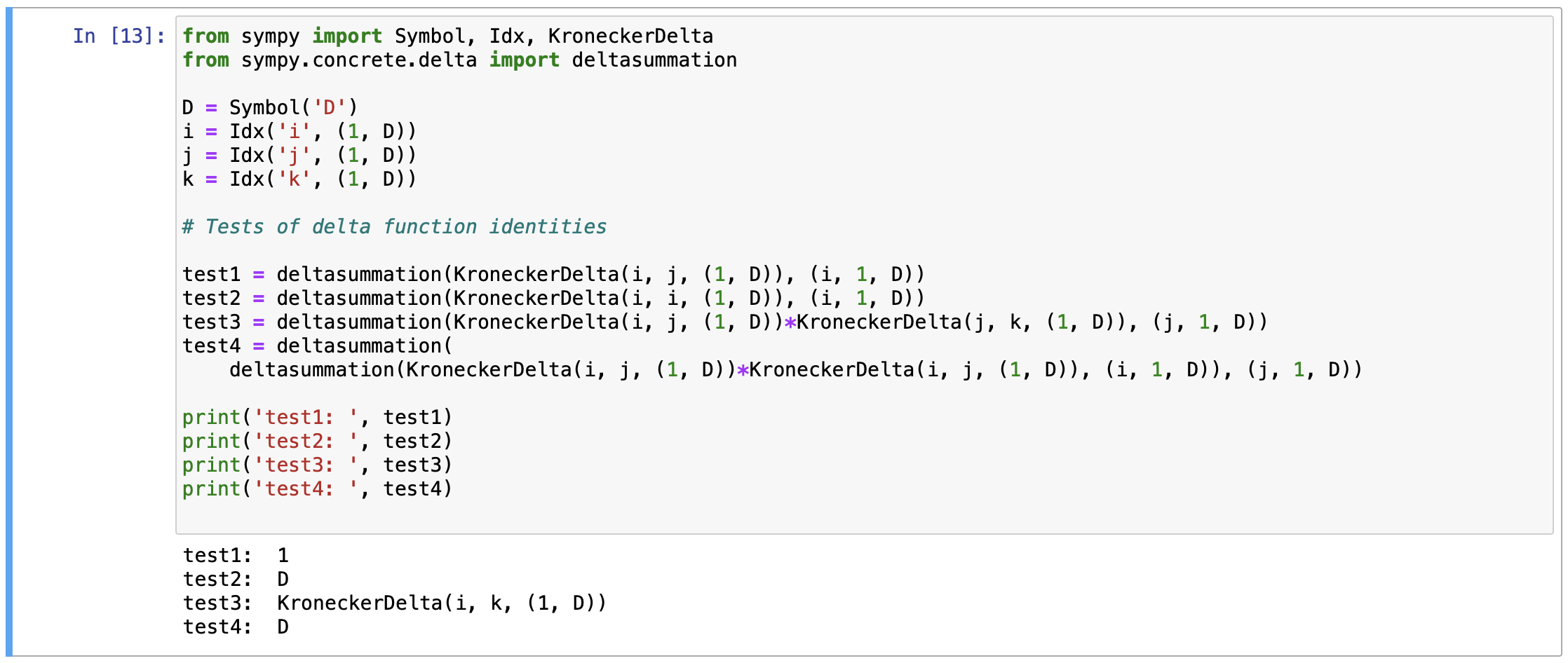}
\end{flushleft}
\end{figure}
\noindent This allows us to reduce the output we were left with at the end of the previous subsection to an analytic expression in $D$ - all of the dependence on the matrix indices having been summed out.

\subsection{Cubic expectation values}
Below we list analytic expressions in $D$ for all cubic expectation values:
\begin{align} \label{eq: cubic exp vals} \nonumber
\sum_{i,j} \langle &M^{\text{phys}}_{ij} M^{\text{phys}}_{ij} M^{\text{phys}}_{ij} \rangle \\ 
&= \frac{\widetilde{\mu}_{V_0}}{2 \sqrt{D(D-1)}} \big( 2 \widetilde{\mu}_{V_0}^2 + 6 \tau^{-1}_{V_0} + 6(D-1) \tau^{-1}_{V_H} + D (D-3) \tau^{-1}_{V_2} \big), \\ \nonumber
\sum_{i,j,k} \langle &M^{\text{phys}}_{ij} M^{\text{phys}}_{ij} M^{\text{phys}}_{jk} \rangle \\
&= \sqrt{\frac{D-1}{D}} \frac{\widetilde{\mu}_{V_0}}{2} \big( 2 \widetilde{\mu}_{V_0}^2 + 6 \tau^{-1}_{V_0} + (4D - 6) \tau^{-1}_{V_H} + D (D-3) \tau^{-1}_{V_2} \big), \\ \nonumber
\sum_{i,j,k} \langle &M^{\text{phys}}_{ij} M^{\text{phys}}_{jk} M^{\text{phys}}_{ki} \rangle = \frac{\widetilde{\mu}_{V_0}}{2D\sqrt{D(D-1)}}  \big( 2D (D-2) \widetilde{\mu}_{V_0}^2 + 6 (D^2 -D -1)\tau^{-1}_{V_0} \\ 
&+ 3(D^3 - 5D^2 + 2D + 2) \tau^{-1}_{V_H} - 3D^2 (D-3) \tau^{-1}_{V_2} \big), \\ \nonumber
\sum_{i,j,k,l} \langle &M^{\text{phys}}_{ij} M^{\text{phys}}_{ij} M^{\text{phys}}_{kl} \rangle \\
&= \sqrt{\frac{(D - 1)}{D}} \frac{\widetilde{\mu}_{V_0}}{2} \big( 2D  \widetilde{\mu}_{V_0}^2 + 6 D\tau^{-1}_{V_0} + 2D(D-1) \tau^{-1}_{V_H} + D^2 (D-3) \tau^{-1}_{V_2} \big), \\ \nonumber
\sum_{i,j,k,l} \langle &M^{\text{phys}}_{ij} M^{\text{phys}}_{jk} M^{\text{phys}}_{kl} \rangle \\
&= \sqrt{\frac{(D - 1)}{D}} \frac{\widetilde{\mu}_{V_0}}{2} \big( 2(D-1)  \widetilde{\mu}_{V_0}^2 + 6 (D-3)\tau^{-1}_{V_0} + (2D-5)(D-1) \tau^{-1}_{V_H} \big), \\ 
\sum_{i,j,k,l} \langle &M^{\text{phys}}_{ij} M^{\text{phys}}_{ik} M^{\text{phys}}_{il} \rangle =  \sqrt{\frac{(D - 1)^3}{D}} \frac{\widetilde{\mu}_{V_0}}{2} \big(2 \widetilde{\mu}^2_{V_0} + 6 \tau^{-1}_{V_0} + 3 (D-2) \tau^{-1}_{V_H} \big), \\
\sum_{i,j,k,l,m} \langle &M^{\text{phys}}_{ij} M^{\text{phys}}_{jk} M^{\text{phys}}_{lm} \rangle = \sqrt{D (D - 1)^3} \frac{\widetilde{\mu}_{V_0}}{2} \big(2 \widetilde{\mu}^2_{V_0} + 6 \tau^{-1}_{V_0} + (D-2) \tau^{-1}_{V_H} \big), \\ 
\sum_{i,j,k,l,m,n} \langle &M^{\text{phys}}_{ij} M^{\text{phys}}_{kl} M^{\text{phys}}_{mn} \rangle = \sqrt{D^3 (D - 1)^3} \widetilde{\mu}_{V_0}  \big(\widetilde{\mu}^2_{V_0} + 3 \tau^{-1}_{V_0} \big).
\end{align}

%

\bibliographystyle{unsrt}
\bibliography{LMTF-Refs}

\end{document}